\documentclass[aps,pra,showpacs,notitlepage,twocolumn,superscriptaddress,nofootinbib]{revtex4-2}

\usepackage{amsmath}
\usepackage{amssymb}
\usepackage{amsthm}
\usepackage{mathtools}
\usepackage{yhmath}
\usepackage{bbold}
\usepackage{bm}
\usepackage{graphicx}
\usepackage{caption}
\usepackage{subcaption}
\usepackage{xcolor}
\usepackage{physics}
\usepackage{chemformula}
\usepackage{tikz}
\usepackage{multirow}
\usepackage{float}
\usepackage{url}

\newtheorem{theorem}{Theorem}[subsection]
\newtheorem{lemma}[theorem]{Lemma}

\newtheorem{innercustomlemma}{Lemma}
\newenvironment{customlemma}[1]{\renewcommand\theinnercustomlemma{#1}\innercustomlemma}{\endinnercustomlemma}

\begin{document}

\title{More quantum chemistry with fewer qubits}

\begin{abstract}
Quantum computation is one of the most promising new paradigms for the simulation of physical systems composed of electrons and atomic nuclei, with applications in chemistry, solid-state physics, materials science, and molecular biology. 
This requires a truncated representation of the electronic structure Hamiltonian using a finite number of orbitals. While it is, in principle, obvious how to improve on the representation by including more orbitals, this is usually unfeasible in practice (e.g., because of the limited number of qubits available) and severely compromises the accuracy of the obtained results. 
Here, we propose a quantum algorithm that improves on the representation of the physical problem by virtue of second-order perturbation theory. In particular, our quantum algorithm evaluates the second-order energy correction through a series of time-evolution steps under the unperturbed Hamiltonian. 
An important application is to go beyond the active-space approximation, allowing to include corrections of virtual orbitals, known as multireference perturbation theory. 
Here, we exploit that the unperturbed Hamiltonian is diagonal for virtual orbitals and show that the number of qubits is independent of the number of virtual orbitals. 
This gives rise to more accurate energy estimates without increasing the number of qubits. 
Moreover, we demonstrate numerically for realistic chemical systems that the total runtime has highly favorable scaling in the number of virtual orbitals compared to previous work. 
Numerical calculations confirm the necessity of the multireference perturbation theory energy corrections to reach accurate ground state energy estimates. 
Our perturbation theory quantum algorithm can also be applied to symmetry-adapted perturbation theory. 
As such, we demonstrate that perturbation theory can help to reduce the quantum hardware requirements for quantum chemistry. 
\end{abstract}

\author{Jakob G\"unther}
\email{jmg@math.ku.dk}
\affiliation{Quantum for Life Center, Department of Mathematical Sciences, University of Copenhagen, Universitetsparken 5, 2100 Copenhagen, Denmark}

\author{Alberto Baiardi}
\email{alberto.baiardi@phys.chem.ethz.ch}
\affiliation{Quantum for Life Center, Department of Chemistry and Applied Biosciences, ETH Zurich, Vladimir-Prelog-Weg 2, 8093 Zurich, Switzerland}

\author{Markus Reiher}
\email{mreiher@ethz.ch}
\affiliation{Quantum for Life Center, Department of Chemistry and Applied Biosciences, ETH Zurich, Vladimir-Prelog-Weg 2, 8093 Zurich, Switzerland}

\author{Matthias Christandl}
\email{christandl@math.ku.dk}
\affiliation{Quantum for Life Center, Department of Mathematical Sciences, University of Copenhagen, Universitetsparken 5, 2100 Copenhagen, Denmark}

\maketitle

\date{\today}

\section{Introduction}
Among the proposed applications of quantum computing, the simulation of physical systems (such as molecules or materials) of increasing size that are unreachable by traditional approaches seems to be one of the most promising ones \cite{AspuruGuzik2011_Review,AspuruGuzik2019_Review,McArdle2020_Review,bauer_quantum_2020,baiardi_quantum_2023,Goings2022_CostEstimate-P450}.
In the context of quantum chemistry, a particularly important task is the delivery of highly accurate ground state energies, as, for example, accurate electronic energies are required to yield a reliable electronic contribution to the free energies of chemical transformations.
On future fault-tolerant quantum computers, quantum phase estimation (QPE) is a rigorous approach to approximate the exact energy of a physical system within a specified error range, provided that the state preparation step can be accomplished with a sufficiently high overlap of the initial state with the target state.
Although it is debatable whether good trial states can be found for all chemically relevant systems \cite{lee_evaluating_2023}, recent complexity theoretic results suggest practical quantum advantage for the ground state energy estimation problem in the presence of a good trial state \cite{gharibianImprovedHardnessResults2022a}.

However, any quantum computer will provide only a limited hardware framework (e.g., in terms of the number of qubits available for the representation of the quantum state) for the encoding of the discretized Hamiltonian that describes the physical system.
Usually, the discretization is considered in occupation number vector space (rarely in real space \cite{chan_grid-based_2023}) and, hence, determined by the number of one-particle functions (e.g., orbitals of electronic structure or modals of vibrational structure) chosen for the construction of the many-particle basis states.
The more orbitals are used to construct the second-quantized Hamiltonian, the more accurate this representation becomes and, hence, the more accurate the approximation of the energy of the physical system will be. 

In view of the finite size of a quantum computer, the number of orbitals that they can represent is likely to be always too small, except for very small systems. 
This requires one to use a reduced-dimensional representation of the physical system that, in turn, then compromises accuracy. 
For instance, active-orbital-space methods in electronic structure theory such as complete active space self-consistent field (CASSCF) \cite{Roos2005_Book-CASSCF,Olsen2011_CASSCF-Review,lischka_multireference_2018,Gonzalez2020-Book}, density matrix renormalization group (DMRG) \cite{Chan2008_Review,Zgid2009_Review,Marti2010_Review-DMRG,Schollwoeck2011_Review,chan2011density,Wouters2013_Review,Kurashige2014_Review,Olivares2015_DMRGInPractice,szalay2015tensor,Yanai2015,Baiardi2020_Review}, or full configuration interaction quantum Monte Carlo (FCIQMC) \cite{booth_fermion_2009,Alavi2011-InitiatorFCIQMC} introduce a reduced space of orbitals in which the many-electron wave function is constructed, leaving aside the far larger number of remaining virtual orbitals (describing the so-called dynamic correlation).
However, these virtual orbitals eventually account for a huge number of tiny energy contributions that, taken together, cannot be neglected in a calculation of the total energy.
The standard traditional procedure to assess this dynamic correlation energy is multi-reference perturbation theory (MRPT), usually taken to the second-order (MRPT2) for feasibility reasons \cite{Roos1990_CASPT2,Roos1992_CASPT2,Angeli2001_NEVPT2}.

It is natural to pursue a similar strategy on a quantum computer in order to deal with the fact that, although a number of $K$ orbitals is required for the proper description of a physical system, only $k$ orbitals with $k \ll K$ can be used on the machine [note that $k$ spin orbitals will usually be mapped onto $\mathcal{O}(k)$ qubits].
It has been argued in this context that the dynamic correlation energy may be added by classical computation of traditional approaches \cite{Reiher2017_PNAS,von_burg_quantum_2021}. 
However, instead of relying on classical approaches, we derive here a quantum algorithm that computes perturbative energy corrections.

In this paper, we consider a perturbative treatment subsequent to an energy measurement directly on the quantum computer in a ``diagonalize-then-perturb'' fashion \cite{Reiher2017_PNAS} to yield accurate electronic energies.
Specifically, we introduce a quantum algorithm based on perturbation theory for estimating consecutive energy corrections to the ground state energy of the total Hamiltonian $H+V$ of a physical system, without solving the full, $K$-orbital quantum many-body problem on a quantum computer, but instead the model $k$-orbital Hamiltonian $H$.
In regimes, where perturbation theory is expected to be a good approximation, this drastically reduces the quantum hardware requirements for quantum chemistry calculations.

We note that the algorithm we propose may also be considered a sanity check of the quantum model that is supposed to represent the physical model.
That is, it may serve as a way to assess the discretization error. 

The algorithm that we propose is general with respect to the specific perturbation theory application: It can be employed as an approach to parallel traditional MRPT2 calculations, but it also allows one to assess intermolecular interaction energies through symmetry-adapted perturbation theory (SAPT) \cite{jeziorski_perturbation_1994}.

In traditional computing, the unperturbed problem can be of mean-field (Hartree-Fock) type (the second-order energy correction is then an estimate for the correlation energy in M\o{}ller-Plesset PT2) or of complete-active-space type as in CASSCF (where the energy correction then approximates the lacking dynamic correlation energy).
In any such case, the zeroth-order model in traditional computing is limited by known drawbacks (such as mean-field ansatz in Hartree--Fock and limited active space size of CASSCF).
We emphasize that quantum computing will allow one to produce a more accurate zeroth-order model as it does not suffer from the curse of dimensionality of CASSCF.

The MRPT2 energy corrections computed in our approach are based on the Dyall-Hamiltonian as the unperturbed Hamiltonian, and the resulting energy corrections are equivalent to those of the totally uncontracted NEVPT2 method in the frozen core approximation.
As such it is size-extensive and free of the intruder-state problem \cite{angeliNelectronValenceState2002}.
Computing the totally uncontracted NEVPT2 energy is classically very expensive, therefore only approximations to it (partially- and strongly-contracted NEVPT2 energies) are computed in practice.

Previous studies using quantum computing for perturbative methods in quantum chemistry include a variational quantum eigensolver (VQE) method for the first-order SAPT corrections as described by Malone \textit{et al.} \cite{malone_towards_2022}.
Their method was extended to second-order corrections by Loipersberger \textit{et al.} \cite{loipersberger_accurate_2023}, while Tammaro \textit{et al.} \cite{tammaro_n-electron_2023} and Krompiec \textit{et al.} \cite{krompiec_strongly_2022} show how MRPT2 energies can be obtained with the help of a VQE subroutine. 
By contrast, we consider algorithms for fault-tolerant quantum computers.

The first steps towards first- and second-order perturbation theory for fault-tolerant quantum computers were taken by Mitarai \textit{et al.} \cite{mitarai_perturbation_2023}.
Key steps in their approach, such as reference state preparation and implementation of the reduced resolvent, is achieved with a quantum signal processing (QSP) approach.
In comparison, we leverage an alternative implementation of the reduced resolvent and exploit certain features of the unperturbed Hamiltonian to reduce the complexity of the method.
Specifically, for MRPT, we exploit the fact that $H$ is diagonal for orbitals not included in the active space while, for SAPT, we use that $H$ is the sum of two commuting Hamiltonians $H_A$ and $H_B$.
The authors of \cite{mitarai_perturbation_2023} give an extensive account of resources estimates for MRPT calculations on polyacenes.
However, their partitioning of the full Hamiltonian into an unperturbed part and a perturbation is done in way that is not standard in quantum chemistry, and numerical evidence for the accuracy of this choice of partitioning is lacking. 
Furthermore, the runtime dependence on $K$, determined by the norm of perturbation coefficients - $\norm{v}_{2/3}^2$ - suffers from poor scaling, leading to highly impractical resource estimates in \cite{mitarai_perturbation_2023}.
Details regarding this issue are discussed in Sec. \ref{section:MRPT2}.

In conclusion, although work on MRPT approaches for quantum computing has been done previously, they either rely on variational quantum eigensolver (VQE) subroutines that lack performance guarantees, or they do not take into account the specifics of the unperturbed MRPT Hamiltonian, thus exhibiting a poor scaling with $K$, both in the number of qubits and the overall runtime.

Lastly, we note that recently a fault-tolerant quantum algorithm for computing the first-order SAPT corrections was proposed \cite{cortes2023faulttolerant}.
This approach is based on a QSP-algorithm for expectation-value estimation, tailored to the specific form of the SAPT interaction operator.
To obtain an at least qualitatively correct description of intermolecular interactions, it is necessary to include dispersion and induction terms, which appear only in the second-order SAPT correction.
The second-order correction is computationally significantly more demanding as it involves the perturbation $V$ twice, and additionally, the appearance of the reduced resolvent operator poses a non-trivial technical challenge.
Our proposed quantum algorithm applied to SAPT can overcome these challenges.

This paper is organized as follows.
In Sec.~\ref{section:PT}, we set the stage 
by reviewing the essential background for Rayleigh-Schr\"{o}dinger perturbation theory and, in Sec.~\ref{section:QPT2}, we introduce our quantum algorithm for calculating perturbative second-order energy corrections.
In Sec.~\ref{section:MRPT2} we apply our algorithm to MRPT and illustrate the accuracy and scaling with realistic chemical systems. 
The application to SAPT is discussed in Sec.~\ref{section:SAPT}.
Lastly, in Sec.~\ref{section:conclusions}, we finish with conclusions and an outlook.

\section{Perturbation Theory}
\label{section:PT}

To introduce some basic notation, we briefly review standard (Rayleigh-Schr\"{o}dinger) perturbation theory, whose energy corrections are of general applicability to different perturbation theory frameworks.
The goal of perturbation theory is to estimate eigenstates and eigenenergies of a Hamiltonian $H + V$, using the eigenstates and eigenenergies of $H$ as a starting point.
The underlying Hilbert space is finite dimensional and has dimension $D$.
We will only focus on the energies in our discussion.

The eigenenergies $E_j(\lambda)$ of the Hamiltonian $H + \lambda V$ are formally expanded in a Taylor series around $\lambda=0$ \cite{kato_perturbation_1995,shavitt_many-body_2009}
\begin{equation}
  E_j(\lambda) = \sum_{k=0}^{\infty} \lambda^k E_j^{(k)} \, .
  \label{eq:expansion_energy_rspt}
\end{equation}
We will solely focus on the ground state ($j=0$) of the total system $H + V$ and drop the index $j$ from now on.
The zeroth-order term $E^{(0)}$ is the energy of the unperturbed ground state $\ket{\Phi_0}$ of $H$, which we assume to be unique.
Although the expansion is not guaranteed to converge for $\lambda=1$ \cite{marie_perturbation_2021}, low-order corrections, in particular $E^{(1)}$ and $E^{(2)}$, often yield valuable improvements to $E^{(0)}$.
The $k$th order correction, $E^{(k)}$, can be expressed as a sum of expectation values containing products of operators including $V$ and the so-called reduced resolvent operator $R_0$ that is defined as
\begin{equation}
\begin{gathered}
    R_0 = \Pi_0 (E_0\mathbb{1} - H)^{-1} \Pi_0 = \sum_{i=1}^{D-1}\frac{\dyad{\Phi_i}}{E_0-E_i} \quad \\
    \text{with} \quad \Pi_0 = (\mathbb{1} - \dyad{\Phi_0}) \, .
\end{gathered}
\label{eq:ReducedResolvent}
\end{equation}
Here, $E_i$ is the energy of the $i$th excited state of $H$, $\ket{\Phi_i}$.
The expressions for $E^{(k)}$ can symbolically be generated by the bracketing technique \cite{brueckner_many-body_1955}.
The three lowest-order terms are:
\begin{equation}
\begin{aligned}
    E^{(1)} &= \expval{V}\equiv \expval{V}{\Phi_0}  \\
    E^{(2)} &= \expval{VR_0V} \\
    E^{(3)} &= \expval{VR_0VR_0V} - \expval{V}\expval{VR_0^2V} 
    \label{eq:defn_E(n)}
\end{aligned}
\end{equation}
where all expectation values are calculated with respect to the zeroth-order ground-state wave function $\ket{\Phi_0}$.
The $k$th order energy correction is a sum over expectation values in which $V$ appears at most $k$ times and $R_0$ at most $k$-1 times.

Lastly, we note that it is possible to calculate perturbative corrections also for approximate ground states of $H$. 
For an approximate ground state $\ket*{\widetilde{\Phi}_0}$ with $\widetilde{E}_0 = \expval*{H}{\widetilde{\Phi}_0}$ we need the projectors 
\begin{equation*}
    P = \dyad*{\widetilde{\Phi}_0}, \quad Q = \mathbb{1}-P
\end{equation*}
and partition the unperturbed Hamiltonian as 
\begin{gather*}
    H = H' + V' \qquad \text{with}\\
    H' = PHP + QHQ \qquad \text{and} \qquad V'=PHQ + QHP\, .
\end{gather*}
Then $\ket*{\widetilde{\Phi}_0}$ is an eigenstate of $H'$ with eigenvalue $\widetilde{E}_0$ and the perturbation is taken to be $V' + V$.

\section{Quantum Algorithm for Perturbation Theory}
\label{section:QPT2}

In this section, we introduce our quantum algorithm for computing Rayleigh-Schr\"{o}dinger perturbation theory energy corrections up to second order. However, we note that the techniques we will present can be generalized to higher order energy corrections.
For simplicity of presentation we assume that $D=2^d$, where $d$ is the number of qubits. 
We consider a decomposition of $H$ and $V$ into a linear combination of Pauli strings $\sigma_i = \sigma_{i,1}\otimes\sigma_{i,2}\otimes\cdots\otimes\sigma_{i,d}$ acting on the qubits, 
\begin{equation}
  V = \sum_{i=1}^{L_V} v_i \sigma_i, \qquad H = \sum_{i=1}^{L_H} h_i \sigma_i \, ,\qquad v_i,h_i\in\mathbb{R}\, .
  \label{eq:PauliDecomposition}
\end{equation}
We note that for many physical models mapped to such a qubit Hamiltonian, the number of terms $L_V$ and $L_H$ is significantly smaller than $4^d$, the total number of Pauli strings on $d$ qubits.
For example, the states of a spin-1/2 lattice system with $d$ sites can be represented by $d$ qubits, and a lattice Hamiltonian that includes only nearest-neighbor interaction has a number of terms that scales linearly in $d$.
A relevant example for our study is a many-electron system expressed in $d$ spin orbitals.
It can be represented by $d$ qubits based, for instance, on the Jordan-Wigner encoding, and the number of terms of the full electronic Hamiltonian with pairwise Coulomb-interaction scales as $\mathcal{O}(d^4)$.

Based on the decomposition of $V$ into Pauli operators, $E^{(1)}$ can be easily calculated by summing expectation values of the $\sigma_i$,
\begin{equation*}
  E^{(1)} = \sum_{i=1}^{L_V} v_i \expval{\sigma_i},
  \label{eq:FirstOrderEnergyCorrectionPauli}
\end{equation*}
which reduces to standard measurements of Pauli operators. 
The expression of the second-order energy correction $E^{(2)} = \expval{VR_0V}$ implies that $E^{(2)}$ can be computed as the expectation value of $VR_0V$ over $\ket{\Phi_0}$.
The key challenge here is to find an efficient implementation of $R_0$.
One could break $R_0$ into terms as done for $V$ and $H$ in Eq. \eqref{eq:PauliDecomposition}, and then measure $\expval{VR_0V}$ term by term.
To find such a decomposition one could expand $(E_0\mathbb{1} - H)^{-1}$ as a power series in $H$. 
However, since the number of terms in $H^k$ grows as $L_H^k$ this approach suffers from a poor scaling.
By regarding $R_0$ as a function of the spectrum of $H$, it is natural to implement $R_0$ through a block-encoding via quantum signal processing, which is the strategy Mitarai \textit{et al.} \cite{mitarai_perturbation_2023} followed.
While their technique could be employed for the applications we consider in Secs.~\ref{section:MRPT2} and~\ref{section:SAPT}, it is in general not obvious how it can be used to exploit an underlying simplified structure that the unperturbed Hamiltonian might possess, e.g. if $H$ is the sum of two commuting Hamiltonians.
We therefore take a different path and express $R_0$ as a linear combination of time-evolution unitaries $U(t_n)$ under the unperturbed Hamiltonian for a series of time-steps $t_n$.
The decomposition of $R_0$ into unitaries $U(t_n)$ is achieved by utilizing the Fourier series of $R_0$, with respect to the spectral variable $E$.
$E^{(2)}$ can then be evaluated as a sum of expectation values $\expval{\sigma_iU(t)\sigma_j}$ at a series of time-steps $t_n$, where $\sigma_i$ and $\sigma_j$ are Pauli strings arising from $V$ acting on the left and on the right of the resolvent, respectively.
The quantities $\expval{\sigma_iU(t)\sigma_j}$ will be evaluated using a quantum computer.
In Secs~\ref{section:MRPT2} and \ref{section:SAPT}, we will show how the terms $\expval{\sigma_iU(t)\sigma_j}$ simplify for specific unperturbed electronic Hamiltonians.

We will now explain our approach by first discussing the Fourier-transformed resolvent, then applying it to the expression of the second-order energy correction, and afterwards presenting the quantum circuit for computing $\expval{\sigma_iU(t)\sigma_j}$.
Key parameters in the discussion are the spectral range $R = E_{D-1} - E_0$ and the spectral gap $\Delta=E_1 - E_0$ of $H$.
First, we write the resolvent as:
\begin{equation}
  R_0  = \sum_{i=1}^{D-1}\dyad{\Phi_i}\frac{1}{E_0 - E}\Bigg\rvert_{E=E_i}
  \label{eq:UnperturbedResolvent}
\end{equation}
i.e., as a function of the spectral variable $E$ evaluated at the energies of the excited states $E_i$.
The general idea is to define a function $g_{\Delta,R}(E)$ that evaluates to $-\frac{1}{E}$ in the interval $[\Delta, R]$ and that is zero at $E=0$ such that
\begin{equation}
  R_0 = \sum_{i=0}^{D-1}g_{\Delta,R}(E_i - E_0)\dyad{\Phi_i}\, ,
  \label{eq:Resolvent-GFunction}
\end{equation}
and then express $g_{\Delta,R}(E)$ in the time domain via a Fourier series.
In Appendix \ref{appendix:def_g} we design a function $g_{\Delta,R}(E)$ which itself and with all its derivatives vanishes at $E=0$ and at $E=R+\Delta$.
With these properties, $g_{\Delta,R}(E):[0,R+\Delta]\to\mathbb{R}$ can be extended to a smooth periodic function with period $R+\Delta$.
Because of smoothness and periodicity, the Fourier series of $g_{\Delta,R}(E)$ converges uniformly by the Dirichlet-Jordan test \cite{zygmund_trigonometric_2003}.
The Fourier series of $g_{\Delta,R}(E)$ is the ($R+\Delta$)-periodic function
\begin{equation}
  \begin{gathered}
    \sum_{n\in\mathbb{Z}}\beta_{\Delta,R,n} e^{-iEt_n}
    \quad \text{with}\quad t_n := \frac{2\pi n}{R+\Delta}\\ 
    \text{and} \quad \beta_{\Delta,R,n} := \frac{1}{R+\Delta}\int_{0}^{R+\Delta} g_{\Delta,R}(E) e^{iEt_n}dE\, ,
  \end{gathered}
  \label{eq:ResolventFourier}
\end{equation}
such that the resolvent can be expressed as
\begin{equation}
  R_0 =  \sum_{n\in \mathbb{Z}}\sum_{i=0}^{D-1}\beta_{\Delta,R,n} e^{-i(E_i-E_0)t_n} \dyad{\Phi_i} .
  \label{eq:Resolvent-SumOverStates}
\end{equation}
We introduce the propagator under $H$, $U(t) = e^{-iHt}$, to rewrite~(\ref{eq:Resolvent-SumOverStates}) as
\begin{equation}
  R_0 = \sum_{n\in \mathbb{Z}}\beta_{\Delta,R,n} e^{iE_0t_n} U(t_n)\, ,
  \label{eq:Resolvent-Propagator}
\end{equation}
which leads to the following expression for $E^{(2)}$,
\begin{equation}
\begin{aligned}
  E^{(2)} &= \expval{VR_0V}\\
  &= \sum_{n\in \mathbb{Z}}\beta_{\Delta,R,n}e^{-iE_0t_n} \expval{VU(t_n)V}\, .
\end{aligned}
\label{eq:SecondOrder-Propagator}
\end{equation}
In practice, we truncate the series to a finite number of terms $N$, and define
\begin{align}
    R_{0,N} &:= \sum_{\abs{n}\leq N}\beta_{\Delta,R,n} e^{iE_0t_n} U(t_n) \label{eq:truncatedR0}\\
    E_N^{(2)} &:= \expval{VR_{0,N}V} \label{eq:sec_QPT:E2N_LCU} \\ 
    &= \sum_{\abs{n}\leq N}\beta_{\Delta,R,n} e^{iE_0t_n} \expval{VU(t_n)V} \nonumber\,.
\end{align}

Hence, we decomposed an approximation of $R_0$ into a linear combination of $N$ unitaries, and found a corresponding approximation $E_N^{(2)}$ to $E^{(2)}$.
It is worth noting that
\begin{equation*}
    e^{iE_0t}\expval{VU(t)V} = \expval{e^{iHt}Ve^{-iHt}V} = \expval{V(t)V}
\end{equation*}
is simply the autocorrelation function of $V$ at time $t$.
We employ the decomposition of $V$ into unitary operators in~\eqref{eq:PauliDecomposition}, so that $E_N^{(2)}$ is expressed as
\begin{equation}
    \begin{aligned}
        E_N^{(2)} &= \sum_{\abs{n}\leq N}\beta_{\Delta,R,n} e^{iE_0t_n} \sum_{i,j=1}^{L_V} v_i v_j \expval{\sigma_i U(t_n) \sigma_j}\\
    \end{aligned}
    \label{eq:E2N_lin_comb_expval_unitaries}
\end{equation}
Equation \eqref{eq:E2N_lin_comb_expval_unitaries} is the essential expression for our method, and describes how an approximation to $E^{(2)}$ can be obtained by computing the overlaps $\expval{\sigma_i U(t_n) \sigma_j}$ on a quantum computer.

The overall accuracy $\epsilon$ of our algorithm is limited by the order of truncation, $N$, and the finite accuracy of the overlap estimation.
The required truncation order for a desired accuracy $\epsilon$ depends on the convergence rate of the Fourier series of $g_{\Delta,R}$.
In Appendix \ref{appendix:fourier_bound}, Lemma \ref{lemma:fouriercoeff_bound}, we show that the Fourier coefficients decay super-algebraically (i.e. faster than any polynomial) in $n$,
\begin{equation*}
    \abs*{\beta_{\Delta,R,n}} \leq \frac{61 R}{\Delta^2}\exp\left(-\sqrt{\frac{\displaystyle \abs{n}\Delta}{\displaystyle 256R}}\right)\, .
\end{equation*}
This result is then used to prove that the truncation error is bounded by 
\begin{equation}
\begin{gathered}
    \abs*{E^{(2)} - E_N^{(2)}} \leq \frac{\epsilon}{2}\\
    \text{if}\quad  N\geq \left\lceil\frac{2132 R}{\Delta}\ln^2 \left(\frac{5.3\cdot10^4 R\norm{v}_1^2}{\Delta^3\epsilon}\right)\right\rceil\, ,
    \label{eq:sec_QPT:trunc_error_bound}
\end{gathered}
\end{equation}
where $\norm{v}_1 = \sum_{i=1}^{L_V}\abs{v_i}$. The proof is provided in Appendix \ref{appendix:trunc_error}.

Equation \eqref{eq:sec_QPT:trunc_error_bound} shows that, on top of the global accuracy $\epsilon$, knowledge of the parameters $\Delta$, $R$ and $\norm{v}_1$ is required in order to get the truncation order $N$. 
Note that $R\leq 2\norm{h}_1$, as shown in Eq. \eqref{eq:appdx_complexity:bound_R} in Appendix \ref{appendix:complexity_circ1}.
With the definition of $H$ and $V$ [see Eq. \eqref{eq:PauliDecomposition}] $\norm{h}_1$ and $\norm{v}_1$ can readily be computed.
For the gap $\Delta$, a realistic lower bound needs to be supplied.
Once the truncation order $N$ is known, the target quantity to compute is given by Eq. \eqref{eq:E2N_lin_comb_expval_unitaries}.

The key step of our method is the estimation of the overlaps using a quantum computer. 
To perform this estimation efficiently, we build on quantum amplitude estimation.
Quantum overlap estimation is closely related to the quantum amplitude estimation problem, and in Appendix \ref{appendix:overlap_estimation}, we describe how a recent version of the quantum amplitude estimation algorithm, the iterative quantum amplitude estimation algorithm (IQAE) \cite{grinko_iterative_2021,fukuzawa_modified_2023} can be employed for quantum overlap estimation.
The results in this part of the Appendix build up to Lemma \ref{lemma:LCU_overlap_estimation_term_by_term}:

\begin{customlemma}{4.5}
    Given a confidence level $1-\alpha \in (0,1)$, a target accuracy $\epsilon>0$, a linear combination of d-qubit unitaries $V_U = \sum_{i=1}^L u_i U_i$ and a d-qubit state-preparation unitary $U_0$, it is possible to obtain an estimate $X$ to the overlap $\expval{V_U}{\Phi_0}$
    such that
    \begin{equation*}
        \mathbb{P}[\abs*{X - \expval{V_U}{\Phi_0}} < \epsilon] \geq 1 - \alpha
    \end{equation*}
    with $1$ ancillary qubit,  a total $\mathcal{O}\left(\epsilon^{-1} \norm{u}_{2/3} \ln\frac{L}{\alpha}\sqrt{\ln\frac{1}{\alpha}}\right)$ applications of $U_0$ and $\mathcal{O}\left(\epsilon^{-1} \abs{u_i}^{2/3}\norm{u}_{2/3}^{1/3} \ln\frac{L}{\alpha}\sqrt{\ln\frac{1}{\alpha}}\right)$ applications of controlled-$U_i$ for each i=1,\dots,L, where $\norm{u}_{2/3} = \left(\sum_{i=1}^L \abs{u_i}^{2/3}\right)^{3/2}$.
\end{customlemma}

The most expensive subroutines of our quantum algorithm are the time evolution $U(t)$ and the state-preparation $U_0$.
Apart from QPE, various algorithms have been proposed to prepare the ground state of a Hamiltonian \cite{mitarai_perturbation_2023,lin_near-optimal_2020,poulin_preparing_2009,ge_faster_2019} , e.g. by eigenvalue filtering.
Furthermore, as argued at the end of Sec. \ref{section:PT}, perturbative corrections can be calculated also for approximate eigenstates, opening up the possibilities to different types of state preparation subroutines $U_0$. 
Therefore, in order to keep the algorithm general, we treat $U_0$ as a black-box, and report the total number of calls to $U_0$ required to compute an estimate to $E^{(2)}$ with $\epsilon$ precision.
Moreover, we will state the total time for which the evolution under $H$ needs to implemented in order to compute the estimate to $E^{(2)}$.

\vspace{5mm}
The overall cost of our method is shown with Lemma \ref{lemma:complexity_algorithm_termbyterm}, with the result that for $H$ and $V$ as in Eq. \eqref{eq:PauliDecomposition} computing an estimate to $E^{(2)}$ with precision $\epsilon$ requires implementing the Hamiltonian evolution under $H$ for a total time upper bounded by

\begin{equation}
    \widetilde{\mathcal{O}}\left(\frac{\norm{h}_1^{3/2} \norm{v}_{2/3}^2}{\Delta^{7/2}\epsilon}\right)\, ,
    \label{eq:sec_qpt:complexity_time}
\end{equation}
and a total number of calls to $U_0$ upper bounded by

\begin{equation}
    \widetilde{\mathcal{O}} \left(\frac{\norm{h}_1^{3/2} \norm{v}_{2/3}^2}{\Delta^{5/2} \epsilon}\right)\, .
    \label{eq:sec_qpt:complexity_U_0}
\end{equation}
Here, the notation $\widetilde{\mathcal{O}}$ denotes the suppression of logarithmic factors, and $\norm{v}_{2/3} = (\sum_{i=1}^{L_V}\abs{v_i}^{2/3})^{3/2}$. 
Furthermore, Lemma \ref{lemma:complexity_algorithm_termbyterm} implies that the quantum algorithm requires $d+1$ qubits.

We note that our approach may be optimized in terms of $\Delta$ and $\norm{h}_1$.
For example, our implementation of $g(E)$ as a trigonometric polynomial is not unique and could be optimized in order to reduce the dependence on $\Delta$ and $R$.
Furthermore, for Hamiltonians $H$ considered in practice, the upper bound $R\leq 2\norm{h}_1$ might turn out to be very loose.

\section{Application to Multireference Perturbation Theory}
\label{section:MRPT2}

So far, we kept our derivation entirely generic, so that it can be applied to any method based on second-order perturbation theory.
In this section we shift the focus to the problem of finding the electronic ground state energy of a molecular system.
Multi-electron states are in practice constructed from a discrete and finite set of single-electron states, i.e., a set of so-called molecular spin orbitals, and via fermion-to-qubit mappings of molecular spin orbitals to qubits the electronic structure problem can be tackled using a quantum computer.

If the number of qubits available is much smaller than the number of molecular spin orbitals one desires to include in the description of a chemical system, a subset of so-called ``active orbitals'' must be chosen which fit the number of qubits, yielding a reduced-dimensional problem.
This reduction in problem size necessarily incurs some errors, however, as we will explain in this section, it is possible to use our quantum algorithm for second-order perturbation theory to reduce this error.
In this regard, our approach parallels classical multi-reference perturbation theory methods for adding dynamical correlation effects on top of a complete active space solution.

Classical active space methods include complete active space configuration interaction (CASCI), CASSCF and DMRG, and well-established methods for including dynamical correlation effects based on second-order perturbation theory exist, such as complete active space and N-electron valence perturbation theory of second order, CASPT2 and NEVPT2, respectively.
These perturbation theory methods differ in the choice of the zeroth-order Hamiltonian.

\sloppy The starting point is an orthonormal single-electron basis of dimension $K$, $\mathcal{I}=\{\phi_1,\dots,\phi_{K}\}$, spanning the single-electron Hilbert space $\mathcal{H}_1(\mathcal{I})$.
We consider a system of $N_{el}$ electrons, so that the full Hilbert space is given by $\Lambda^{N_{el}}(\mathcal{H}_1(\mathcal{I}))$. 
The quantities of interest are the eigenenergies of the full electronic Hamiltonian

\begin{equation}
  H_{el} = \sum_{p,q\in\mathcal{I}} h_{pq} a_p^{\dag} a_q 
         + \frac{1}{2}\smashoperator{\sum_{p,q,r,s\in \mathcal{I}}} g_{pqrs} a_p^{\dag} a_q^{\dag} a_r a_s \,
  \label{eq:ElectronicSchroedinger}
\end{equation}
for a fixed number of $N_{el}$ electrons, and where the fermionic creation and annihilation operators corresponding to $\phi_i$ are denoted by $a_i^{\dag}$ and $a_i$.
The coefficients $h_{pq}$ and $g_{pqrs}$ are the standard one- and two-electron integrals in Hartree atomic units\cite{helgaker_molecular_2014}
\begin{equation}
  \begin{aligned}
    h_{pq} =& \int \, \overline{\phi_p}(\vb{x})\left( -\frac{\nabla^2}{2} - \sum_{I=1}^{N_{nuc}}\frac{Z_I}{\abs{\vb{R}_I - \vb{x}}}\right) \phi_q(\vb{x}) d\vb{x}, \\
    g_{pqrs} =& \int \int \, \frac{\overline{\phi_p}(\vb{x}_1)\overline{\phi_q}(\vb{x}_2)\phi_r(\vb{x}_2)\phi_s(\vb{x}_1)}{\abs{\vb{x}_1-\vb{x}_2}} d\vb{x}_1 d\vb{x}_2 \, ,
  \end{aligned}
  \label{eq:sec_QPT:electron_integrals}
\end{equation}
where $\mathbf{x}$ denotes spatial and spin variables, $N_{nuc}$ is the number of nuclei, and $Z_I$ and $\vb{R}_I$ are their charge numbers and positions, respectively.
We assume that the one-electron basis $\mathcal{I}$ is constructed from molecular spin orbitals that diagonalize the Fock operator $F$, i.e.
\begin{equation*}
  F = \sum_{p\in\mathcal{I}} \epsilon_p a_p^{\dag} a_p \, ,
  \label{eq:FockOperator}
\end{equation*}
where $\epsilon_p$ are the orbital energies.
The Fock operator results from a mean-field approximation of the electronic Hamiltonian of ~\eqref{eq:ElectronicSchroedinger}.
Although Hartree-Fock does not yield quantitatively accurate predictions, it is considered a good starting point for more accurate methods \cite{helgaker_molecular_2014}.

Methods aiming at improving upon Hartree-Fock results face the problem that the full Hilbert space has dimension $\binom{K}{N_{el}}$.
When employing standard atom-centered basis sets, the size of the single-particle basis, $K$, is roughly proportional to, and at least an order of magnitude larger than, the number of electrons.
Therefore, the dimension of the Hilbert space scales exponentially in the number of electrons, which is the reason why representing general electronic wave functions on a classical computer becomes unfeasible.

\begin{figure}
    \centering
    \includegraphics[width=0.4\linewidth]{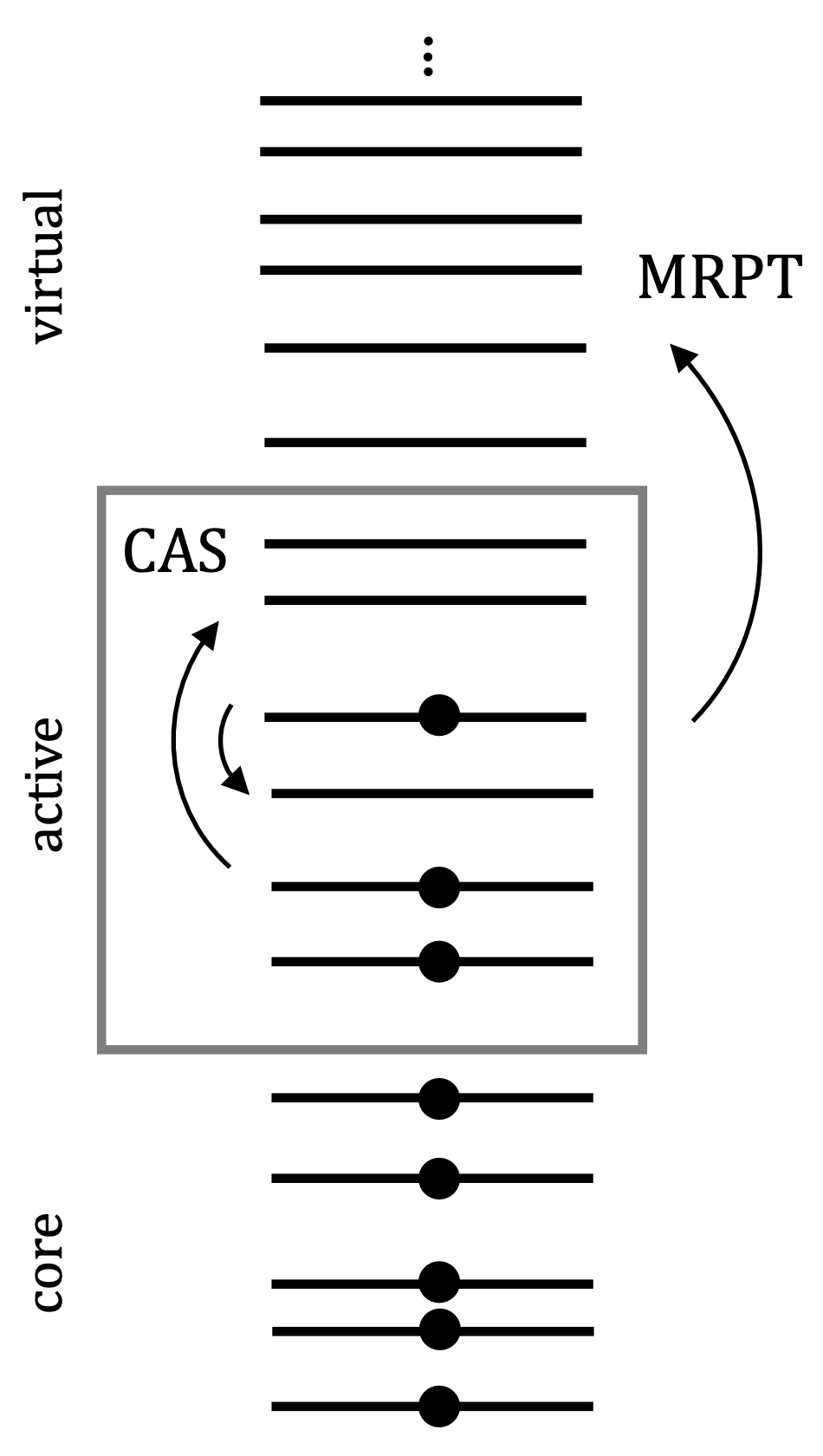}
    \caption{Illustration of occupations of core, active and virtual orbitals. Lines and circles symbolize orbitals and electrons, respectively.}
    \label{fig:active_space}
\end{figure}

Many methods in quantum chemistry deal with this issue by considering only a small, manageable subset of the full, many-electron Hilbert space.
In the class of active-space methods this smaller subset is defined on the level of the single-electron basis.
Active-space methods partition the single-particle basis into three parts, the inactive (or core) space $\mathcal{I}_{core}$, the space of active orbitals $\mathcal{I}_{act}$ and the virtual space $\mathcal{I}_{virt}$, such that $\mathcal{I} = \mathcal{I}_{core}\cup\mathcal{I}_{act}\cup\mathcal{I}_{virt}$ (see Fig. \ref{fig:active_space}).
We will use indices $a,b,c,d,e,f$ for active orbitals, $v,w,x,y$ for virtual orbitals and $p,q,r,s$ for arbitrary orbitals.
The number of active orbitals will be denoted by $k$.
The active space is defined as the span of $N_{el}$-electron Slater determinants where all virtual orbitals are unoccupied and all core orbitals are occupied.
This leads to a reduced problem size of $N_{el}^{act} = N_{el} - \abs{\mathcal{I}_{core}}$ electrons in $k$ molecular spin orbitals.
On this subspace, the Hamiltonian $H_{el}$ (see~\eqref{eq:ElectronicSchroedinger}) can be associated with the CAS Hamiltonian $H_{CAS}$, which acts on the Hilbert space $\Lambda^{N_{el}^{act}}(\mathcal{H}_1(\mathcal{I}_{act}))$ and reads
\begin{equation*}
    H_{CAS} = \smashoperator{\sum_{a,b\in\mathcal{I}_{act}}} h_{ab}' a_a^{\dag} a_b 
        + \frac{1}{2}\smashoperator{\sum_{\quad a,b,c,d\in \mathcal{I}_{act}}} g_{abcd} a_a^{\dag} a_b^{\dag} a_c a_d \, .
    \label{eq:CASHamiltonian}
\end{equation*}
Here, $h_{ab}'$ includes the mean-field potential of the core electrons, which the active electrons are subjected to in the active space approximation.
Note that the energy of the core electrons themselves is not part of $H_{CAS}$.
The ground state wave function and energy of $H_{CAS}$ are denoted by $\ket*{\Phi_0^{CAS}}$ and $E_0^{CAS}$, respectively.
For increasing $k$ and $N_{el}^{act}$, classical implementations of active space methods face the same dimensionality issue as before, because the dimension of $\Lambda^{N_{el}^{act}}(\mathcal{H}_1(\mathcal{I}_{act}))$ is $\binom{N_{el}^{act}}{k}$.
The promise of fault-tolerant quantum computers for chemistry is to target larger active spaces with the help of quantum algorithms like QPE.
Nonetheless, every active space computation, whether classical or quantum, is necessarily an approximation, and it is desirable to improve on that.

Multireference perturbation theory (MRPT) methods are employed to go beyond the active space approximation, accounting for Slater determinants not included in the active space.
The second-order MRPT energy correction will contain contributions from Slater determinants associated with variable occupation numbers of core, active and virtual orbitals.
In fact, it will contain contributions associated with so-called single and double excitations from the core and active spaces to the active and virtual spaces (by single and double excitations we refer here to a Slater determinant that differs from the reference one by the occupation of one or two spin orbitals, respectively).
For the sake of clarity, we introduce an approximation and keep the core orbitals occupied.
We refer to this approximation as ``frozen-core approximation''.
In this setting the Hilbert space for the perturbed and unperturbed Hamiltonians is $\Lambda^{N_{el}^{act}}(\mathcal{H}_1(\mathcal{I}_{act}\cup\mathcal{I}_{virt}))$.
We note that in many situations this approximation is justified and used in practice.
Extending our methods to include also electron excitations from the core orbitals is conceptually trivial.
The full Hamiltonian in the frozen-core approximation is
\begin{equation}
  H'_{el} = \smashoperator{\sum_{p,q\in\mathcal{I}_{act}\cup\mathcal{I}_{virt}}} h'_{pq}a_p^{\dag}a_q
    + \frac{1}{2}\smashoperator{\sum_{p,q,r,s\in \mathcal{I}_{act}\cup\mathcal{I}_{virt}}} g_{pqrs} a_p^{\dag}a_q^{\dag}a_ra_s + E_{core} \, .
  \label{eq:FrozenCore}
\end{equation}
Compared to the full Hamiltonian Eq. \eqref{eq:ElectronicSchroedinger}, the core electrons enter Eq. \eqref{eq:FrozenCore} only through the single electron coefficients $h'_{pq}$ and the energy-shift $E_{core}$, the energy of the core electrons.
We define the unperturbed Hamiltonian $H$ as the sum of $H_{CAS}$, the Fock operator on the virtual space, and $E_{core}$ (note that this choice yields the Dyall Hamiltonian in the frozen-core approximation~\cite{dyall_choice_1995}) as
\begin{equation}
\begin{gathered}
    H = H_{CAS} + F_{virt} + E_{core}\\ 
    F_{virt} = \smashoperator{\sum_{v\in\mathcal{I}_{virt}}} \epsilon_v a_v^{\dag} a_v \, .
\end{gathered}
\label{eq:Dyall}
\end{equation}
The unperturbed reference function can easily be constructed from the ground-state wave function of $H_{CAS}$, $\ket*{\Phi_0^{CAS}}$, as $\ket{\Phi_0} = \ket*{\Phi_0^{CAS}}\wedge \ket*{0\dots 0}$, which is an eigenfunction of $H$ with energy $E_0^{CAS} + E_{core}$.
In practice, the active space is chosen such that the virtual orbitals all have sufficiently high orbital energies to ensure that $\ket{\Phi_0}$ is also the ground-state wave function of $H$. 
The expression for $V$ follows from $V = H'_{el} - H$,
\begin{equation*}
    V = \smashoperator{\sum_{\substack{p,q\in\mathcal{I}_{act}\cup\mathcal{I}_{virt}\\ \{p,q\}\not\subset \mathcal{I}_{act}}}} h'_{pq}a_p^{\dag} a_q + \frac{1}{2} \smashoperator{\sum_{\substack{p,q,r,s\in \mathcal{I}_{virt}\cup\mathcal{I}_{act}\\ {\{p,q,r,s\}\not\subset \mathcal{I}_{act}}}}} g_{pqrs} a_p^{\dag} a_q^{\dag} a_r a_s - F_{virt}\, .
\end{equation*}
One could directly apply the general perturbation theory algorithm to $H$ and $V$ and obtain the asymptotic cost from Eqs \eqref{eq:sec_qpt:complexity_time} and \eqref{eq:sec_qpt:complexity_U_0}.
We briefly discuss the scaling of this approach with respect to the large number of molecular spin orbitals $K$, and we assume that $K \gg k$.
It is evident that the number of terms in $V$ scales as $\mathcal{O}(K^4)$.
Since $V$ appears once in the first-order correction $E^{\text{MRPT}(1)}$ and twice in the second-order correction $E^{\text{MRPT}(2)}$, the calculations of $E^{\text{MRPT}(1)}$ and $E^{\text{MRPT}(2)}$ involve $\mathcal{O}(K^4)$ and $\mathcal{O}(K^8)$ expectation values, respectively.
More specifically, the dependence on $K$ in the cost of the computation of $E^{\text{MRPT}(2)}$ in Eqs. \eqref{eq:sec_qpt:complexity_time} and \eqref{eq:sec_qpt:complexity_U_0} is determined by $\norm{v}_{2/3}^2$, apart from logarithmic factors.
For the approach of using the full difference $V = H'_{el} - H$ as the perturbation we will denote this quantity by $\norm{v_{\text{diff}}}_{2/3}^2$.
As described later in this section, for a realistic chemical system we numerically observe that $\norm{v_{\text{diff}}}_{2/3}^2$ scales as $\mathcal{O}(K^{6.2})$, which would incur considerable cost for the method.

We note that directly using the full difference $V = H'_{el} - H$ as the perturbation is very similar to the approach by Mitarai \textit{et al.} \cite{mitarai_perturbation_2023} when they applied their QSP perturbation theory algorithm to multireference perturbation theory.
Their unperturbed, active space Hamiltonian is defined after the fermion-to-qubit mapping and is therefore different compared to the CAS- or Dyall-Hamiltonian. 
Nonetheless,  all terms of the full Hamiltonian $H'_{el}$ are taken into account, either in the unperturbed Hamiltonian or in the perturbation.
The perturbation norm $\norm{v}_{2/3}^2$, as defined for their partitioning into $H$ and $V$, obeys the same scaling in $K$ as $\norm{v_{\text{diff}}}_{2/3}^2$.
Highly impractical numbers for $\norm{v}_{2/3}^2$ lead \cite{mitarai_perturbation_2023} to conclude that their perturbation theory method does not reduce the computational cost of solving chemical systems.
Noteworthy, their method is not accompanied by a reduction in the number of qubits.
 
In this light we set the following two criteria: First, computing $E^{\text{MRPT}(2)}$ should not require more qubits than is necessary to represent the active space orbitals plus a constant number of ancillary qubits. 
Second, we want to overcome the issue of impractically large values of $\norm{v_{\text{diff}}}_{2/3}^2$.
In the following we explain how to accomplish these goals.

A first simplification of $V$ is possible by noting that $a_v\ket{\Phi_0} = 0$ if $a_v$ corresponds to a virtual orbital.
Consequently, the expectation values $\expval{a_p^{\dag}a_q}$ and $\expval{a_p^{\dag}a_q^{\dag}a_ra_s}$ calculated over the reference wave function will be non-zero only if $p,q,r,s\in\mathcal{I}_{act}$.
However, none of the terms in $V$ act only on the active orbitals, therefore the first-order energy correction is (within the frozen-core approximation) zero, i.e.,
\begin{equation}
  E^{\text{MRPT}(1)} = \expval{V} = 0.
  \label{eq:FirstOrderCorrection}
\end{equation}
Extending this reasoning to the (truncated) second-order energy correction, 
\begin{equation*}
\begin{aligned}
    E_N^{\text{MRPT}(2)} &= \sum_{\abs{n} \leq N} \beta_{\Delta,R,n} e^{iE_0t} \expval{VU(t_n)V}\, ,
\end{aligned}    
\end{equation*}
by taking into account only the terms of $V$ that are non-vanishing when acting on $\ket{\Phi_0}$, we can write
\begin{equation}
\begin{aligned}
    \langle V & U(t_n) V\rangle
    = \smashoperator{\sum_{v,a,w,b}} h'_{va} h'_{wb} \expval*{a_b^{\dag} a_w U(t_n) a_v^{\dag} a_a}\\
    &+ \frac{1}{2}\smashoperator{\sum_{v,a,p,q,b,c}} h'_{va} g_{pqbc} \expval*{a_c^{\dag} a_b^{\dag}a_q a_p U(t_n) a_v^{\dag}a_a}\\ 
    &+ \frac{1}{2}\smashoperator{\sum_{p,q,a,b,v,c}} g_{pqab} h'_{vc} \expval*{a_c^{\dag} a_v U(t_n) a_p^{\dag} a_q^{\dag} a_a a_b}\\
    &+ \frac{1}{4}\smashoperator{\sum_{p,q,a,b,r,s,c,d}} g_{pqab} g_{rscd} \expval*{a_d^{\dag} a_c^{\dag} a_s a_r U(t_n) a_p^{\dag} a_q^{\dag} a_a a_b}\\
    &=: T_1(t_n) + T_2(t_n) + T_3(t_n) + T_4(t_n)
\end{aligned}
\label{eq:sec_mrpt:defn_sums_Tn}
\end{equation}
with
\begin{equation*}
\begin{gathered}
    v,w \in \mathcal{I}_{virt}\, ,\quad a,b,c,d \in \mathcal{I}_{act}\\
    p,q,r,s \in \mathcal{I}_{act} \cup \mathcal{I}_{virt}\, , \quad \{p,q\},\{r,s\}\not\subset\mathcal{I}_{act}\, .
\end{gathered}
\end{equation*}
To keep the discussion of further reductions of these terms simple, here in the main text we will just consider the first of the four sums, $T_1(t_n)$.
The full derivation for all four sums is analogous and is given in Appendix \ref{appendix:<VU(t)V> for MRPT}.

$H$ commutes with the number operators $n_v = a_v^{\dag}a_v$ of the virtual orbitals, $[H,n_v] = [\epsilon_v n_v,n_v] = 0$, therefore it conserves the number of electrons in each virtual orbital.
The same holds for the propagator $U(t_n)$.
Consequently, $\expval*{a_b^{\dag} a_w U(t_n) a_v^{\dag} a_a}$ is non-vanishing only if creation and annihilation operators appear in pairs,
\begin{equation*}
    T_1(t_n) = \smashoperator{\sum_{\substack{v\in\mathcal{I}_{virt}\\a,b\in\mathcal{I}_{act}}}} h'_{va} h'_{vb} \expval*{a_b^{\dag} a_v U(t_n) a_v^{\dag} a_a}\, .
\end{equation*}
Next, we employ the time-dependent creation and annihilation operators in the Heisenberg picture, $a_p^{\dag}(t) \equiv e^{iHt}a_p^{\dag} e^{-iHt}$ and $a_p(t) \equiv e^{iHt}a_p e^{-iHt}$.
Since $H$ is diagonal for the virtual orbitals, the time dependence of their creation and annihilation operators is particularly simple, $a_v(t) = a_ve^{-i\epsilon_vt}$ and $a_v^{\dag}(t) = a_v^{\dag}e^{i\epsilon_vt}$.
Using this insight, we find
\begin{align*}
        T_1(t_n) &= \smashoperator{\sum_{\substack{v\in\mathcal{I}_{virt}\\a,b\in\mathcal{I}_{act}}}} h'_{va} h'_{vb} \expval*{a_b^{\dag} e^{-iHt_n} e^{iHt_n} a_v e^{-iHt_n} a_v^{\dag} a_a}\\
        &= \smashoperator{\sum_{\substack{v\in\mathcal{I}_{virt}\\a,b\in\mathcal{I}_{act}}}} h'_{va} h'_{vb} e^{-i\epsilon_vt_n} \expval*{a_b^{\dag} e^{-iHt_n} a_v a_v^{\dag} a_a}\\
        &= \smashoperator{\sum_{\substack{v\in\mathcal{I}_{virt}\\a,b\in\mathcal{I}_{act}}}} h'_{va} h'_{vb} e^{-i(\epsilon_v + E_{core})t_n} \expval*{a_b^{\dag} e^{-iH_{CAS}t_n} a_a}\, ,
\end{align*}
where, in the last line, we exploited the anticommutation relations between fermionic second-quantization operators, as well as $a_v\ket{\Phi_0} =0$.
Furthermore, since the last expectation value only contained active orbitals, we rewrote $e^{-iHt_n}$ as $e^{-i(H_{CAS} + E_{core})t_n}$.

Following these steps for all terms (see Appendix \ref{appendix:<VU(t)V> for MRPT}), we get
\begin{align}
    T_1(t_n) &= \smashoperator{\sum_{a,b\in\mathcal{I}_{act}}} v_{a,b}(t_n) \expval*{a_b^{\dag} e^{-iH_{CAS}t_n} a_a} \nonumber\\
    T_2(t_n) &= \smashoperator{\sum_{a,b,c,d\in\mathcal{I}_{act}}} v_{abc,d}(t_n) \expval*{a_d^{\dag} e^{-iH_{CAS}t_n} a_a^{\dag} a_b a_c} \nonumber \\
    T_3(t_n) &= \smashoperator{\sum_{a,b,c,d\in\mathcal{I}_{act}}} v_{bcd,a}(t_n) \expval*{a_d^{\dag} a_c^{\dag} a_b e^{-iH_{CAS}t_n} a_a} \label{eq:E2N_MRPT} \\
    T_4(t_n) &= \smashoperator{\sum_{a,b,c,d\in\mathcal{I}_{act}}} v_{ab,cd}(t_n) \expval*{a_d^{\dag} a_c^{\dag} e^{-iH_{CAS}t_n} a_a a_b} \nonumber \\
    &\phantom{{}=\:\: }+ \smashoperator{\sum_{a,b,c,d,e,f\in\mathcal{I}_{act}}} v_{abc,def}(t_n) \expval*{a_f^{\dag} a_e^{\dag} a_d e^{-iH_{CAS}t_n} a_a^{\dag} a_b a_c}\, ,\nonumber 
\end{align}
where the coefficients are defined by
\begin{equation}
    \begin{aligned}
        v_{a,b}(t) &= \smashoperator{\sum_{v\in\mathcal{I}_{virt}}} h'_{va} h'_{vb} e^{-i(\epsilon_v + E_{core})t} \\
        v_{abc,d}(t) &= \smashoperator{\sum_{v\in\mathcal{I}_{virt}}} h'_{vd} g_{vabc} e^{-i(\epsilon_v + E_{core})t} \\
        v_{ab,cd}(t) &= \tfrac{1}{2} \smashoperator{\sum_{v,w\in\mathcal{I}_{virt}}} g_{vwab} g_{vwcd} e^{-i(\epsilon_v+\epsilon_w + E_{core})t}\\
        v_{abc,def}(t) &= \smashoperator{\sum_{v\in\mathcal{I}_{virt}}} g_{vabc} g_{vdef}e^{-i(\epsilon_v + E_{core})t}\, .
    \end{aligned}
    \label{eq:sec_MRPT:defn_coeffs_MRPT}
\end{equation}
This reduction shares similarities with classical perturbation theory methods based on the Laplace-transform of the resolvent \cite{haserMollerPlessetMP2Perturbation1993, ayalaLinearScalingSecondorder1999}.
In particular, the method of \cite{sokolovTimedependentFormulationMultireference2016} also achieves a reduction in the number of terms to be computed for MRPT2 energies by requiring only 3-RDM like terms instead of 4-RDM.

Finally, the creation and annihilation operators in Eq. \eqref{eq:E2N_MRPT} need to be mapped to Pauli strings acting on qubits resulting in an expression of the form
\begin{equation}
\begin{aligned}
    E_N^{\text{MRPT}(2)} &= \smashoperator{\sum_{\abs{n}\leq N}} \beta_{\Delta,R,n} \smashoperator{\sum_{i=1}^{L^{\text{MRPT}(2)}} } v_{ni}^{\text{MRPT(2)}}\\
    &\phantom{{}\quad} \times \expval{\sigma_{i,1} e^{-iH_{CAS}t_n} \sigma_{i,2}}\, .\label{eq:sec_MRPT2:E2N_lincomb_unitaries}
\end{aligned}
\end{equation}
Equation \eqref{eq:sec_MRPT2:E2N_lincomb_unitaries} is not of the form of \eqref{eq:E2N_lin_comb_expval_unitaries}, since the perturbation coefficients $v_{ni}^{\text{MRPT(2)}}$ depend on $n$, too.
Following the complexity analysis of the general algorithm (Appendix \ref{appendix:complexity_circ1}), the coefficients enter the cost as
\begin{align*}
    \norm*{v^{\text{MRPT(2)}}\beta}_{2/3} = \left(\sum_{\abs{n}\leq N} \smashoperator{\sum_{i=1}^{L^{\text{MRPT}(2)}}} \abs{\beta_{\Delta,R,n} v_{ni}^{\text{MRPT}(2)}}^{2/3}\right)^{3/2}\, .
\end{align*}
The perturbation coefficients $v_{ni}^{\text{MRPT}(2)}$ derive directly from the coefficients in Eq. \eqref{eq:sec_MRPT:defn_coeffs_MRPT}, and by moving the absolute values into the sums over the virtual orbitals we obtain an upper on the perturbation coefficients that is independent of $n$:
\begin{align*}
    &\left(\sum_i \abs*{v_{ni}^{\text{MRPT(2)}}}^{2/3}\right)^{3/2}\\
    &\leq \Bigg( 2^{2/3} \sum_{a,b\in \mathcal{I}_{act}} \left(\sum_{v\in\mathcal{I}_{virt}} \abs{h'_{va} h'_{vb}}\right)^{2/3}\\
    &+ 2^{7/3} \sum_{a,b,c,d\in\mathcal{I}_{act}} \left(\sum_{v\in\mathcal{I}_{virt}} \abs{h'_{va} g_{vbcd}} \right)^{2/3}\\
    &+ 2^{2/3} \sum_{a,b,c,d\in\mathcal{I}_{act}} \left(\sum_{vw\in\mathcal{I}_{virt}}\abs{g_{vwab} g_{vwcd}}\right)^{2/3}\\
    &+ 4 \sum_{a,b,c,d,e,f\in\mathcal{I}_{act}} \left(\sum_{v\in\mathcal{I}_{virt}} \abs{g_{vabc}g_{vdef}}\right)^{2/3} \Bigg)^{3/2}\\
    &\equiv \norm*{v^{\text{MRPT(2)}}}_{2/3}\, .
\end{align*}
See Appendix \ref{appendix:mrpt2_norm} for details.
As a consequence we can decouple $\norm*{v^{\text{MRPT(2)}}\beta}_{2/3}$ and use the upper bound

\begin{equation*}
    \norm*{v^{\text{MRPT(2)}}\beta}_{2/3} \leq \norm*{v^{\text{MRPT(2)}}}_{2/3} \norm{\beta_{\Delta,R}}_{2/3}
\end{equation*}
so that in the bounds of the general perturbation theory algorithm in Eqs. \eqref{eq:sec_qpt:complexity_time} and \eqref{eq:sec_qpt:complexity_U_0}   $\norm{v^{\text{MRPT(2)}}}_{2/3}$ plays the role of $\norm{v}_{2/3}^2$ and $L^{\text{MRPT(2)}}$ that of $L_V^2$.

Another parameter that entered the complexity analysis of the algorithm is $R$, the spectral range of $H$.
In Eqs. \eqref{eq:sec_qpt:complexity_time} and \eqref{eq:sec_qpt:complexity_U_0} this parameter has already been replaced by the upper bound $2\norm{h}_1$.
The Dyall-Hamiltonian [see Eq. \eqref{eq:Dyall}], which we use as the unperturbed Hamiltonian, is acting on the virtual orbitals with a Fock operator, so its spectral range will depend on $K$.
Therefore, if we were to simply use the spectral range of the Dyall-Hamiltonian, this would incur additional $K$ dependence in the complexity of the method.
However, we argue that in this specific application, $R$ can be replaced by $R_{CAS} + \text{const}$, where $R_{CAS}$ is the spectral range of $H_{CAS}$.
As a consequence, we will get rid of the $K$-dependence caused by $R$.
To see that this replacement is justified, consider the exponents in Eq. \eqref{eq:E2N_MRPT}, which encode the energy via the Fourier transform, see Eqs. \eqref{eq:ResolventFourier} - \eqref{eq:SecondOrder-Propagator}. 
The operator $e^{-iH_{CAS}t}$ is acting in the expectation values of Eq. \eqref{eq:E2N_MRPT}, which, together with the prefactor $e^{i(E_0-E_{core})t} = e^{iE_0^{CAS}t}$, results in an energy in the exponent of at most $R_{CAS}$.
Moreover, the energies of at most two virtual orbitals contribute via the coefficients, so the largest energy that occurs is upper bounded by $R_{CAS} + 2\underset{v\in\mathcal{I}_{virt}}{\text{max}}\epsilon_v$.
Thus, since $R_{CAS} \leq 2\norm*{h^{CAS}}_1$, the 1-norm $\norm{h}_1$ in Eqs. \eqref{eq:sec_qpt:complexity_time} and \eqref{eq:sec_qpt:complexity_U_0} can be replaced by $\norm*{h^{CAS}}_1$,
and we arrive at our main result:

\vspace{3mm}
\textit{Main Result:} Computing $E^{\text{MRPT}(2)}$ with precision $\epsilon$ requires $k+1$ qubits and implementing the evolution under $H_{CAS}$ for a total time upper bounded by 
\begin{equation*}
    \widetilde{\mathcal{O}}\left(\frac{\norm{h^{CAS}}_1^{3/2} \norm{v^{\text{MRPT(2)}}}_{2/3}}{\Delta^{7/2}\epsilon}\right)\, ,
\end{equation*}
and a total number of calls to $U_0$ upper bounded by
\begin{equation*}
    \widetilde{\mathcal{O}} \left(\frac{\norm{h^{CAS}}_1^{3/2} \norm{v^{\text{MRPT(2)}}}_{2/3}}{\Delta^{5/2} \epsilon}\right)\, .
\end{equation*}

Explicit expressions for upper bounds on $\norm{h^{CAS}}_{1}$ and on $\norm{v^{\text{MRPT}(2)}}_{2/3}$ are provided in Appendix \ref{appendix:mrpt2_norm}. 

\vspace{3mm}

\textit{Numerical Results:}

We compute the energy corrections $E^{\text{MRPT}(2)}$ for multiple chemical systems to investigate the quality of our perturbation theory method.
For computational feasibility, we did not run a full quantum algorithm by simulating the time-evolution classically, instead we computed $E^{\text{MRPT}(2)}$ by implementing the resolvent of $H_{CAS}$ directly.
Moreover, we systematically study the behaviour of the perturbation coefficients norms $\norm{v^{\text{MRPT(2)}}}_{2/3}$ and $\norm{v_{\text{diff}}}_{2/3}^2$ for increasing system sizes.
In the computation of all results the quantum chemistry software PySCF \cite{PySCF} was employed.
Code and instructions to reproduce our results are shared in a GitHub repository at \url{github.com/jagunther/more_quant_chem_fewer_qub}.

\begin{figure}[!t]
    \includegraphics[width=\linewidth]{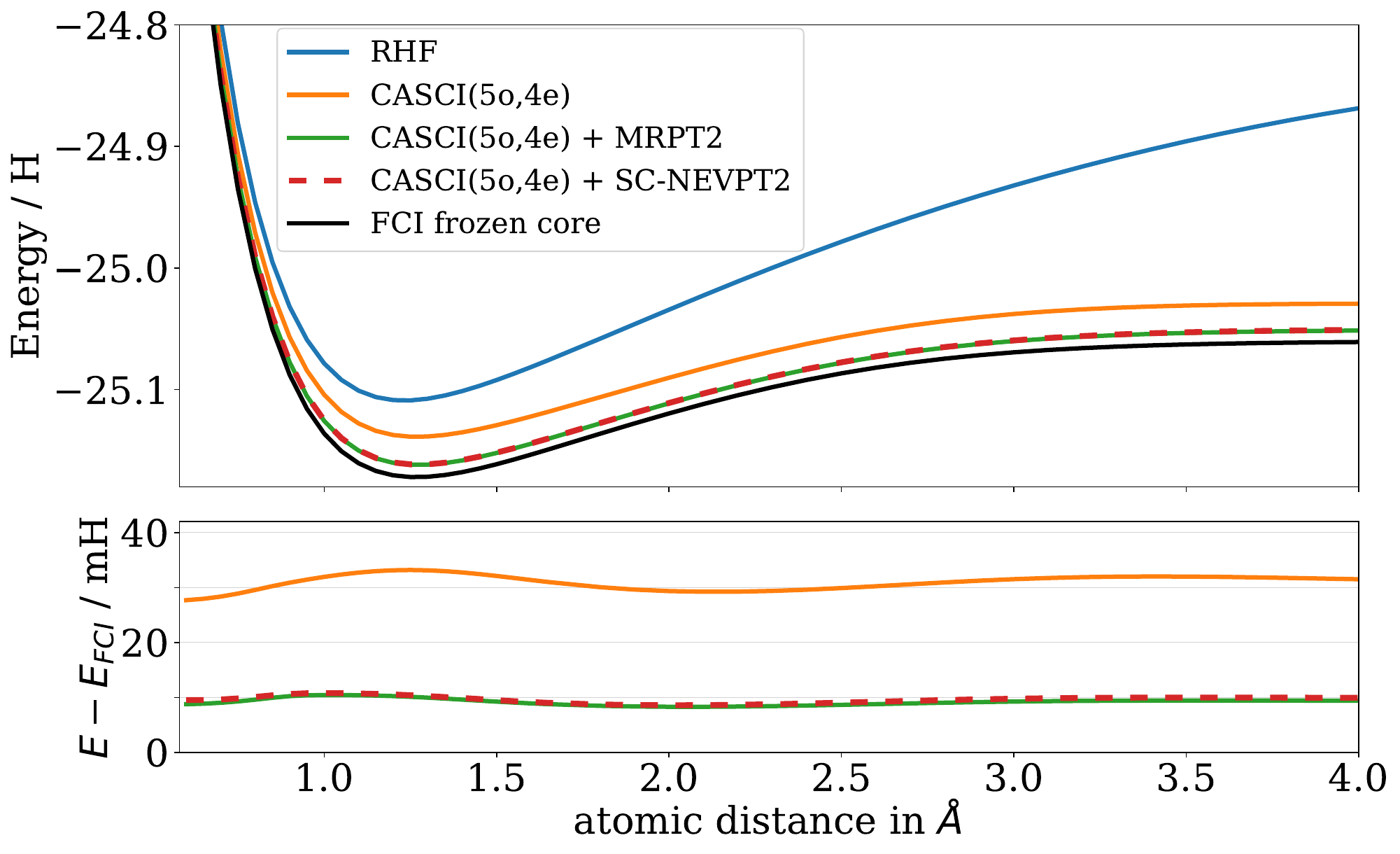}
    \caption{Potential-energy surface scan of the B-H molecule with different methods: Restricted Hartree-Fock (blue), CASCI with four electrons in five spatial orbitals with and without the MRPT2 correction (orange and green, respectively) and Full-CI with frozen core electrons. (Upper) Absolute energies, (lower) energies differences with respect to the Full-CI frozen core energies.}
    \label{fig:BH_curve}
\end{figure}

First, the impact of the MRPT2 corrections is demonstrated numerically with the dissociation curve of the singlet ground state of the B-H molecule in the 6-31G basis set, see Fig. \ref{fig:BH_curve}.
Based on Hartree-Fock orbitals, a full-valence space active space with five spatial orbitals and four electrons was chosen, yielding the CASCI(5o,4e) curve.
This corresponds to $k=10$ active spin-orbitals and with the frozen-core approximation (the 1$s$ orbital of B is frozen) the full number of spin-orbitals is $K=20$.
Considering this relatively small system allows us to do a full-CI calculation as an exact reference (red curve).
The green curve shows how the MRPT2 correction improves the active space calculation by including contributions from the virtual orbitals.
In fact it recovers about two thirds of the missing correlation energy of CASCI(5o,4e) compared to the frozen core Full-CI result, and as seen in Table \ref{tab:dissociation_energies}, the error in the dissociation energy $D_e$ is reduced from 3.1 mH to 0.8 mH.
Despite the relatively small size of the basis set, this result highlights an important fact: To achieve ``chemical accuracy'' ($1.6$\;mH) it is necessary to go beyond the active space approximation and take into account dynamical correlation effects.

\begin{table}[!b]
    \centering
    \begin{tabular}{l|c|c}
        Method & $D_e$ in mH & $E_{min}$ in H \\ \hline
        CASCI(5o,4e) & 108.7 & -25.1388\\
        CASCI(5o,4e) + SC-NEVPT2 & 111.1 & -25.1616 \\
        CASCI(5o,4e) + MRPT2 & 111.0 & -25.1619\\
        Full CI frozen core & 111.8 & -25.1720 
    \end{tabular}
    \caption{Dissociation and minimum energies of \ch{BH}.}
    \label{tab:dissociation_energies}
\end{table}

Furthermore, to investigate the effect of the MRPT2 correction on a larger chemical system with a barrier, we considered the interconversion reaction from oxywater (\ch{H2OO}) to hydrogen peroxide (\ch{HOOH}) via a transition state (TS).
This system has sparked interest among theoretical chemists in the 1990s, because it was debated whether oxywater was a stable minimum on the potential energy surface or not \cite{meredithOxywaterWaterOxide1992,huangCanOxywaterBe1996}.
Starting from a restricted Hartree-Fock (RHF) calculation in the def2-TZVP basis set, we optimized the geometries of \ch{H2OO}, \ch{HOOH} and the transition state at the CASCI(10o, 14e) level.
This active space includes all valence orbitals and valence electrons, translating to the number of active and total spin-orbitals $k=20$ and $K=144$, respectively.
At the resulting geometries, two different second-order perturbative energy corrections were calculated, the MRPT2-correction described in this paper, denoted MRPT2, and a traditional multireference perturbation correction, (strongly contracted)-NEVPT2.
All results are reported in Table \ref{tab:HOOH}.
To put the change of barrier height of the transition state into perspective, the reaction rate $k_{rate}$ for the reaction $\ch{H2OO}\to\ch{HOOH}$ predicted by transition state theory is dominated by the scaling $k_{rate} \propto \exp{-\Delta G_{TS}/(k_B T)}$, where $\Delta G_{TS}$ is the difference in Gibbs free energy between the transition state and oxywater and $k_B$ is the Boltzmann-constant.
The change in Gibbs free energy is dominated by the change in electronic energy $\Delta E_{TS} = E_{TS} - E_{\ch{H2OO}}$, therefore we approximate $k_{rate} \propto \exp{-\Delta E_{TS}/(k_B T)}$.
According to this scaling, the reduction in barrier height from $44.89$ mHartree to $16.11$ mHartree because of the MRPT2 correction increases $k_{rate}$ by a factor of $10^5$ at $T=300$ K.
Despite the approximations involved in our argument, it is evident that an exact solution in the active-space approximation is not sufficient to obtain quantitative results.
Although this is not a new insight in the field of traditional quantum chemistry, this fact is often disregarded in the literature on quantum computing for chemistry, where the active-space approximation has become a standard.
The difference between MRPT2 correction and the SC-NEVPT2 energy correction is non-negligible for this bigger system.
As mentioned before, the MRPT2 correction computed with our method is equivalent to the totally-uncontracted NEVPT2 correction. 
Therefore, the observed difference between MRPT2 and SC-NEVPT2 can be attributed to the approximations made in the definition of the SC-NEVPT2 correction.

\begin{figure}
     \centering
     \begin{subfigure}[b]{0.3\linewidth}
         \centering
         \includegraphics[width=\textwidth]{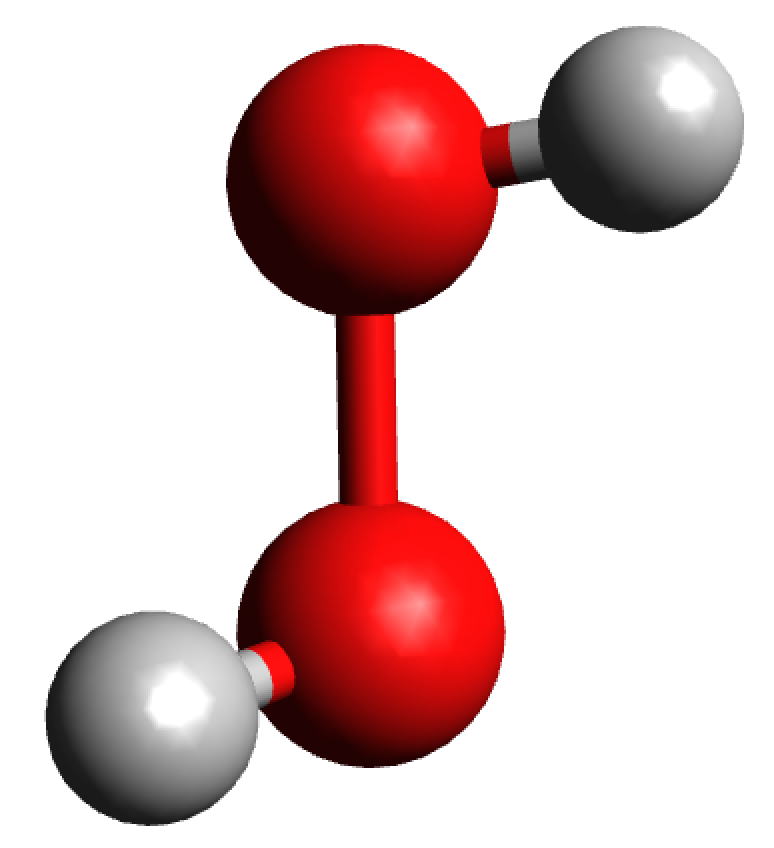}
         \caption{\ch{HOOH}}
         \label{fig:y equals x}
     \end{subfigure}
     \hfill
     \begin{subfigure}[b]{0.27\linewidth}
         \centering
         \includegraphics[width=\textwidth]{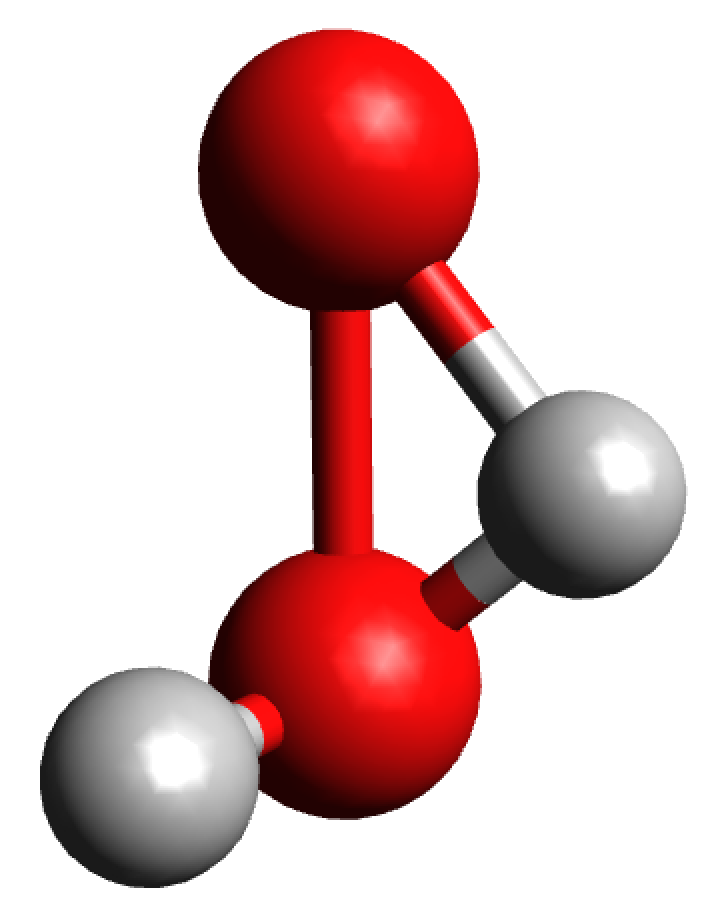}
         \caption{TS}
         \label{fig:three sin x}
     \end{subfigure}
     \hfill
     \begin{subfigure}[b]{0.3\linewidth}
         \centering
         \includegraphics[width=\textwidth]{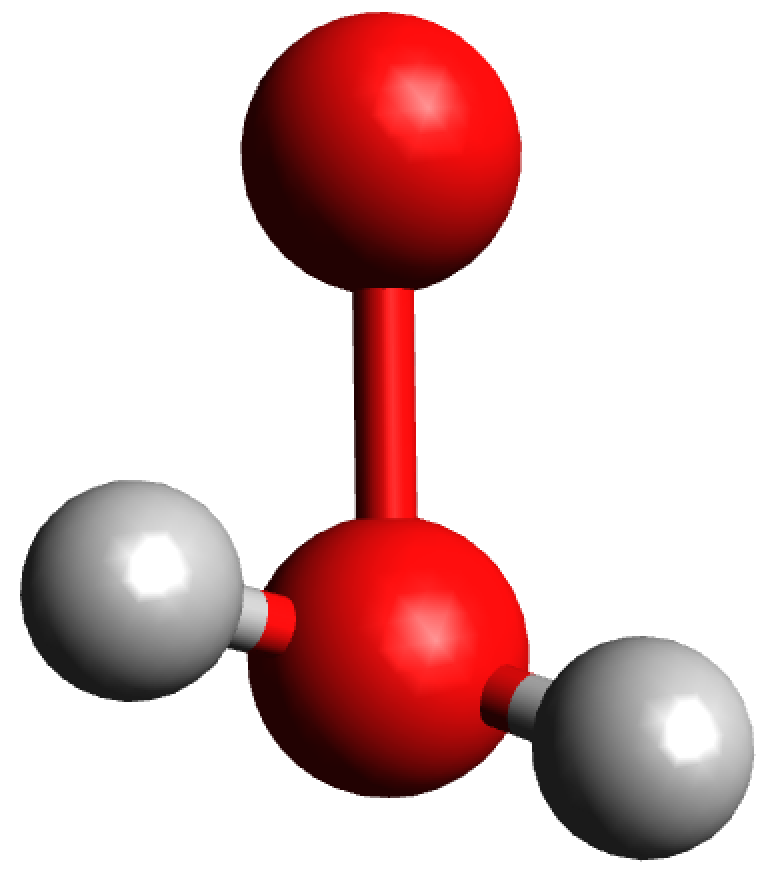}
         \caption{\ch{H2OO}}
         \label{fig:five over x}
     \end{subfigure}
        \caption{Geometries of (a) hydrogen peroxide (b) the transition state and (c) oxywater.}
        \label{fig:HOOH_geometries}
\end{figure}

\begin{table*}
    \centering
    
    \begin{tabular}{l|c|c|c|c|c}
        method & \ch{HOOH} & TS & \ch{H2OO} & $\Delta E_{TS}$  \\ \hline
        RHF & -396015.58 & -395792.72 & -395869.61 & 76.89 \\
        CASCI(10o,14e) & -396117.08 & -395891.00 & -395935.89 & 44.89 \\
        CASCI(10o,14e) + SC-NEVPT2 & -397181.21 & -396951.40 & -396971.30 & 19.89 \\
        CASCI(10o,14e) + MRPT2 & -397201.63 & -396975.35 & -396991.46 & 16.11 \\
    \end{tabular}
    \caption{Energies in kJ/mol of hydrogen peroxide, the transition state and oxywater in the basis set def2-TZVP.
    All geometries are optimized at the CASCI(10o,14e) level, and for all methods the frozen-core approximation is used.}
    \label{tab:HOOH}
\end{table*}

\begin{figure}
    \centering
    \includegraphics[width=\linewidth]{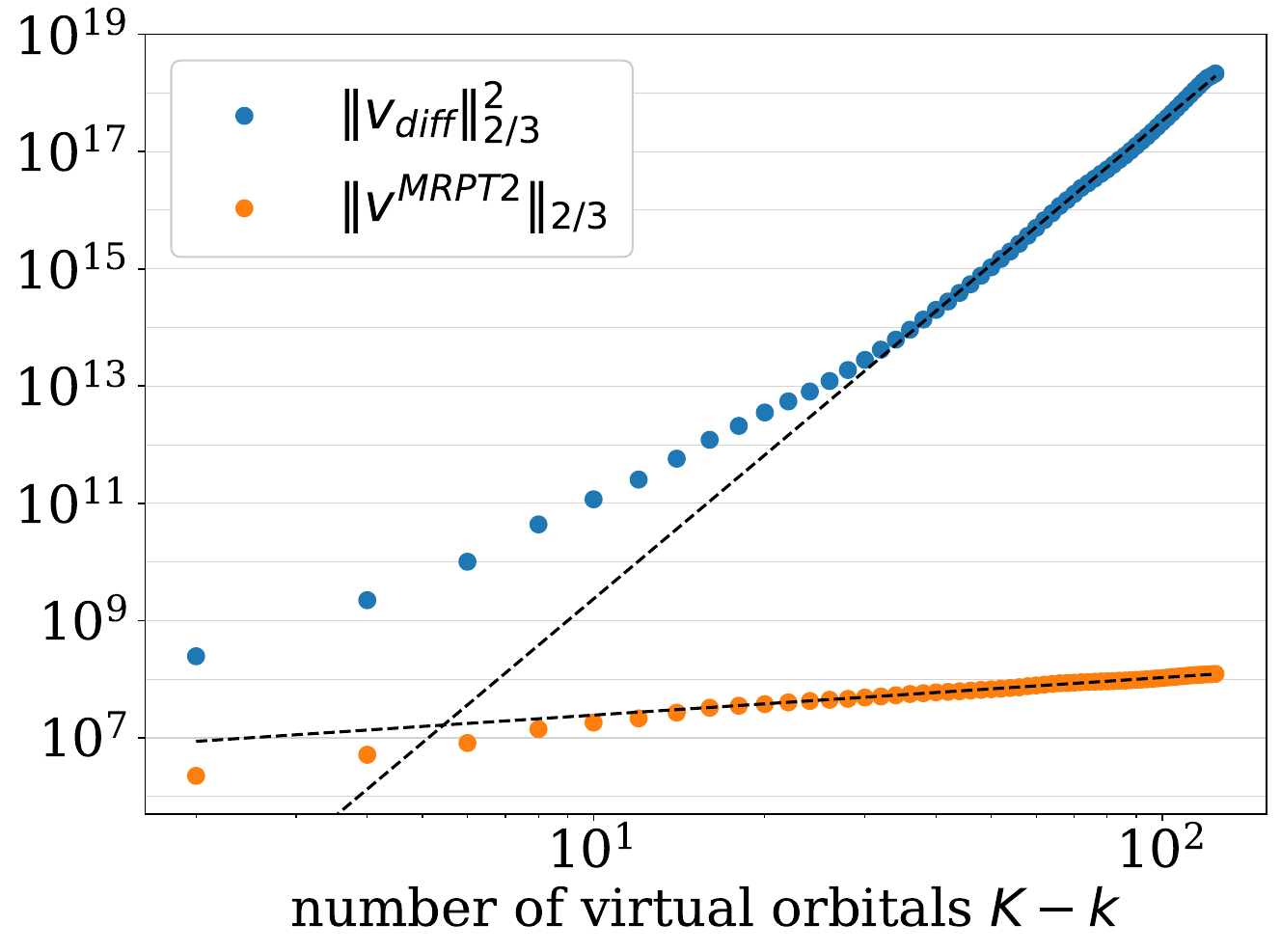}
    \caption{Perturbation coefficient norms $\norm{v_{\text{diff}}}_{2/3}^2$ and $\norm{v^{\text{MRPT}2}}_{2/3}$ at the \ch{H2O2} geometry with increasing number of virtual orbitals. The dashed lines are power-law fits $\alpha (K-k)^{\beta}$ to the data points, the obtained values $\alpha$ and $\beta$ are given in Table \ref{tab:power_law_fit}.}
    \label{fig:norm_v_mrpt2_kconst}
\end{figure}

\begin{figure}
    \centering
    \includegraphics[width=\linewidth]{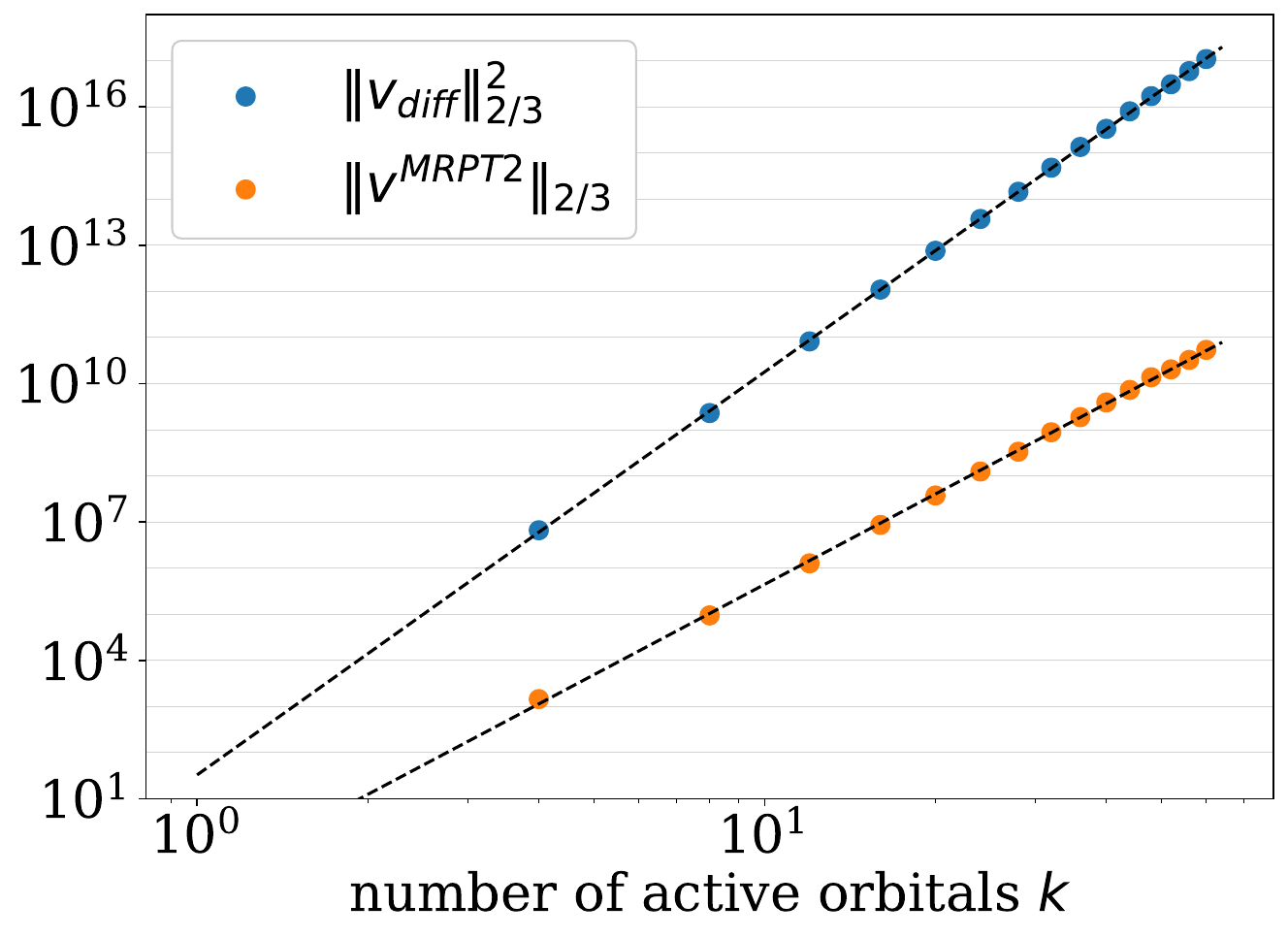}
    \caption{Perturbation coefficient norms $\norm{v_{\text{diff}}}_{2/3}^2$ and $\norm{v^{\text{MRPT}2}}_{2/3}$ for the hydrogen chain with increasing number of active orbitals while $k/K = const.$ The dashed lines are power-law fits $\alpha k^{\beta}$ to the data points, the obtained values $\alpha$ and $\beta$ are given in Table \ref{tab:power_law_fit}.}
    \label{fig:norm_v_mrpt2_K/kconst}
\end{figure}

Furthermore, for \ch{H2O2} at the TS geometry we have investigated the scaling of $\norm*{v^{\text{MRPT2}}}_{2/3}$ and $\norm{v_{\text{diff}}}_{2/3}^2$ with increasing number of virtual orbitals, the results are shown in Fig. \ref{fig:norm_v_mrpt2_kconst}.
The data points we compute are upper bounds to the relevant coefficient norms, but the bounds are tight up to a constant factor. Details of this technical aspect are provided in Appendix \ref{appendix:mrpt2_norm}.
The difference in order of magnitude between the two quantities is tremendous, for the highest number of virtual orbitals considered, $K-k =122$, $\norm*{v^{\text{MRPT2}}}_{2/3}$ and $\norm*{v_{\text{diff}}}_{2/3}^2$ differ by about 10 orders of magnitude.
Fitting a power law to the data points resulted in exponents $0.8$ and $6.2$ for $\norm*{v^{\text{MRPT2}}}_{2/3}$ and $\norm*{v_{\text{diff}}}_{2/3}^2$, respectively, underlining the difference in scaling.
In this regard, we conclude that our approach to compute dynamical correlation effects on a quantum computer constitutes a significant improvement over the previous algorithm of Mitarai \textit{et al} \cite{mitarai_perturbation_2023}.
Additionally, the scaling of $\norm*{v_{\text{diff}}}_{2/3}^2$ and $\norm*{v^{\text{MRPT2}}}_{2/3}$ with increasing number of active orbitals was examined by studying hydrogen chains in the 6-31G basis set.
The ratio of active to virtual orbitals remains constant as the number of hydrogen atoms (and therefore of active orbitals) is increased.
Data points and fits are shown in Fig. \ref{fig:norm_v_mrpt2_K/kconst}, and the fitted power-law coefficients are given in Table \ref{tab:power_law_fit}. 
We observe that $\norm*{v^{\text{MRPT2}}}_{2/3}$ outperforms $\norm*{v_{\text{diff}}}_{2/3}^2$ by scaling lower by a quadratic factor in $k$.
However, $\norm*{v^{\text{MRPT2}}}_{2/3}$ still behaves as $k^{6.5}$, leaving open room for further improvements in the scaling with $k$.

\begin{table}
    \centering
    
    \begin{tabular}{c|c|c|c|c|}
         & \multicolumn{2}{c}{Scaling with $(K-k)$} & \multicolumn{2}{|c|}{Scaling with $k$} \\[1mm]
        & $\alpha$ & $\beta$ & $\alpha$ & $\beta$ \\[1mm] 
        \hline $\norm{v_{\text{diff}}}_{2/3}^2$ &$16$ & $8.2$ & $33$ & 8.7 \\[1mm]
        $\norm{v^{\text{MRPT}2}}_{2/3}$ & $5.6\cdot 10^6$ & $0.64$ & $0.13$ & 6.5
    \end{tabular}
    \caption{Fitted power-law coefficients for $\norm{v_{\text{diff}}}_{2/3}^2$ and $\norm{v^{\text{MRPT}2}}_{2/3}$ as a function of $k$ and $K-k$ as presented in Figs. \ref{fig:norm_v_mrpt2_kconst} and \ref{fig:norm_v_mrpt2_K/kconst}.}
    \label{tab:power_law_fit}
\end{table}

The true value of quantum computing in chemistry is for routine applications of multiconfigurational electronic structures, often found in 3d transition metal compounds. As an extreme case for such applications, we consider the FeMo-cofactor, the active site of the metalloenzyme nitrogenase \cite{Reiher2017_PNAS,lee_even_2021}. 
To describe the important static electron correlation in FeMoco, an active space of on the order of $k\sim 100$ active spin orbitals will be needed.
However, about $K\sim 4000$ spin orbitals would be required in a triple-zeta basis set to describe the entire electronic structure (static and dynamic correlation) of FeMoco, where the $K-k$ inactive orbitals could be taken into account via perturbation theory.
In this regard, the reduction in the number of qubits from a priori $\mathcal{O}(K)$ to $\mathcal{O}(k)$ qubits is significant and implies that even more accurate quantum chemical results could be obtained with a limited number of qubits.
While FeMoco is just one prominent example, the role of dynamical correlation is,  without a doubt, a general problem in computational chemistry for active-space based approaches \cite{fiedlerEnergeticalStructuralProperties1993, pierlootCASPT2MethodInorganic2003,steinDelicateBalanceStatic2016}.
Including dynamical correlation effects on top of active-space ground state energies obtained on a quantum computer is necessary in order for quantum computing to be of value for routine applications in quantum chemistry.

Finally, as we assumed $K\gg k$, we have focused on $K$ in our analysis, and our result has not been optimized with respect to the number of active orbitals, $k$. 
An improvement in the scaling of $\norm*{v^{\text{MRPT}(2)}}_1$, both in $K$ and in $k$, might be achieved by working out factorizations of the coefficient-tensors in Eq. \eqref{eq:sec_MRPT:defn_coeffs_MRPT}, similar to those presented in references \cite{von_burg_quantum_2021} and \cite{lee_even_2021}.
To account for the overall scaling of our method with $k$ it would not only be necessary to consider $\norm{\beta_{\Delta,R}}_1$, but also it would be relevant to inspect the implementation of the ground state preparation unitary $U_0$ and the time-evolution $U(t)$ under the unperturbed Hamiltonian, which we kept as black-boxes here.
Since we focused on an asymptotic complexity analysis without taking too much care of prefactors, the upper bounds in our scaling are likely to be very loose.
Moreover, our cost model is in terms of access to the ground state preparation unitary $U_0$ and the time-evolution $U(t)$, which do not directly translate to gate counts.
Including the explicit implementation of these subroutines and working out additional improvements for the scaling in $k$ is an important step to make multireference perturbation theory on a quantum computer even more practical, however, it is beyond the scope of this paper and will be subject of a future study.

\sloppy \section{Application to Symmetry-Adapted Perturbation Theory}
\label{section:SAPT}

Symmetry-adapted perturbation theory (SAPT) is a method for calculating the interaction energy between two molecules A and B perturbatively \cite{jeziorski_perturbation_1994}.
The molecular complex formed by A and B together is called ``dimer'', while A and B are referred to, individually, as ``monomers''.
The interaction energy $E_{int}$ is defined as the difference between the energy of the dimer, $E_{AB}$, and the sum of energies of the monomers, $E_A + E_B$:
\begin{equation}
  E_{int} = E_{AB} - E_A - E_B \, .
  \label{eq:def_E_int}
\end{equation}
This difference in energy is computed in the Born-Oppenheimer approximation, so each of the three energies corresponds to an electronic energy calculated with clamped nuclei. 
Furthermore, the monomers are treated as rigid, i.e. to obtain the dimer configuration, the monomer configurations of the nuclei (used to compute $E_A$ and $E_B$) are kept frozen and are simply placed in proximity to each other.
This way, for given monomer configurations of the nuclei, the interaction energy $E_{int}$ is a function only of the distance between the monomers and a set of angles describing their mutual orientation.
Additional energy contributions caused by relaxation of the nuclei in the dimer configuration are typically small \cite{szalewicz_comment_1998}.

When applying SAPT, particular care must be paid to the antisymmetry properties of the electronic wave function.
In the standard description of many-electron wave functions, a state is an element of the antisymmetrized Hilbert space $\Lambda^{N}(\mathcal{H}_1)$, which is spanned by all $N$-electron Slater determinants constructed from molecular spin orbitals in the single-particle space $\mathcal{H}_1$.
By contrast, SAPT treats the $N_A$ electrons of A and $N_B$ electrons of B as two distinguishable sets of particles, described by two different single-particle Hilbert spaces $\mathcal{H}_{1,A}$ and $\mathcal{H}_{1,B}$.
The full many-electron Hilbert space is then constructed as $\mathcal{H}_{AB} = \mathcal{H}_A \otimes \mathcal{H}_B = \Lambda^{N_A}(\mathcal{H}_{1,A})\otimes \Lambda^{N_B}(\mathcal{H}_{1,B})$ \cite{rybak_manybody_1991}.
The corresponding two sets of annihilation and creation operators, $a_{\mu}, a_{\mu}^{\dag}$ for $\mathcal{H}_A$ and $b_{\nu}, b_{\nu}^{\dag}$ for $\mathcal{H}_B$, fulfill the usual fermionic commutation relations within each set.
However, operators belonging to different sets commute.
The two molecular spin orbital basis sets are denoted by $\mathcal{I}_A = \{\phi_1^A,\dots,\phi_{K_A}^A\}$ and $\mathcal{I}_B = \{\phi_1^B,\dots,\phi_{K_B}^B\}$. They are usually chosen to be ``dimer-centered bases'' i.e., both bases describe a set of molecular spin orbitals spanning the whole dimer configuration and typically expressed in terms of the same atom-centered Gaussians \cite{patkowski_recent_2020}.
Other basis set choices are viable, for example monomer-centered basis sets with additional basis functions spanning the interaction region \cite{williams_effectiveness_1995}.

The unperturbed Hamiltonian $H$ is defined as the sum of Hamiltonians $H_A$ and $H_B$, they describe the monomers individually on some level of electronic structure theory, and their ground state energies are $E_A$ and $E_B$.
Consequently, the ground state $\ket{\Phi_0}$ of $H$ is the product of the monomer ground states, $\ket*{\Phi_0^A}$ and $\ket*{\Phi_0^B}$, with energy given by

\begin{equation}
\begin{aligned}
      H \ket{\Phi_0} &= (H_A \otimes \mathbb{1} + \mathbb{1} \otimes H_B)\ket*{\Phi_0^A}\otimes\ket*{\Phi_0^B}\\
      &= (E_A + E_B)\ket{\Phi_0}.
  \label{eq:SAPT_SchrodingerEquation}
\end{aligned}
\end{equation}

The perturbation operator $V$ contains the interaction between the monomers, i.e. the electron-electron, electron-nuclei and nuclei-nuclei interactions between particles associated with A and B.
With this choice of $H$ and the resulting perturbative expansion of $E_{AB}$, the energies of the monomers cancel in the expression of the interaction energy in~\eqref{eq:def_E_int}, which is then given by the perturbative energy corrections only.
However, the Rayleigh-Schrödinger energy-corrections (see Sec. \ref{section:PT}) are modified in order to mitigate the errors caused by the unphysical partitioning into A and B electrons.
This is achieved by applying \textit{a-posteriori} an antisymmetrization operator $\mathcal{A}$ onto the wave function, yielding the so-called Symmetrized Rayleigh-Schrödinger (SRS) corrections $E_{SRS}^{(k)}$, which are summed to yield the interaction energy.

In practice the antisymmetrizer $\mathcal{A}$ is approximated by a one-electron exchange operator \cite{moszynski_manybody_1994,patkowski_recent_2020}
\begin{gather*}
    \mathcal{P}_1 = -\sum_{ijkl} S_{il}S_{kj} a_i^{\dag} a_j b_k^{\dag}b_l\\
    \text{where} \qquad S_{il} = \int \overline{\phi_i^A(\vb{x})} \phi_l^B(\vb{x}) d\vb{x}
\end{gather*}
is the overlap between molecular orbitals $\phi_i^A(\vb{x})$ and $\phi_l^B(\vb{x})$.
This approximation, usually referred to as the $S^2$-approximation \cite{jeziorski_perturbation_1994},
yields the following expression for the first- and second-order corrections:

\begin{equation}
  \begin{aligned}
    E_{SRS}^{(1)}(S^2) &= \expval{V} + \expval{(\expval{V} - V)\mathcal{P}_1}\\
    &=: E_{pol}^{(1)} + E_{exch}^{(1)}\\
    E_{SRS}^{(2)}(S^2) &= \expval{VR_0V} + \expval{(\expval{V}-V)(\expval{\mathcal{P}_1} - \mathcal{P}_1)R_0V}\\
    &=: E_{pol}^{(2)} + E_{exch}^{(2)}\, ,
  \end{aligned}
  \label{eq:FirstSecondOrderCorrection}
\end{equation}
where we have used the usual partitioning into the so-called polarization and exchange energies.
The inter-monomer interaction is given by \cite{moszynski_mollerplesset_1993}

\begin{equation}
\begin{aligned}
  V &= V_{elA-elB} + V_{elA-nucB} + V_{nucA-elB} - V_0\\
  &= \sum_{pqrs} g_{pqrs}^{AB} a_p^{\dag} a_s b_q^{\dag}b_r 
    + \sum_{jk} h_{jk}^{B} a_k^{\dag}a_j\\
    &\phantom{{}\quad }+ \sum_{lm}h_{lm}^{A} b_{m}^{\dag}b_l - V_0,
  \label{eq:PerturbationOperator}
\end{aligned}
\end{equation}

where
\begin{align*}
    g_{pqrs}^{AB} &= \int \int \, \frac{\overline{\phi^A_p}(\vb{x}_1)\overline{\phi^B_q}(\vb{x}_2)\phi_r^B(\vb{x}_2)\phi_s^A(\vb{x}_1)}{\abs{\vb{x}_1-\vb{x}_2}} d\vb{x}_1 d\vb{x}_2\\
    h_{jk}^{B} &= -\sum_{i=1}^{N_{nuc}^B} \mel{\phi_j}{\frac{Z_i}{\abs{\mathbf{r} - \mathbf{R}_{B_i}}}} {\phi_k}\\
    h_{lm}^{A} &= -\sum_{i=1}^{N_{nuc}^A} \mel{\phi_l}{\frac{Z_i}{\abs{\mathbf{r} - \mathbf{R}_{A_i}}}} {\phi_m}\\
    V_0 &= \sum_{i,j=1}^{N_{nuc}^A,N_{nuc}^B}\frac{Z_{A_i}Z_{B_j}}{\abs{\mathbf{R}_{A_i}-\mathbf{R}_{B_j}}}\, .
\end{align*}
We will now illustrate how our method enables us to compute contributions to first- and second-order corrections while only having access to one of the states.
As an example, we consider the contribution $\expval{V_{elA-nucB}\mathcal{P}_1R_0V_{elA-elB}}$ to the second-order correction:
\begin{align*}
    &\expval{V_{elA-nucB}\mathcal{P}_1R_0V_{elA-elB}}\\
    &= \sum_{\substack{ijkl\\mn\\pqrs}} S_{il} S_{kj} h_{mn}^B g_{pqrs}^{AB} \expval{a_n^{\dag} a_m a_i^{\dag} a_j b_k^{\dag}b_l R_0 a_p^{\dag} a_s b_q^{\dag}b_r} \\
    &= \sum_{n \in \mathbb{Z}} \sum_{\substack{ijkl\\mn\\pqrs}} S_{il} S_{kj} h_{mn}^B g_{pqrs}^{AB}\beta_{\Delta,R,n} e^{iE_0t_n}\\
    &\phantom{{}=\:\:} \times \expval{a_n^{\dag} a_m a_i^{\dag} a_j b_k^{\dag}b_l e^{-iHt_n} a_p^{\dag} a_s b_q^{\dag}b_r}\\
    &= \sum_{n \in \mathbb{Z}} \sum_{\substack{ijkl\\mn\\pqrs}} S_{il} S_{kj} h_{mn}^B g_{pqrs}^{AB}\beta_{\Delta,R,n} e^{iE_0t_n}\\ 
    &\phantom{{}=\:\:} \times \expval{a_n^{\dag} a_m a_i^{\dag} a_j e^{-iH_A t_n}a_p^{\dag} a_s}_A
    \expval{b_k^{\dag}b_l e^{-iH_Bt_n} b_q^{\dag}b_r}_B\, ,
\end{align*}
where $\expval{\dots}_X = \expval{\dots}{\Phi_0^X}$ involves only one of the monomer.
Consequently, $\expval{V_{elA-nucB}\mathcal{P}_1R_0V_{elA-elB}}$ can be computed from quantities that are obtained from the monomers' states $\ket{\Phi_0^A}, \ket{\Phi_0^B}$ individually.
The procedure is analogous for the other contributions to the energy corrections.

\section{Conclusions and Outlook}
\label{section:conclusions}

In this paper, we presented a general quantum algorithm for the calculation of second-order perturbative energy corrections, and specified it for molecular electronic energies.

For the application to multi-reference perturbation theory, we found that no qubits are required to represent the virtual orbitals, and in general the quantum algorithm requires only a single ancilla qubit.
We have analyzed our quantum algorithm in terms of calls to $U_0$ and total Hamiltonian evolution time under $H$ with a focus on minimzing the $K$ dependence since it is the one deemed the most impactful, in contrast to the number of active space orbitals $k$.

In terms of $K$, we observed numerically for a chemically realistic system that with our approach for a fixed active space (i.e. fixed $k$), we achieve sublinear scaling in the number of virtual orbitals by taking the structure of the unperturbed Hamiltonian into account explicitly.
By contrast, we also showed that naively computing the second-order correction results in a runtime that scales to the eighth power in the number of virtual orbitals, which lead the previous paper on fault-tolerant MRPT to the conclusion that their method was impractical \cite{mitarai_perturbation_2023}.
Consequently, our method opens up the way to treat a significant number of virtual orbitals, which is a key improvement for multi-reference perturbation theory.
Furthermore, we demonstrated that the energy correction computed with our method significantly improves on the result of the active space calculation by including dynamical correlation effects of virtual orbitals. 

With this, we have laid the groundwork for multi-reference perturbation theory on a quantum computer.
While the dependence on $k$ is polynomial, we have not optimized it. 
The asymptotic analysis we performed might be improved by lowering the upper bounds through a more careful analysis of the prefactors, which will allow for realistic resource estimates for molecular systems of interest.
For the application to symmetry-adapted perturbation theory, we found that it is possible to compute the second-order intermolecular interaction energy while only having access to one of the monomers' states at a time.

Our algorithm can be considered a method to be implemented on a quantum computer to refine the energy (e.g., evaluated by quantum phase estimation) of a discretized representation of a physical model.
The discretization may be achieved by activating only a limited number of one-particle functions (i.e. orbitals), the so-called active space.
In turn, the reduction in the number of orbitals reduces the accuracy of the energy obtained (as is also the case for state-of-the-art traditional approaches such as CASSCF, DMRG, and FCIQMC) and requires a procedure that allows one to assess the importance of  the many neglected orbitals for a quantitatively correct total energy of the physical system.
Our approach delivers perturbative energy corrections for this assessment that can be viewed as an approximation of the error introduced by switching to a reduced-dimensional discretization (the active space) from a more fine-grained discretization with far more orbitals.

In terms of applications, the algorithm can be used to improve on QPE results and hence provides a quantum MRPT version of traditional approaches such as CASPT2 and NEVPT2.
As such, it can be a key component of endeavors to elucidate reaction mechanisms as proposed for enzymatic and catalytic reactions in Refs. \citenum{Reiher2017_PNAS,von_burg_quantum_2021} that would otherwise require switching to a classical computer for this task. 

However, our algorithm is also applicable in the field of intermolecular interactions, where the SAPT approach is known to deliver  accurate interaction energies for systems that cannot be treated as a whole, but only in terms of its parts (for instance, in host-guest chemistry).
As such, our algorithm can become an integral part of drug-discovery and drug-design attempts, where accurate intermolecular interaction energies are
key to the free energy of a docking process.
Moreover, we note that our setting and algorithm might be of value also to other fields, not only those that rely on electronic or vibrational structure theory applied in chemistry, solid state physics, materials science, and biochemistry.

We have focused our presentation here to second-order perturbation theory as this a relevant and widely used method in practical applications. Still, higher-order perturbative corrections can, in a straightforward fashion, be considered within our framework.

\section*{Acknowledgements}
\sloppy We acknowledge financial support from the Novo Nordisk Foundation (Grant No. NNF20OC0059939 ‘Quantum for Life’). 
Part of the work conducted by M.C., J.G. and M.R. within the frame of this collaboration is part of the Research Project ``Molecular Recognition from Quantum Computing''. Work on Molecular Recognition from Quantum Computing'' is supported by Wellcome Leap as part of the Quantum for Bio (Q4Bio) Program.
M.C. and J.G. also acknowledge support from the European Research Council (ERC Grant Agreement No. 81876) and VILLUM FONDEN via the QMATH Centre of Excellence (Grant No.10059).

\section*{Appendix}
\appendix
\addcontentsline{toc}{section}{Appendices}

\subsection{Definition of $g_{\Delta,R}(E)$}
\label{appendix:def_g}
In this Appendix we provide the definition of the function $g_{\Delta,R}(E)$, which is used to implement the resolvent $R_0$ in Sec. \ref{section:QPT2}.
We start from
\begin{equation*}
    g'_{\Delta,R}(E) = -\frac{\mathbb{1}_{[\frac{\Delta}{4},R+\Delta\frac{3}{4}]}}{E}\, ,
\end{equation*}
where $\mathbb{1}_{[a,b]}$ is the indicator function of the interval $[a,b]\subset \mathbb{R}$.
The function $g_{\Delta,R}(E)$ can be smoothened by multiplying it with another indicator function ($h_{\Delta,R}$ below) that smoothly decays at its edges and whose support is inside the support of $g'_{\Delta,R}$.
To construct $h_{\Delta,R}$, we employ a standard bump function
\begin{equation*}
    \eta(E) = \begin{cases}
        c_{\eta}e^{-(1-E)^{-1}}e^{-(1+E)^{-1}},\quad \abs{E} < 1\\
        0, \quad \text{else}
    \end{cases}
\end{equation*}
where $c_{\eta}$ is a normalizing constant such that 
\begin{equation}
    \int_{-1}^1 \eta(E) dE = 1\, . \label{eq:c_eta}
\end{equation}
The function $\eta(E)$ is smooth but not analytic, and the derivatives $\frac{d^n}{dE^n}\eta$ at $E=1$ and $E=-1$ are zero for all $n$ \cite{nestruev_smooth_2003}.
We define the narrowed mollifier 
\begin{equation*}
    \eta_{\Delta}(E) = \frac{4}{\Delta}\eta\Big(\frac{4E}{\Delta}\Big)\, ,
\end{equation*}
whose support is $[-\frac{\Delta}{4},\frac{\Delta}{4}]$ and which is still normalized. 
The function 
\begin{equation}
    h_{\Delta,R}(E) = (\mathbb{1}_{[\frac{3}{4}\Delta, R+\frac{\Delta}{4}]} \ast \eta_{\Delta})(E)
\end{equation}
\begin{figure}
    \centering
    \includegraphics[width=\linewidth]{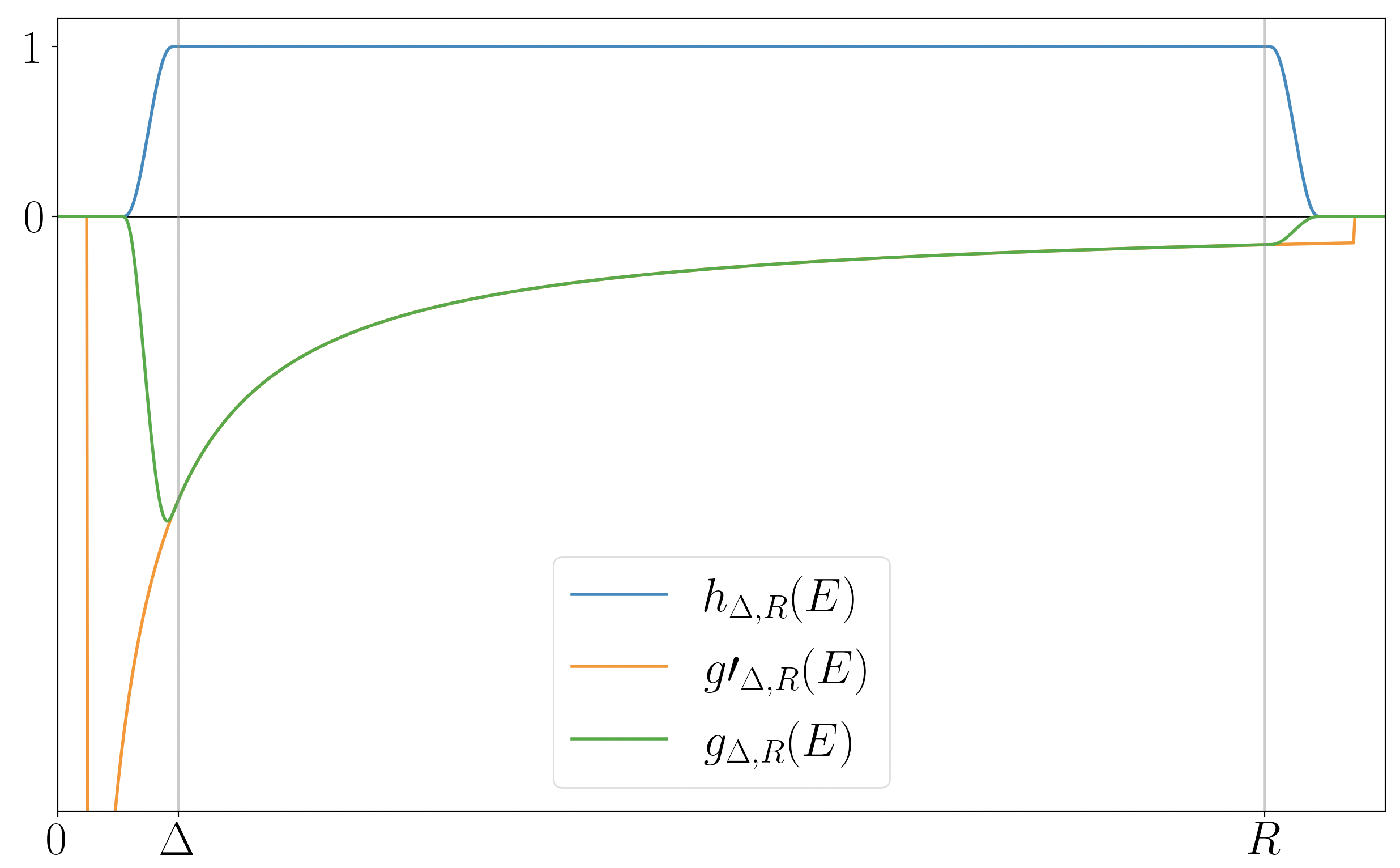}
    \caption{Illustration of the functions used in the construction of $g_{\Delta,R}(E)$, the function playing the role of the resolvent.}
    \label{fig:approx_resolvent_3fcts}
\end{figure}
evaluates to $1$ on the interval $[\Delta,R]$ and goes smoothly to $0$ at $E=\frac{\Delta}{2}$ and $E=R+\frac{\Delta}{2}$, see also Fig. \ref{fig:approx_resolvent_3fcts}.

Consequently, we can define the function $g_{\Delta, R}(E)$ with the desired properties by
\begin{equation}
    g_{\Delta,R}(E) = (g'_{\Delta,R} \cdot h_{\Delta,R})(E)\, . \label{eq:defn_g}
\end{equation}

\subsection{Bounds on Fourier Coefficients of $g_{\Delta,R}(E)$}
\label{appendix:fourier_bound}
The Fourier coefficients of $g_{\Delta,R}(E)$ are defined as
\begin{gather*}
    \beta_{\Delta,R,n} = \frac{1}{R+\Delta}\int_{0}^{R+\Delta} g_{\Delta,R}(E)e^{-iEt_n}dE\, ,\\
    t_n = \frac{2\pi n}{R+\Delta}\, , \quad n\in \mathbb{Z}\, ,
\end{gather*}
and in this section we prove an upper bound on the Fourier coefficients $\beta_{\Delta,R,n}$ in terms of $n, \Delta$ and $R$.
We use a standard approach and relate the decay properties of the Fourier coefficients to the smoothness of $g_{\Delta,R}(E)$. 
In particular, we prove the following lemma

\begin{lemma}
    \label{lemma:fouriercoeff_bound}
    The Fourier coefficients $\beta_{\Delta,R,n}$ of the function $g_{\Delta,R}(E)$ on the interval $[0,R+\Delta]$ satisfy
    \begin{equation*}
        \abs*{\beta_{\Delta,R,n}} \leq \frac{61R}{\Delta^2} 2^{-(\frac{\abs{n}\Delta}{256R})^{1/2}}
    \end{equation*}
    for all $\abs{n}\geq 0$ and $R\geq \Delta > 0$.
\end{lemma}

\noindent\textit{Proof:} 
The Fourier coefficients relate to the Fourier transform $\widehat{g_{\Delta,R}}(t)$ of $g_{\Delta,R}(E)$ by
\begin{gather*}
    \beta_{\Delta,R,n} = \frac{1}{R+\Delta} \widehat{g_{\Delta,R}}(t_n)\\
    \text{where} \quad \widehat{g_{\Delta,R}}(t) = \smashoperator{\int_{0}^{R+\Delta}} g_{\Delta,R}(E)e^{-iEt}dE\, .
\end{gather*}
Since the Fourier transform of the $\ell$th derivative (which we denote by $D^{\ell}$) is given by \cite{grubb_distributions_2009}
\begin{equation*}
    \widehat{D^{\ell} g_{\Delta,R}}(t) = (-it)^{\ell}\widehat{g_{\Delta,R}}(t)
\end{equation*}
we can take absolute values to get
\begin{equation*}
    \abs*{\beta_{\Delta,R,n}} = \frac{1}{R+\Delta} \abs*{\widehat{g_{\Delta,R}}(t_n)} = \frac{1}{R+\Delta}\frac{\abs*{\widehat{D^{\ell}g_{\Delta,R}}(t_n)}}{\abs{t_n}^{\ell}}\, .
\end{equation*}
By construction, $g_{\Delta,R}(E)$ itself and with all its derivatives vanishes outside the interval $[\frac{\Delta}{2},R + \frac{\Delta}{2}]$.
Therefore we can employ the bound
\begin{equation*}
    \abs*{\widehat{D^{\ell}g_{\Delta,R}}(t)} \leq \int_{\frac{\Delta}{2}}^{R+\frac{\Delta}{2}} \abs{D^{\ell}g_{\Delta,R}(E)}dE = R\norm{D^{\ell} g_{\Delta,R}}_{\infty}\, ,
\end{equation*}
which results in
\begin{equation}
    \abs*{\beta_{\Delta,R,n}} \leq \frac{R}{R+\Delta} \frac{\norm{D^{\ell}g_{\Delta,R}}_{\infty}}{\abs{t_n}^{\ell}} \leq \frac{\norm{D^{\ell}g_{\Delta,R}}_{\infty}}{\abs{t_n}^{\ell}}\, .
    \label{eq:appdx_boundsFourier:ineq_beta_1}
\end{equation}
Next, by means of the Leibniz product rule the numerator of inequality Eq. \eqref{eq:appdx_boundsFourier:ineq_beta_1} is upper bounded by
\begin{align}
    &\norm{D^{\ell}g_{\Delta,R}}_{\infty} = \underset{E\in[\frac{\Delta}{2},R + \frac{\Delta}{2}]}{\text{sup}}{\abs{D^{\ell}(g'_{\Delta,R}\cdot h_{\Delta,R})(E)}} \label{ineq:Dl_g_DeltaR} \\
    &= \underset{E\in[\frac{\Delta}{2},R + \frac{\Delta}{2}]}{\text{sup}}\abs{\sum_{\ell'=0}^{\ell}\binom{\ell}{\ell'}\big(D^{\ell - \ell'}g'_{\Delta,R}(E)\big)\big(D^{\ell'}h_{\Delta,R}(E)\big)} \nonumber \\
    &\leq \underset{E\in[\frac{\Delta}{2},R + \frac{\Delta}{2}]}{\text{sup}}\sum_{\ell'=0}^{\ell}\binom{\ell}{\ell'} \abs*{D^{\ell - \ell'}g'_{\Delta,R}(E)}\abs*{D^{\ell'}h_{\Delta,R}(E)} \, .\nonumber
\end{align}
The derivatives of $h_{\Delta,R}(E)$ can be bounded by
\begin{align}
    \abs*{D^{\ell'}h_{\Delta,R}(E)} &= \abs*{D^{\ell'}(\mathbb{1}_{[\frac{3}{4}\Delta, R+\frac{\Delta}{4}]} \ast \eta_{\Delta})(E)} \nonumber \\
    &= \abs*{(\mathbb{1}_{[\frac{3}{4}\Delta, R+\frac{\Delta}{4}]} \ast D^{\ell'}\eta_{\Delta}) (E)} \nonumber \\
    &= \abs*{\int_{\frac{3}{4}\Delta}^{R + \frac{\Delta}{4}}D^{\ell'}\eta_{\Delta}(E-E')dE'} \nonumber \\
    &\leq R \norm*{D^{\ell'}\eta_{\Delta}}_{\infty} \label{ineq:1 D h_Delta,R}
\end{align}
Lemma 2 in the study of A. Israel \cite{israel_eigenvalue_2015} provides a bound for the derivatives of the bump function:
\begin{equation*}
    \norm*{D^{\ell'}\eta}_{\infty} \leq c_{\eta}(16)^{\ell'}\ell'^{2\ell'}\, ,
\end{equation*}
where $c_{\eta}$ is the normalization constant of $\eta(E)$, see Eq. \eqref{eq:c_eta}.
Since the derivatives of $\eta(E)$ and $\eta_{\Delta}(E)$ are related by $D^{\ell'}\eta_{\Delta}(E)  = \big(\frac{4}{\Delta}\big)^{\ell'+1}D^{\ell'}\eta(\frac{4E}{\Delta})$, we find, via Eq. \eqref{ineq:1 D h_Delta,R}, the bound
\begin{equation}
    \abs*{D^{\ell'}h_{\Delta,R}(E)} \leq \frac{4c_{\eta} R}{\Delta}\left(\frac{64}{\Delta}\right)^{\ell'}\ell'^{2\ell'}\, .\label{ineq:Dl_h}
\end{equation}
On the interval $[\frac{\Delta}{2}, R + \frac{\Delta}{2}]$ the function $g'_{\Delta,R}(E)$ evaluates to $-\frac{1}{E}$, so that we can bound
\begin{equation}
\begin{aligned}
    &\underset{E\in[\frac{\Delta}{2},R + \frac{\Delta}{2}]}{\text{sup}} \abs*{D^{\ell - \ell'}g'_{\Delta,R}(E)}\\
    &= \underset{E\in[\frac{\Delta}{2},R + \frac{\Delta}{2}]}{\text{sup}} \abs*{D^{\ell - \ell'}\frac{1}{E}}\\
    &= (\ell - \ell')!\Big(\frac{2}{\Delta}\Big)^{\ell - \ell' + 1}\, . \label{ineq:Dll'_g'}
\end{aligned}
\end{equation}
Inserting the bounds Eqs. \eqref{ineq:Dl_h} and \eqref{ineq:Dll'_g'} into Eq. \eqref{ineq:Dl_g_DeltaR} gives
\begin{align*}
    &\norm{D^{\ell}g_{\Delta,R}}_{\infty}\\
    & \leq \frac{4c_{\eta}R}{\Delta} \sum_{\ell'=0}^{\ell}\binom{\ell}{\ell'} (\ell - \ell')! \Big(\frac{2}{\Delta}\Big)^{\ell - \ell' + 1} \big(\frac{64}{\Delta}\big)^{\ell'}\ell'^{2\ell'}\\
    & = \frac{8c_{\eta}R2^{\ell}\ell!}{\Delta^{\ell + 2}}\sum_{\ell'=0}^{\ell} (32)^{\ell'}\frac{\ell'^{2\ell'}}{\ell'!}\\
    & \leq \frac{8c_{\eta}R2^{\ell}}{\Delta^{\ell + 2}}(\ell + 1)(32\ell^2)^{\ell}\\
    & \leq \frac{8c_{\eta}R}{\Delta^2}\Big(\frac{128\ell^2}{\Delta}\Big)^{\ell}\, .
\end{align*}
Consequently, by virtue of Eq. \eqref{eq:appdx_boundsFourier:ineq_beta_1} and the definition of $t_n$ we get
\begin{align}
    \abs*{\beta_{\Delta,R,n}} & \leq \frac{8c_{\eta}R}{\Delta^2}\Big(\frac{128\ell^2}{\Delta\abs{t_n}}\Big)^{\ell} \nonumber \\
    & \leq \frac{8c_{\eta}R}{\Delta^2}\Big(\frac{64 R\ell^2}{\Delta\abs{n}}\Big)^{\ell}\, . \label{ineq:beta_DeltaRn_2}
\end{align}
Now the idea is to choose the integer $\ell$ in such a way that the right-hand side of this inequality decays exponentially in $\abs{n}^{1/2}$. 
We do this by first considering large $\abs{n}$ and then small $\abs{n}$.
For $\abs{n}$ large enough it is possible to find $\ell$ such that
\begin{equation}
    \Big(\frac{\abs{n}\Delta}{256R}\Big)^{1/2} \leq \ell \leq \Big(\frac{\abs{n}\Delta}{128R}\Big)^{1/2}\, .\label{eq:condition_l}
\end{equation}
Inserting the upper bound of $\eqref{eq:condition_l}$ into the parenthesis of Eq. \eqref{ineq:beta_DeltaRn_2} results in
\begin{equation*}
    \abs*{\beta_{\Delta,R,n}} \leq \frac{8c_{\eta}R}{\Delta^2} 2^{-\ell}\, ,
\end{equation*}
and by inserting the lower bound of $\eqref{eq:condition_l}$ into the exponent we obtain
\begin{equation}
    \abs*{\beta_{\Delta,R,n}} \leq \frac{8c_{\eta}R}{\Delta^2} 2^{-(\frac{\abs{n}\Delta}{256R})^{1/2}}\, . \label{ineq:beta_DeltaRn_3}
\end{equation}
Condition Eq. \eqref{eq:condition_l} is fulfilled if $\big(\frac{\abs{n}\Delta}{128R}\big)^{1/2}$ and $\big(\frac{\abs{n}\Delta}{256R}\big)^{1/2}$ are at least an integer apart, i.e. for 
\begin{equation}
    \Big(\frac{\abs{n}\Delta}{128R}\Big)^{1/2} - \Big(\frac{\abs{n}\Delta}{256R}\Big)^{1/2} \geq 1\, ,
\end{equation}
which is equivalent to the condition $\abs{n} \geq \frac{256R}{(\sqrt{2}-1)^{2}\Delta}$. 
Now we will show that bound Eq. \eqref{ineq:beta_DeltaRn_3} holds also for smaller values of $\abs{n}$, i.e. for $\abs{n} \leq \frac{256R}{(\sqrt{2}-1)^{2}\Delta}$.
From the definition of $\beta_{\Delta,R,n}$ one obtains the upper bound
\begin{equation}
    \abs{\beta_{\Delta,R,n}} \leq \norm{g_{\Delta,R}}_{\infty} = \frac{4}{\Delta}\, .
    \label{ineq:beta_1overDelta}
\end{equation}
Then, for $\abs{n} \leq \frac{256R}{(\sqrt{2}-1)^{2}\Delta}$ and using $R\geq\Delta$ and $c_{\eta} \approx 7.514$, we get
\begin{align*}
    \abs{\beta_{\Delta,R,n}} &\leq \frac{4R}{\Delta^2}\\
    &\leq \frac{4c_{\eta}R}{\Delta^2}2^{-\frac{1}{\sqrt{2}-1}}\\
    &\leq \frac{4c_{\eta}R}{\Delta^2}2^{-(\frac{\abs{n}\Delta}{256R})^{1/2}}\, .
\end{align*}
This shows that the bound of Eq. \eqref{ineq:beta_DeltaRn_3} holds for all $n$, and by inserting the numerical value for $c_{\eta}$ we obtain the bound as stated in the lemma. \hfill $\square$

\subsection{Truncation Error of Second-Order Energy}
\label{appendix:trunc_error}
In this Appendix we show how the error $\abs*{E^{(2)} - E_N^{(2)}}$ depends on the order of truncation $N$ of the Fourier series of $g_{\Delta,R}$.
We prove the following Lemma:

\begin{lemma}
    \label{lemma:truncation_error_E2N}
    For $R$, $\Delta$ and $\norm{v}_1$ as defined in Sec. \ref{section:QPT2} and for $\epsilon>0$, there is an integer $\widetilde{N} = \widetilde{N}(R,\Delta,\epsilon,\norm{v}_1)$ such that the truncation error $\abs*{E^{(2)} - E_N^{(2)}}$ is bounded by 
    \begin{equation*}
        \abs*{E^{(2)} - E_N^{(2)}} \leq \frac{\epsilon}{2}\quad \text{if} \quad N\geq \widetilde{N}\, ,
    \end{equation*}
    and furthermore, $\widetilde{N}\leq \left\lceil\frac{2132R}{\Delta}\ln^2\left(\frac{5.3\cdot10^4 R^2 \norm{v}_1^2}{\Delta^3\epsilon}\right)\right\rceil$\, .
\end{lemma}

\noindent\textit{Proof:} With Eqs. \eqref{eq:SecondOrder-Propagator} and \eqref{eq:sec_QPT:E2N_LCU} we have 
\begin{equation*}
    \abs*{E^{(2)} - E_{N}^{(2)}} = \abs*{\expval{VR_0V} - \expval{VR_{0,N}V}}
\end{equation*}
an by inserting Eqs. \eqref{eq:Resolvent-Propagator} and \eqref{eq:truncatedR0} we get
\begin{align*}
    \abs*{E^{(2)} - E_{N}^{(2)}} &\leq \sum_{\abs{n}>N}\abs*{\beta_{\Delta,R,n}}\abs*{\expval*{VU(t_n)V}}\\
    &\leq \sum_{\abs{n}>N}\abs*{\beta_{\Delta,R,n}} \norm{V\ket{\Phi_0}}_2 \norm{U(t_n)V\ket{\Phi_0}}_2\\
    &= \norm{V\ket{\Phi_0}}_2^2 \sum_{\abs{n}>N} \abs*{\beta_{\Delta,R,n}} \\
    &= \expval*{V^2}\sum_{\abs{n}>N}\abs*{\beta_{\Delta,R,n}}\, ,
\end{align*}
where in the second line the Cauchy-Schwarz inequality was used.
With the decomposition of $V$ into Pauli operators Eq. \eqref{eq:PauliDecomposition} $\expval*{V^2}$ is upper bounded by
\begin{equation*}
    \expval{V^2} = \sum_{i,j=1}^{L_V} v_iv_j\expval{\sigma_i\sigma_j} \leq \norm{v}_1^2\, ,
\end{equation*}
such that
\begin{equation*}
    \abs*{E^{(2)} - E_{N}^{(2)}} \leq \norm{v}_1^2\sum_{\abs{n}>N}\abs*{\beta_{\Delta,R,n}}\, .
\end{equation*}
We use constants $M_{\Delta, R} = \frac{61 R}{\Delta^2}$ and $q_{\Delta,R} = \ln(2)\big(\frac{\Delta}{256R}\big)^{\frac{1}{2}}$ to write the bound on the Fourier coefficients $\beta_{\Delta,R,n}$ of Lemma \ref{lemma:fouriercoeff_bound} in a more compact form,
\begin{equation*}
    \abs*{\beta_{\Delta,R,n}} \leq M_{
    \Delta,R}e^{-q_{\Delta,R}\abs{n}^{1/2}}\, .
\end{equation*}
Now we bound:
\begin{align*}
    \sum_{\abs{n}>N}\abs{\beta_{\Delta,R,n}}
    &\leq M_{\Delta,R}\sum_{\abs{n}>N}e^{-q_{\Delta,R}\abs{n}^{\frac{1}{2}}}\\
    &\leq 2M_{\Delta,R}\int_{N}^{\infty} e^{-q_{\Delta,R}x^{\frac{1}{2}}}dx\\
    &=\frac{4M_{\Delta,R}}{q_{\Delta,R}^2}\int_{q_{\Delta,R}N^{\frac{1}{2}}}^\infty ye^{-y}dy\\
    &=\frac{4M_{\Delta,R}}{q_{\Delta,R}^2} (q_{\Delta,R}N^{\frac{1}{2}} + 1) e^{-q_{\Delta,R}N^{\frac{1}{2}}}\\
    &\leq \frac{8M_{\Delta,R}}{q_{\Delta,R}^2} e^{-\frac{q_{\Delta,R}}{2}N^{\frac{1}{2}}}\, ,
\end{align*}
where in the second line $xe^{-x} \leq e^{-\frac{x}{2}}$ was used.
As a result, the truncation error is bounded by
\begin{equation}
    \abs*{E^{(2)} - E_{N}^{(2)}} \leq \frac{8M_{\Delta,R}\norm{v}_1^2}{q_{\Delta,R}^2} e^{-\frac{q_{\Delta,R}}{2}N^{\frac{1}{2}}}\, . \label{ineq:truncError_E2N}
\end{equation}
Now define
\begin{equation*}
    \widetilde{N}\equiv \Bigg\lceil\frac{4}{q_{\Delta,R}^2}\ln^2\Big(\frac{16M_{\Delta,R}\norm{v}_1^2}{q_{\Delta,R}^2 \epsilon} \Big)\Bigg\rceil\, ,
\end{equation*}
then, for $N\geq \widetilde{N}$, and via Eq. \eqref{ineq:truncError_E2N}, we get 
\begin{equation}
    \abs*{E^{(2)} - E_{N}^{(2)}} \leq \frac{\epsilon}{2}\, . \label{ineq:bound_E2N}
\end{equation}
Lastly, we insert the definitions of $M_{\Delta,R}$ and $g_{\Delta,R}$ to get the final result
\begin{equation*}
    \widetilde{N} \leq \Bigg\lceil \frac{2132 R}{\Delta}\ln^2(\frac{5.3\times10^4 R^2 \norm{v}_1^2}{\Delta^3\epsilon}) \Bigg\rceil\, \eqno\square
\end{equation*}

\subsection{Overlap Estimation with Iterative Quantum Amplitude Estimation}
\label{appendix:overlap_estimation}
As explained in Sec. \ref{section:QPT2}, a key aspect of our algorithm is to compute overlaps using a quantum computer.
The following lemma provides upper bounds on the costs for computing such overlaps:

\begin{lemma}
    \label{lemma:overlap}
    Given a confidence level $1-\alpha \in (0,1)$, a target accuracy $\epsilon>0$ and d-qubit unitaries $U$ and $U_0$ with $U_0\ket{0} = \ket{\Phi_0}$, it is possible to obtain an estimate $X'$ to the overlap $X = \expval{U}{\Phi_0}$
    such that
    \begin{equation*}
        \mathbb{P}[\abs*{X' - X} < \epsilon] \geq 1 - \alpha
    \end{equation*}
    with $d+1$ qubits, and at most $\frac{1116}{\epsilon}\ln\frac{63}{\alpha}$ applications of $U_0$ and at most $\frac{558}{\epsilon}\ln\frac{63}{\alpha}$ applications of controlled-$U$.
\end{lemma}

The algorithm behind Lemma \ref{lemma:overlap} is amplitude estimation, in particular the iterative quantum amplitude estimation algorithm (IQAE) \cite{grinko_iterative_2021}\cite{fukuzawa_modified_2023} .
Before we prove Lemma \ref{lemma:overlap}, we give a brief summary of amplitude estimation and present key aspects of IQAE.
With Lemma \ref{lemma:amp_est_adapted} we will cast the query-complexity of IQAE as shown in Ref. \cite{fukuzawa_modified_2023} into a suitable form to prove Lemma \ref{lemma:overlap}.

In the general setting of amplitude estimation a state $\ket{\psi}$ is given, and one is interested in the component of $\ket{\psi}$ along another state $\ket{\psi'}$. Access is given to a unitary $U_{\psi}$ preparing
\begin{gather*}
    U_{\psi}\ket{0} = \ket{\psi} = \sqrt{a}\ket{\psi'} + \sqrt{1-a}\ket*{\psi'^{\perp}}\, ,\\
    \text{with}\quad \braket*{\psi'^{\perp}}{\psi'} = 0\, ,
\end{gather*}
and the challenge is to estimate the amplitude $a = \abs{\braket{\psi'}{\psi}}^2$ with as few applications of $U_{\psi}$ as possible.
It is convenient to introduce an angle $\theta_a\in[0,\frac{\pi}{2}]$ such that $\sin^2(\theta_a) = a$.
A unitary operator $Q = -S_{\psi}S_{\psi'}$ is defined, where 
\begin{align*}
    S_{\psi} &= \mathbb{1} - 2\dyad{\psi}\\
    S_{\psi'} &= \mathbb{1} - 2\dyad{\psi'}
\end{align*}
are reflection operators along $\ket{\psi}$ and $\ket{\psi'}$, respectively.
Note that with access to the state-preparation unitary $U_{\psi}$ one can implement $S_{\psi}$ as 
\begin{equation}
    S_{\psi} = U_{\psi}Z_0U_{\psi}^{\dag}\quad \text{where} \quad Z_0 = \mathbb{1} - 2\dyad{0}\, .\label{eq:QAE_reflection}
\end{equation}
The key insight of amplitude estimation is that the probability of measuring $\ket{\psi'}$ in the state $Q^k\ket{\psi}$ is \cite{brassard_quantum_2002}
\begin{equation}
    \mathbb{P}[\ket{\psi'}] = \abs*{\mel{\psi'}{Q^k}{\psi}}^2 = \sin^2((2k+1)\theta_a)\, . \label{eq:AE_Qk}
\end{equation}
As other amplitude estimation algorithms, IQAE estimates $\sin^2((2k+1)\theta_a)$ for different powers of $k$ and uses that information to infer the value of $a$.
The algorithm iteratively calculates estimates to $\theta_a$ and stops once a specified precision is attained.
Compared to the textbook amplitude estimation algorithm by Brassard et al. \cite{brassard_quantum_2002}, IQAE has the advantage that no quantum Fourier transform is needed.
IQAE was first published in 2021 by Grinko and coworkers \cite{grinko_iterative_2021}, and this year Fukuzawa \textit{et al.} \cite{fukuzawa_modified_2023} presented an updated version of IQEA (``modified IQAE'') for which it was shown that optimal query complexity for amplitude estimation is achieved. Their main result is the following theorem:

\begin{theorem}[Theorem 3.1 \cite{fukuzawa_modified_2023}]
    \label{thm:modIQAE}
    Given a confidence level $1-\alpha\in(0,1)$, a target accuracy $\epsilon>0$, and a (d+1)-qubit unitary $U_{\psi,flag}$ satisfying
    \begin{equation*}
        U_{\psi,flag}\ket{0}\ket{0}_{\text{flag}} = \sqrt{a}\ket{\psi'}\ket{0}_{\text{flag}} + \sqrt{a-1}\ket{\psi'^{\perp}}\ket{1}_{\text{flag}}
    \end{equation*}
    where $\ket{\psi'}$ and $\ket{\psi'^{\perp}}$ are arbitrary d-qubit states and $a\in[0,1]$, modified IQAE outputs a confidence interval for $a$ that satisfies $\mathbb{P}[a\notin[a_l,a_u]]\leq \alpha$, where $a_u-a_l<2\epsilon$, leading to an estimate $\hat{a}$ for $a$ such that $\abs{a-\hat{a}}<\epsilon$ with a confidence of $1-\alpha$, using $\mathcal{O}(\frac{1}{\epsilon}\ln\frac{1}{\alpha})$ applications of $U_{\psi,flag}$.
\end{theorem}

\noindent More specifically, Lemma 3.7. of the same reference gives an explicit bound:

\begin{lemma}[Lemma 3.7 \cite{fukuzawa_modified_2023}]
    \label{lemma:modIQAE}
    Modified IQAE requires at most $\frac{62}{\epsilon}\ln\frac{21}{\alpha}$ applications of $U_{\psi,flag}$, achieving the desired query complexity of $\mathcal{O}(\frac{1}{\epsilon}\ln\frac{1}{\alpha})$
\end{lemma}

In our setting, $U_{\psi}$ does not include the action on an ancillary flag qubit, instead we have access to a unitary $U_{\psi'}$  preparing $\ket{\psi'}$.
Therefore, we adapt Lemma \ref{lemma:modIQAE} to our setting and show

\begin{lemma}
    \label{lemma:amp_est_adapted}
    Given a confidence level $1-\alpha\in(0,1)$, a target accuracy $\epsilon>0$, a d-qubit state preparation unitary $U_{\psi'}$ with $U_{\psi'}\ket{0} = \ket{\psi'}$ and another unitary $V$, modified IQAE leads to an estimate $\hat{a}$ for
    \begin{equation*}
        a = \abs{\expval{V}{\psi'}}^2
    \end{equation*}
    such that $\abs{a-\hat{a}}<\epsilon$ with a confidence of $1-\alpha$, using at most $\frac{62}{\epsilon}\ln\frac{21}{\alpha}$ applications of $V$ and at most $\frac{124}{\epsilon}\ln\frac{21}{\alpha}$ applications of $U_{\psi'}$.
\end{lemma}

\noindent \textit{Proof:} The idea is to define $U_{\psi} = V U_{\psi'}$ with $U_{\psi}\ket{0} = \ket{\psi}$ such that $a = \abs*{\braket{\psi'}{\psi}}^2$, and apply the results of modified IQAE.
First, we explain how to run IQAE without the ancillary flag qubit.
The quantum computing part of IQAE consists of measuring the flag qubit of the circuit $Q^kU_{\psi,flag}\ket{0}\ket{0}_{\text{flag}}$ to get
\begin{equation*}
    \mathbb{P}[\ket{0}_{\text{flag}}] = \sin^2((2k+1)\theta_a)\, .
\end{equation*}

However, with access to the state-preparation unitary $U_{\psi'}$ one can perform the measurement of Eq. \eqref{eq:AE_Qk} directly by preparing the circuit $U_{\psi'}^{\dag}Q^kU_{\psi}\ket{0}$ and measuring in the computational basis,
\begin{equation*}
    \mathbb{P}[\ket{0}] = \abs*{\expval{U_{\psi'}^{\dag}Q^kU_{\psi}}{0}}^2 = \sin^2((2k+1)\theta_a)\, .
\end{equation*}
Thus, IQAE can readily be applied to the setting where $U_{\psi}$ does not include the action on a flag qubit.
According to Theorem 3.1 of Ref. \cite{fukuzawa_modified_2023}, IQAE will lead to an estimate $\hat{a}$ for $a = \abs*{\expval{V}{\psi'}}^2 = \abs*{\braket{\psi'}{\psi}}^2$ such that $\abs*{\hat{a}-a}<\epsilon$ with confidence $1-\alpha$.
It remains now to prove the upper bounds on the number of applications of $U_{\psi'}$ and $V$.

As stated earlier in Eq. \eqref{eq:QAE_reflection}, the reflection operators, and therefore the operator $Q$, can be implemented through the state preparation unitaries $U_{\psi}$ and $U_{\psi'}$
\begin{equation*}
    Q = -S_{\psi}S_{\psi'} = -U_{\psi}Z_0U_{\psi}^{\dag}U_{\psi'}Z_0U_{\psi'}^{\dag}\, .
\end{equation*}
IQAE runs the circuits $U_{\psi'}^{\dag}Q^kU_{\psi}\ket{0}$ for different values of $k$.
In each circuit, the number of applications of $U_{\psi'}$ equals the number of applications of $U_{\psi}$, 2$k$ + 1 applications each. 
Thus, we can use the upper bound $\frac{62}{\epsilon}\ln\frac{21}{\alpha}$ on the number of applications of $U_{\psi}$ given in Lemma \ref{lemma:modIQAE} both for $U_{\psi}$ and $U_{\psi'}$.
Lastly, with the definition $U_{\psi} = VU_{\psi'}$, the upper bounds $\frac{124}{\epsilon}\ln\frac{21}{\alpha}$ for $U_{\psi'}$ and $\frac{62}{\epsilon}\ln\frac{21}{\alpha}$ for $V$ follow.\hfill $\square$
\vspace{1cm}

\noindent We now have all ingredients to prove Lemma \ref{lemma:overlap}.
\vspace{1cm}

\noindent\textit{Proof Lemma \ref{lemma:overlap}: }
First, the overlap is written in terms of amplitude and phase, and in terms of real and imaginary part:
\begin{equation*}
    X = re^{i\theta} = r\cos(\theta) + ir\sin(\theta), \quad r\in\mathbb{R_{\geq 0}},\quad \theta\in[0,2\pi)\, .
\end{equation*}
We consider the following amplitudes
\begin{align}
    x_0 &= \abs*{\expval{U}{\Phi_0}}^2 = r^2 \label{eq:OEA_x0} , \\
    x_1 &= \abs*{\bra{\Phi_0}\bra{+}cU\ket{\Phi_0}\ket{+}}^2 = \frac{1}{2}\abs{1 + re^{i\theta}}^2 , \label{eq:OEA_x1} \\
    x_2 &= \abs*{\bra{\Phi_0}\bra{+}cU(\mathbb{1}\otimes e^{iZ\frac{\pi}{4}})\ket{\Phi_0}\ket{+}}^2 = \frac{1}{2}\abs{1 - ire^{i\theta}}^2\, ,\label{eq:OEA_x2}
\end{align}
where $cU = \mathbb{1}\otimes\dyad{0} + U\otimes\dyad{1}$. 
As noted in Ref. \cite{knill_optimal_2007}, from $x_0, x_1$ and $x_2$ we can extract
\begin{align}
    r &= \sqrt{x_0} \label{eq:OEA_r} , \\
    \cos\theta &= \frac{4x_1 - 1 - x_0}{2\sqrt{x_0}} , \label{eq:OEA_cos}\\
    \sin\theta &= -\frac{4x_2 - 1 - x_0}{2\sqrt{x_0}} \label{eq:OEA_sin}\, ,
\end{align}
so the amplitudes $x_0, x_1$ and $x_2$, lead to the value of $X$.
The strategy is to compute estimates $\hat{x}_0, \hat{x}_1$ and $\hat{x}_2$ to $x_0, x_1$ and $x_2$ via IQAE, in particular the IQAE version laid out in Lemma 3.
For the estimation of $x_0, x_1$ and $x_2$, the unitaries $V$ and $U_{\psi'}$ of Lemma \ref{lemma:amp_est_adapted} correspond to
\begin{equation}
    \begin{aligned}[c]
        x_0:&\\
        x_1:&\\
        x_2:&
    \end{aligned}
    \qquad
    \begin{aligned}[c]
        V &= U\\
        V &= cU\\
        V &= cU(\mathbb{1}\otimes e^{iZ\frac{\pi}{4}})
    \end{aligned}
    \qquad\quad
    \begin{aligned}[c]
        U_{\psi'} &= U_0\\
        U_{\psi'} &= U_0 \otimes \text{H}\\
        U_{\psi'} &= U_0 \otimes \text{H}\, .
    \end{aligned}
    \label{eq:IQAE_xi_V_Upsi}
\end{equation}
The accuracy and confidence parameters for the estimation of $x_i\,(i=0,1,2)$ will be denoted by $\epsilon_i$ and $\alpha_i$ respectively, i.e., with IQAE we get $x'_i$ such that $\abs*{x'_i - x_i} < \epsilon_i$ with confidence $1-\alpha_i$.
The estimate to $X$ we get with our strategy is denoted by $X' = r'e^{i\theta'}$, where $r'$ and $\theta'$ are related to $x'_0, x'_1$ and $x'_2$ analogous to Eqs. \eqref{eq:OEA_r}-\eqref{eq:OEA_sin}.
The estimation error is bounded by
\begin{align*}
    \abs*{X' - X} &= \abs*{r'e^{i\theta'} - re^{i\theta}}\\
    &\leq \abs*{r'\cos\theta' - r\cos\theta} + \abs*{r'\sin\theta' - r\sin\theta}\\
    &\leq \abs*{x'_0 - x_0} + 2\abs*{x'_1 - x_1} + 2\abs*{x'_2 - x_2}\\
    &\leq \epsilon_0 + 2\epsilon_1 + 2\epsilon_2\, .
\end{align*}
We choose $\epsilon_0 = \epsilon_1 = \epsilon_2 = \epsilon/5$, resulting in
\begin{equation*}
    \abs*{X' - X} \leq \epsilon\, .
\end{equation*}
Moreover, since the computations for $x'_0, x'_1$ and $x'_2$ are independent, the probability of obtaining all three estimates with specified accuracies $\epsilon_i$ is given by
\begin{align*}
    &\mathbb{P}[\abs*{x'_i - x_i} < \epsilon_i \,\, \forall i\in\{0,1,2\}] \\
    &\geq (1-\alpha_0)(1-\alpha_1)(1-\alpha_2)\\ 
    & \geq 1 - (\alpha_0 + \alpha_1 + \alpha_2)\, .
\end{align*}
We choose $\alpha_0 = \alpha_1 = \alpha_2 = \alpha/3$ to obtain the desired confidence
\begin{align*}
    \mathbb{P}[\abs*{X' - X} \leq \epsilon] &= \mathbb{P}[\abs*{x'_i - x_i} < \epsilon_i \,\, \forall i\in\{0,1,2\}]\\
    &\geq 1 - \alpha\, .
\end{align*}
Lastly, we show the upper bounds on the number of applications of $cU$ and $U_0$ by employing the upper bounds on the number of calls to $V$ and $U_{\psi'}$ of Lemma \ref{lemma:amp_est_adapted}. 
For simplicity, we count an application of $U$ as an application of $cU$.
From Eq. \eqref{eq:IQAE_xi_V_Upsi} we see that for each case $i=0,1,2$, the unitaries $U_{\psi'}$ and $V$ contain $U_0$ and $cU$ once, respectively.
Consequently, the total number of applications of $U_0$ is at most $\frac{186}{\epsilon_i}\ln\frac{21}{\alpha_i} = \frac{558}{\epsilon}\ln\frac{63}{\alpha}$, and the total number of applications of $cU$ is at most $\frac{378}{\epsilon_i}\ln\frac{21}{\alpha_i} = \frac{1116}{\epsilon}\ln\frac{63}{\alpha}$.\hfill $\square$
\vspace{1cm}

The next lemma applies the overlap estimation to linear combinations of unitaries:

\begin{lemma}
    \label{lemma:LCU_overlap_estimation_term_by_term}
    Given a confidence level $1-\alpha \in (0,1)$, a target accuracy $\epsilon>0$, a linear combination of d-qubit unitaries $V_U = \sum_{i=1}^L u_i U_i$ and a d-qubit state-preparation unitary $U_0$, then it is possible to obtain an estimate $X'$ to the overlap $X= \expval{V_U}{\Phi_0}$
    such that
    \begin{equation*}
        \mathbb{P}[\abs*{X' - X} < \epsilon] \geq 1 - \alpha
    \end{equation*}
    with $1$ ancillary qubit,  a total $\mathcal{O}\left(\epsilon^{-1} \norm{u}_{2/3} \ln\frac{L}{\alpha}\sqrt{\ln\frac{1}{\alpha}}\right)$ applications of $U_0$ and $\mathcal{O}\left(\epsilon^{-1} \abs{u_i}^{2/3}\norm{u}_{2/3}^{1/3} \ln\frac{L}{\alpha}\sqrt{\ln\frac{1}{\alpha}}\right)$ applications of controlled-$U_i$ for each i=1,\dots,L, where $\norm{u}_{2/3} = \left(\sum_{i=1}^L \abs{u_i}^{2/3}\right)^{3/2}$.
\end{lemma}

\noindent\textit{Proof:} The quantum algorithm of Lemma \ref{lemma:overlap} is applied to each $U_i$ and the results are summed to get
\begin{equation*}
    \sum_{i=1}^L u_i X_i' = X'\, .
\end{equation*}
Each term $X_i = \expval{U_i}{\Phi_0}$ is associated with a confidence level $1-\alpha_i \in (0,1)$ and an accuracy $\epsilon_i > 0$, i.e. $\mathbb{P}[\abs*{X_i' - X_i} < \epsilon_i] \geq 1 - \alpha'$.
We choose to have the same confidence level $1-\alpha'$ for each term, but we will optimize the individual accuracies $\epsilon_i$ to reduce the overall runtime.
The probability of all $L$ terms $X_i'$ to lie in their respective interval $(X_i - \epsilon_i, X_i + \epsilon_i)$ is lower bounded by
\begin{equation*}
    \mathbb{P}\big[\abs*{X'_i - X_i} < \epsilon_i, \quad \forall 1\leq i\leq L\big]
    = (1 - \alpha')^L
    \geq 1-L\alpha'\, ,
\end{equation*}
where Bernoulli's inequality, $(1+x)^r\geq 1+xr$ for $x\geq -1$ and integer $r\geq 1$, was used in the last line.
By choosing $\alpha' = \frac{\alpha}{2L}$ we know that with probability $1 - \frac{\alpha}{2}$ all random variables $u_i(X_i' - X_i)$ are bounded by $2\abs{u_i}\epsilon_i$.
Under the assumption that $\abs{X'_i - X_i} \leq \epsilon_i$ for all $i$ we can upper bound the probability that the global error $\abs{X' - X}$ is large:
\begin{align}
    & \mathbb{P}[\abs{X'-X}\geq\epsilon \text{ AND } \abs{X_i' - X_i} \leq \epsilon_i \; \forall i] \nonumber \\
    =& \mathbb{P}\left(\abs{\sum_{i=1}^L u_i (X_i' - X_i) \geq \epsilon}   \text{ AND } \abs{X_i' - X_i} \leq \epsilon_i \; \forall i\right) \nonumber \\
    \leq& \mathbb{P}\left(\sum_{i=1}^L \abs{u_i} \abs{X_i' - X_i} \geq \epsilon \text{ AND } \abs{X_i' - X_i} \leq \epsilon_i \; \forall i\right) \nonumber \\
    \leq& \exp{-\frac{\epsilon^2}{2\sum_i\abs{u_i}^2 \epsilon_i^2}} \leq \frac{\alpha}{2}\, , \label{eq:appdx_complexity:hoeffding}
\end{align}
where Hoeffding's inequality was employed from third to fourth line for the random variables $\abs{u_i}\abs{X_i' - X_i}$.
Together with the probability that there is at least one $i$ for which $\abs{X'_i - X_i} > \epsilon_i$ we can upper bound
\begin{align*}
    \mathbb{P}[\abs{X'-X}\geq\epsilon] &\leq \mathbb{P}[\abs{X'-X} \geq \epsilon \text{ AND } \abs{X_i' - X_i} \leq \epsilon_i \;\forall i]\\
    &+ \mathbb{P}[\abs{X_i' - X_i} > \epsilon_i \text{ for some }i]\\
    & \leq \alpha,
\end{align*}
resulting in the global confidence $1-\alpha$.

Next, Hoeffding's inequality in Eq. \eqref{eq:appdx_complexity:hoeffding} from above is used to relate global accuracy $\epsilon$, individual accuracies $\epsilon_i$ and $\alpha$, so we rewrite the inequality to 
\begin{equation*}
    \epsilon^2 \geq 2 \ln(\frac{2}{\alpha}) \sum_{i=1}^L \abs{u_i}^2 \epsilon_i^2\, .
\end{equation*}
Lemma \ref{lemma:overlap} shows that the cost of computing $X_i$ with accuracy $\epsilon_i$ is proportional to $\frac{c}{\epsilon_i}$ for some constant $c$, so we set up a Lagrangian minimization problem in the variables $\epsilon_i$:
\begin{equation*}
    \mathcal{L} = \sum_{i=1}^L \frac{c}{\epsilon_i} + \lambda \left(2 \ln(\frac{2}{\alpha})\sum_{i=1}^L \abs{u_i}^2 \epsilon_i^2 - \epsilon^2 \right)\, .
\end{equation*}
Solving $\frac{\partial \mathcal{L}}{\partial \epsilon_i} = 0$ and $\frac{\partial\mathcal{L}}{\partial \lambda} = 0$ results in the optimized values
\begin{equation*}
    \epsilon_i = \frac{\epsilon}{\sqrt{2\ln(2/\alpha)}\abs{u_i}^{2/3}\norm{u}_{2/3}^{1/3}}\, ,
\end{equation*}
where $\norm{u}_{2/3} = (\sum_j \abs{u_j}^{2/3})^{3/2}$.
Therefore, by Lemma \ref{lemma:overlap}, for each $i$ the number of applications of $U_0$ and controlled-$U_i$ is
\begin{equation*}
    \mathcal{O}\left(\epsilon^{-1} \abs{u_i}^{2/3}\norm{u}_{2/3}^{1/3} \ln\frac{L}{\alpha}\sqrt{\ln\frac{1}{\alpha}}\right).
\end{equation*}
Consequently, the cumulative number of applications of $U_0$ and controlled-$U_i$ for all $i$ is 
\begin{equation*}
    \mathcal{O}\left(\epsilon^{-1} \norm{u}_{2/3} \ln\frac{L}{\alpha}\sqrt{\ln\frac{1}{\alpha}}\right) .\eqno\square
\end{equation*}

\subsection{Complexity Analysis of Quantum Algorithm} \label{appendix:complexity_circ1}
In this Appendix we discuss the complexity analysis of the quantum algorithm presented in Sec. \ref{section:QPT2} for calculating the second-order perturbation energy $E^{(2)}$.
Lemma \ref{lemma:fouriercoeff_bound} and Lemma \ref{lemma:LCU_overlap_estimation_term_by_term} are the main ingredients of the proof, allowing us put bounds on the truncation error $\abs*{E^{(2)}-E_N^{(2)}}$ and on the error because of finite precision in estimating overlap terms.

\begin{lemma}
    \label{lemma:complexity_algorithm_termbyterm}
    Given a Hamiltonian $H = \sum_{i=1}^{L_H}h_i\sigma_i$ with spectral gap $\Delta$ and a perturbation $V = \sum_{i=1}^{L_V}v_i\sigma_i$, both acting on $d$ qubits, and given access to a unitary $U_0$ preparing the ground state $\ket{\Phi_0}$ of $H$, it is possible to compute an estimate to the second-order perturbation energy $E^{(2)} = \expval{VR_0V}{\Phi_0}$ using $d+1$ qubits and with precision $\epsilon$ and confidence $1-\alpha\in(0,1)$. 
    Computing the estimate to $E^{(2)}$ requires implementing the Hamiltonian evolution under $H$ for a total time upper bounded by

    \begin{align*}
        &\mathcal{O}\Bigg(\frac{\norm{v}^2_{2/3} \norm{h}_1^{3/2}}{\Delta^{5/2}\epsilon} \ln^3\left(\frac{\norm{h}_1^2 \norm{v}_1^2}{\Delta^3 \epsilon}\right)\\
    &\phantom{{}\quad}\times \ln\left(\frac{L_V^2 \norm{h}_1}{\alpha\Delta} \ln^2\left(\frac{\norm{h}_1^2\norm{v}_1^2}{\Delta^3\epsilon}\right)\right)\sqrt{\ln\frac{1}{\alpha}}\Bigg)\, .
    \end{align*}
    Furthermore, the total number of calls to $U_0$ is upper bounded by

    \begin{align*}
        &\mathcal{O}\Bigg(\frac{\norm{v}^2_{2/3} \norm{h}_1^{3/2}}{\Delta^{7/2}\epsilon} \ln^5\left(\frac{\norm{h}_1^2 \norm{v}_1^2}{\Delta^3 \epsilon}\right)\\
        &\phantom{{}\quad}\times \ln\left(\frac{L_V^2 \norm{h}_1}{\alpha\Delta} \ln^2\left(\frac{\norm{h}_1^2\norm{v}_1^2}{\Delta^3\epsilon}\right)\right)\sqrt{\ln\frac{1}{\alpha}}\Bigg)\, .
    \end{align*}
\end{lemma}

\noindent\textit{Proof:} 
We explain how the errors that are involved are bounded, and afterwards work out the resulting cost of the quantum algorithm.
Two types of errors need to be controlled: The truncation error $\abs*{E^{(2)}-E_N^{(2)}}$ in the Fourier series and the error because of the finite precision in the estimation of the individual overlap terms.
The full second-order correction, $E^{(2)}$, and the estimate obtained by the algorithm, denoted by $X_{N}^{(2)}$, are given by [compare also to $E_N^{(2)}$ Eq. \eqref{eq:E2N_lin_comb_expval_unitaries}]
\begin{align*}
    E^{(2)} &= \sum_{n \in \mathbb{Z}}\beta_{\Delta,R,n} e^{iE_0t_n} \sum_{i,j=1}^{L_V} v_i v_j \expval{\sigma_i U(t_n) \sigma_j}\\
    X_{N}^{(2)} &= \sum_{\abs{n}\leq N} \beta_{\Delta,R,n} e^{iE_0t_n} \sum_{i,j=1}^{L_V} v_i v_j X_{nij}\, ,
\end{align*}
where $X_{nij}$ is the finite-sample estimate to $\expval{\sigma_i U(t_n) \sigma_j}$.
For the truncation error, Lemma \ref{lemma:truncation_error_E2N} provides the bound
\begin{equation*}
    \abs*{E^{(2)} - E_N^{(2)}}\leq \frac{\epsilon}{2} \quad \text{if}\quad N\geq \widetilde{N}\, ,
\end{equation*}
where $\widetilde{N}$ satisfies
\begin{equation}
    \widetilde{N} \leq \mathcal{O}\left(\frac{R}{\Delta} \ln^2\left(\frac{R^2 \norm{v}_1^2}{\Delta^3 \epsilon} \right)\right)\, .
    \label{eq:appdx_complexity:bound_tildeN}
\end{equation}
Next, we discuss the error 
\begin{align*}
    \abs*{X_{\widetilde{N}}^{(2)} - E_{\widetilde{N}}^{(2)}} &\leq \sum_{\abs{n}\leq \widetilde{N}} \sum_{i,j=1}^{L_V}\abs*{\beta_{\Delta,R,n} v_i v_j}\\
    &\phantom{{}\qquad} \times \abs*{X_{nij} - \expval{\sigma_i U(t_n) \sigma_j}}\\
\end{align*}
because of finite precision in the estimation of the overlap. This will result in an upper bound on the total number of calls to $U_0$ and an upper bound on the total evolution time.
With Lemma \ref{lemma:LCU_overlap_estimation_term_by_term} we know that with confidence $1-\alpha$ an $\frac{\epsilon}{2}$-approximation $X_{\widetilde{N}}^{(2)}$ to $E_{\widetilde{N}}^{(2)}$ can be obtained, such that 
\begin{equation*}
    \abs*{E^{(2)} - X_{\widetilde{N}}^{(2)}} \leq \abs*{E^{(2)} - E_{\widetilde{N}}^{(2)}} + \abs*{E_{\widetilde{N}}^{(2)} - X_{\widetilde{N}}^{(2)}} \leq \epsilon\, ,
\end{equation*}
as desired.
The cost of computing $X_{\widetilde{N}}^{(2)}$ depends on the total number of overlap terms in $E_N^{(2)}$, which is $L_V^2\widetilde{N}$, and
the coefficients,
\begin{align*}
    \norm{v^2 \beta_{\Delta,R}}_{2/3} \equiv &\left(\sum_{\abs{n}\leq \widetilde{N}}\sum_{i,j=1}^{L_V}\abs{\beta_{\Delta,R,n} v_i v_j}^{2/3}\right)^{3/2}\\
    &= \norm{v}_{2/3}^2 \left( \sum_{\abs{n}\leq \widetilde{N}}\abs{\beta_{\Delta,R,n}}^{2/3}\right)^{3/2}\\
    &= \mathcal{O}\left(\frac{\norm{v}_{2/3}^2 R^{3/2}}{\Delta^{5/2}} \ln^3\left(\frac{R^2 \norm{v}_1^2}{\Delta^3 \epsilon}\right) \right)\, ,
\end{align*}
where Eqs. \eqref{ineq:beta_1overDelta} and \eqref{eq:appdx_complexity:bound_tildeN} were used.
Therefore, by Lemma \ref{lemma:LCU_overlap_estimation_term_by_term}, the total number of calls to $U_0$ is upper bounded by
\begin{align}
    &\mathcal{O} \left(\frac{\norm{v^2\beta_{\Delta,R}}_{2/3}}{\epsilon}\ln \left(\frac{L_V^2 \widetilde{N}}{\alpha}\right)\sqrt{\ln\frac{1}{\alpha}}\right)\, \\
    & = \mathcal{O}\Bigg(\frac{\norm{v}^2_{2/3} R^{3/2}}{\Delta^{5/2}\epsilon} \ln^3\left(\frac{R^2 \norm{v}_1^2}{\Delta^3 \epsilon}\right)\\
    &\phantom{{}\quad}\times \ln\left(\frac{L_V^2 R}{\alpha\Delta} \ln^2\left(\frac{R^2\norm{v}_1^2}{\Delta^3\epsilon}\right)\right)\sqrt{\ln\frac{1}{\alpha}}\Bigg)\, .
    \label{eq:appdx_complexity:termbyterm_tot_U0}
\end{align}
Next, we will upper bound the total evolution time.
If we upper bound the evolution time $t_n$ of each term by $t_{\widetilde{N}}$, then we can multiply $t_{\widetilde{N}} = \frac{2\pi \widetilde{N}}{R + \Delta}$ by the overall number of applications of all terms to obtain an upper bound on the total evolution time (see Lemma \ref{lemma:LCU_overlap_estimation_term_by_term}):
\begin{align}
    &\mathcal{O} \left(\frac{\widetilde{N}\norm{v^2\beta_{\Delta,R}}_{2/3}}{R\epsilon}\ln \left(\frac{L_V^2 \widetilde{N}}{\alpha}\right)\sqrt{\ln\frac{1}{\alpha}}\right)\\
    & = \mathcal{O}\Bigg(\frac{\norm{v}^2_{2/3} R^{3/2}}{\Delta^{7/2}\epsilon} \ln^5\left(\frac{R^2 \norm{v}_1^2}{\Delta^3 \epsilon}\right)\\
    &\phantom{{}\quad}\times \ln\left(\frac{L_V^2 R}{\alpha\Delta} \ln^2\left(\frac{R^2\norm{v}_1^2}{\Delta^3\epsilon}\right)\right)\sqrt{\ln\frac{1}{\alpha}}\Bigg) \label{eq:appdx_complexity:termbyterm_tot_time}
\end{align}

Lastly, the range $R = E_{D-1} - E_0$ of the eigenvalues of $H$ is upper bounded by
\begin{equation}
    \begin{aligned}
    R & \leq 2\underset{i=0,\dots,D-1}{\max}\abs{E_i}\\
    & = 2 \underset{\norm{\ket{\psi}}=1}{\sup} \norm{H\ket{\psi}}_2\\
    & \leq 2 \underset{\norm{\ket{\psi}}=1}{\sup} \sum_{i=1}^{L_H} \abs{h_i} \norm{\sigma_i\ket{\psi}}_2\\
    & \leq 2\norm{h}_1\, .
    \end{aligned}
    \label{eq:appdx_complexity:bound_R}
\end{equation}
Inserting Eq. \eqref{eq:appdx_complexity:bound_R} into Eqs. \eqref{eq:appdx_complexity:termbyterm_tot_U0} and \eqref{eq:appdx_complexity:termbyterm_tot_time} results in the complexities stated in the Lemma.\hfill $\square$
\vspace{1cm}

\subsection{Simplifications for MRPT} \label{appendix:<VU(t)V> for MRPT}

This Appendix completes the derivation of all four terms in Eq. \eqref{eq:E2N_MRPT}, starting from Eq. \eqref{eq:sec_mrpt:defn_sums_Tn}. 
As explained in the main text, the creation and annihilation operators of virtual orbitals must appear in pairs, otherwise a term vanishes. Therefore we can write out 
\begin{align*}         
    T_1(t) &=\smashoperator{\sum_{\substack{v\in\mathcal{I}_{virt}\\a,b\in\mathcal{I}_{act}}}} h'_{va} h'_{vb} \expval*{a_b^{\dag} a_v U(t) a_v^{\dag} a_a} \\
    T_2(t) &=\tfrac{1}{2} \smashoperator{\sum_{\substack{v\in\mathcal{I}_{virt}\\a,b,c,d\in\mathcal{I}_{act}}}} h'_{vd} g_{vabc} \expval*{a_d^{\dag} a_v U(t) a_v^{\dag} a_a^{\dag} a_b a_c}\\
    &\phantom{{}\quad}+\tfrac{1}{2} \smashoperator{\sum_{\substack{v\in\mathcal{I}_{virt}\\a,b,c,d\in\mathcal{I}_{act}}}}  h'_{vd} \underbrace{g_{avbc}}_{g_{vacb}} \expval*{a_d^{\dag} a_v U(t) \underbrace{a_a^{\dag} a_v^{\dag} a_b a_c}_{a_v^{\dag}a_a^{\dag} a_c a_b}}\\
    T_3(t) &= \tfrac{1}{2} \smashoperator{\sum_{\substack{v\in\mathcal{I}_{virt}\\a,b,c,d\in\mathcal{I}_{act}}}} h'_{va} g_{vbcd} \expval*{a_d^{\dag} a_c^{\dag} a_b a_v U(t) a_v^{\dag} a_a}\\
    &\phantom{{}\quad} +\tfrac{1}{2} \smashoperator{\sum_{\substack{v\in\mathcal{I}_{virt}\\a,b,c,d\in\mathcal{I}_{act}}}} h'_{va} \underbrace{g_{bvcd}}_{g_{vbdc}} \expval*{\underbrace{a_d^{\dag} a_c^{\dag} a_v a_b}_{a_c^{\dag} a_d^{\dag} a_b a_v} U(t) a_v^{\dag} a_a}\\
    T_4(t) &= \tfrac{1}{4} \smashoperator{\sum_{\substack{v\in\mathcal{I}_{virt}\\a,b,c,d,e,f\in\mathcal{I}_{act}}}} g_{vabc} g_{vdef} \expval*{a_f^{\dag} a_e^{\dag} a_d a_v U(t) a_v^{\dag} a_a^{\dag} a_b a_c}\\
    &\phantom{{}\quad}+ \tfrac{1}{4} \smashoperator{\sum_{\substack{v\in\mathcal{I}_{virt}\\a,b,c,d,e,f\in\mathcal{I}_{act}}}} g_{vabc} \underbrace{g_{dvef}}_{g_{vdfe}} \expval*{\underbrace{a_f^{\dag}a_e^{\dag} a_v a_d}_{a_e^{\dag} a_f^{\dag} a_d a_v} U(t) a_v^{\dag} a_a^{\dag} a_b a_c}\\
    &\phantom{{}\quad}+ \tfrac{1}{4} \smashoperator{\sum_{\substack{v\in\mathcal{I}_{virt}\\a,b,c,d,e,f\in\mathcal{I}_{act}}}} g_{vdef} \underbrace{g_{avbc}}_{g_{vacb}} \expval*{a_f^{\dag} a_e^{\dag} a_d a_v U(t) \underbrace{a_a^{\dag} a_v^{\dag} a_b a_c}_{a_v^{\dag} a_a^{\dag} a_c a_b}}\\
    &\phantom{{}\quad}+\tfrac{1}{4} \smashoperator{\sum_{\substack{v\in\mathcal{I}_{virt}\\a,b,c,d,e,f\in\mathcal{I}_{act}}}} \overbrace{g_{avbc}}^{g_{vacb}} \underbrace{g_{dvef}}_{g_{vdfe}} \expval*{\underbrace{a_f^{\dag} a_e^{\dag} a_v a_d}_{a_e^{\dag} a_f^{\dag} a_d a_v} U(t) \underbrace{a_a^{\dag} a_v^{\dag} a_b a_c}_{a_v^{\dag} a_a^{\dag} a_c a_b}}\\
    &\phantom{{}\quad}+\tfrac{1}{4} \smashoperator{\sum_{\substack{v,w\in\mathcal{I}_{virt}\\a,b,c,d\in\mathcal{I}_{act}}}} g_{vwab} g_{vwcd} \expval*{a_d^{\dag} a_c^{\dag} a_w a_v U(t) a_v^{\dag} a_w^{\dag} a_a a_b}\\
    &\phantom{{}\quad} + \tfrac{1}{4} \smashoperator{\sum_{\substack{v,w\in\mathcal{I}_{virt}\\a,b,c,d\in\mathcal{I}_{act}}}} g_{vwab} \underbrace{g_{wvcd}}_{g_{vwdc}} \expval*{ \underbrace{a_d^{\dag} a_c^{\dag} a_v a_w}_{a_c^{\dag} a_d^{\dag} a_w a_v} U(t) a_v^{\dag} a_w^{\dag} a_a a_b}
\end{align*}
By anticommuting the operators as indicated (and swapping indices $c\leftrightarrow d$ in the last sum) the expressions can be combined:
\begin{align*}
    T_1(t) &=\smashoperator{\sum_{\substack{v\in\mathcal{I}_{virt}\\a,b\in\mathcal{I}_{act}}}} h'_{va} h'_{vb} \expval*{a_b^{\dag} a_v U(t) a_v^{\dag} a_a} \\
    T_2(t) &= \smashoperator{\sum_{\substack{v\in\mathcal{I}_{virt}\\a,b,c,d\in\mathcal{I}_{act}}}} h'_{vd} g_{vabc} \expval*{a_d^{\dag} a_v U(t) a_v^{\dag} a_a^{\dag} a_b a_c}\\
    T_3(t) &= \smashoperator{\sum_{\substack{v\in\mathcal{I}_{virt}\\a,b,c,d\in\mathcal{I}_{act}}}} h'_{va} g_{vbcd} \expval*{a_d^{\dag} a_c^{\dag} a_b a_v U(t) a_v^{\dag} a_a}\\
    T_4(t) &= \smashoperator{\sum_{\substack{v\in\mathcal{I}_{virt}\\a,b,c,d,e,f\in\mathcal{I}_{act}}}} g_{vabc} g_{vdef} \expval*{a_f^{\dag} a_e^{\dag} a_d a_v U(t) a_v^{\dag} a_a^{\dag} a_b a_c}\\
    &+
    \tfrac{1}{2} \smashoperator{\sum_{\substack{v,w\in\mathcal{I}_{virt}\\a,b,c,d\in\mathcal{I}_{act}}}} g_{vwab} g_{vwcd} \expval*{a_d^{\dag} a_c^{\dag} a_w a_v U(t) a_v^{\dag} a_w^{\dag} a_a a_b}.
\end{align*}
Next, as explained in the main text, the time-evolution of the virtual operators under $H$ is simple, $a_v(t) \equiv e^{iHt}a_v e^{-iHt} = a_ve^{-i\epsilon_vt}$, so we get
\begin{align*}
    T_1(t) &= \smashoperator{\sum_{\substack{v\in\mathcal{I}_{virt}\\a,b\in\mathcal{I}_{act}}}} h'_{va} h'_{vb} e^{-i(E_{core} + \epsilon_v)t} \expval*{a_b^{\dag} e^{-iH_{CAS}t} a_a} \\
    T_2(t) &= \smashoperator{\sum_{\substack{v\in\mathcal{I}_{virt}\\a,b,c,d\in\mathcal{I}_{act}}}} g_{vabc} h'_{vd} e^{-i(E_{core}+\epsilon_v)t} \expval*{a_d^{\dag} e^{-iH_{CAS}t} a_a^{\dag} a_b a_c}\\
    T_3(t) &= \smashoperator{\sum_{\substack{v\in\mathcal{I}_{virt}\\a,b,c,d\in\mathcal{I}_{act}}}} h'_{va} g_{vbcd} e^{-i(E_{core}+\epsilon_v)t} \expval*{a_d^{\dag} a_c^{\dag} a_b e^{-iH_{CAS}t} a_a}\\
    T_4(t) &= \smashoperator{\sum_{\substack{v\in\mathcal{I}_{virt}\\a,b,c,d,e,f\in\mathcal{I}_{act}}}} g_{vabc} g_{vdef} e^{-i(E_{core}+\epsilon_v)t} \expval*{a_f^{\dag} a_e^{\dag} a_d e^{-iH_{CAS}t} a_a^{\dag} a_b a_c}\\
    &+
    \tfrac{1}{2} \smashoperator{\sum_{\substack{v,w\in\mathcal{I}_{virt}\\a,b,c,d\in\mathcal{I}_{act}}}} g_{vwab} g_{vwcd} e^{-i(E_{core}+\epsilon_v+\epsilon_w)t}
    \\
    &\phantom{{}\quad}\times \expval*{a_d^{\dag} a_c^{\dag} e^{-iH_{CAS}t} a_a a_b}.
\end{align*}
Lastly, by defining the complex coefficients
\begin{align*}
    v_{a,b}(t) &= \smashoperator{\sum_{v\in\mathcal{I}_{virt}}} h'_{va} h'_{vb} e^{-i(\epsilon_v + E_{core})t}\\
    v_{abc,d}(t) &= \smashoperator{\sum_{v\in\mathcal{I}_{virt}}} h'_{vd} g_{vabc} e^{-i(\epsilon_v + E_{core})t} \\
    v_{abc,def}(t) &= \smashoperator{\sum_{v\in\mathcal{I}_{virt}}} g_{vabc}g_{vdef} e^{-i(\epsilon_v + E_{core})t}\\
    v_{ab,cd}(t) &= \tfrac{1}{2} \smashoperator{\sum_{v,w\in\mathcal{I}_{virt}}} g_{vwab} g_{vwcd} e^{-i(\epsilon_v+\epsilon_w + E_{core})t}\, ,
\end{align*}
we arrive at the expressions in Eq. \eqref{eq:E2N_MRPT}.

\subsection{MRPT2 norm bounds}
\label{appendix:mrpt2_norm}
First, we discuss the expressions for $\norm{h}_1$ and $\norm{v}_{2/3}$ corresponding to the Dyall-Hamiltonian $H$ [Eq. \eqref{eq:Dyall}] and to the full difference $V = H'_{el} - H$, respectively.

A fermion-to-qubit mapping will map each creation or annihilation operator to the sum of two Pauli strings with prefactor $1/2$.
Therefore, the fermion-to-qubit mapping of the Dyall-Hamiltonian is
\begin{align*}
    H &= \smashoperator{\sum_{a,b\in\mathcal{I}_{act}}} h_{ab}' a_a^{\dag} a_b 
    + \frac{1}{2}\smashoperator{\sum_{\quad a,b,c,d\in \mathcal{I}_{act}}} g_{abcd} a_a^{\dag} a_b^{\dag} a_c a_d\\
    & + \smashoperator{\sum_{v\in\mathcal{I}_{virt}}} \epsilon_v a_v^{\dag}a_v + E_{core}\\
    & \longrightarrow\\
    &\sum_{a,b\in\mathcal{I}_{act}} \sum_{i=1}^4 \frac{h_{ab}'}{4} \sigma_{abi} + \sum_{\quad a,b,c,d\in \mathcal{I}_{act}}\sum_{i=1}^{16} \frac{g_{abcd}}{32}\sigma_{abcdi} \\
    & + \sum_{v\in\mathcal{I}_{virt}} \sum_{i=1}^4 \frac{\epsilon_v}{4} \sigma_{vi} + E_{core}\, ,
\end{align*}
where $\sigma_{abi}, \sigma_{abcdi}$ and $\sigma_{vi}$ are Pauli strings.
The Pauli strings are not necessarily all distinct in the expression above, however, we still obtain an upper bound on $\norm{h}_1$ by calculating
\begin{equation*}
    \norm{h}_1 \leq \smashoperator{\sum_{a,b\in\mathcal{I}_{act}}} \abs{h_{ab}'} + \frac{1}{2}\smashoperator{\sum_{a,b,c,d\in \mathcal{I}_{act}}} \abs{g_{abcd}} + \smashoperator{\sum_{v\in \mathcal{I}_{virt}}} \abs{\epsilon_v} + \abs{E_{core}}\, .
\end{equation*}

Since two strings of creation and annihilation operators with distinct sets of indices are mapped to distinct sets of Paulistrings, this upper bound is tight up to taking permutations of indices into account and collecting identity terms.
In a similar way, the perturbation $V$ is mapped to
\begin{align*}
    V & = \smashoperator{\sum_{\substack{p,q\in\mathcal{I}_{act}\cup\mathcal{I}_{virt}\\ \{p,q\}\not\subset \mathcal{I}_{act}}}} h'_{pq}a_p^{\dag} a_q + \frac{1}{2} \smashoperator{\sum_{\substack{p,q,r,s\in \mathcal{I}_{virt}\cup\mathcal{I}_{act}\\ {\{p,q,r,s\}\not\subset \mathcal{I}_{act}}}}} g_{pqrs} a_p^{\dag} a_q^{\dag} a_r a_s - F_{virt}\\
    & \longrightarrow \\
    & \sum_{\substack{p,q\in\mathcal{I}_{act}\cup\mathcal{I}_{virt}\\ \{p,q\}\not\subset \mathcal{I}_{act}}} \sum_{i=1}^4 \frac{h'_{pq}}{4} \sigma_{pqi} - \sum_{v\in\mathcal{I}_{virt}}\sum_{i=1}^4 \frac{\epsilon_v}{4}\sigma_{vi}\\ 
    & + \sum_{\substack{p,q,r,s\in \mathcal{I}_{virt}\cup\mathcal{I}_{act}\\ {\{p,q,r,s\}\not\subset \mathcal{I}_{act}}}} \sum_{i=1}^{16} \frac{g_{pqrs}}{32}\sigma_{pqrsi}
\end{align*}
and $\norm{v}_{2/3}$ can be upper bound by
\begin{align*}
    \norm{v}_{2/3} \leq 2 \Bigg(& \qquad  \smashoperator{\sum_{\substack{p,q\in\mathcal{I}_{act}\cup\mathcal{I}_{virt}\\ \{p,q\}\not\subset \mathcal{I}_{act}}}} \abs{h'_{pq}}^{2/3} + \sum_{v\in\mathcal{I}_{virt}} \abs{\epsilon_v}^{2/3}\\
    & + \sum_{\substack{p,q,r,s\in \mathcal{I}_{virt}\cup\mathcal{I}_{act}\\ {\{p,q,r,s\}\not\subset \mathcal{I}_{act}}}} \abs{g_{pqrs}}^{2/3} \Bigg)^{3/2}\, .
\end{align*}

Next, we discuss upper bounds on $\norm{h^{CAS}}_1$ and $\norm{v^{\text{MRPT(2)}}}_{2/3}$.
Analogous to the preceding derivation of the upper bound on $\norm{h}_1$, we see that
\begin{equation*}
    \norm{h^{CAS}}_1 \leq \smashoperator{\sum_{a,b\in\mathcal{I}_{act}}} \abs{h_{ab}'} + \frac{1}{2}\smashoperator{\sum_{a,b,c,d\in \mathcal{I}_{act}}} \abs{g_{abcd}}\, .
\end{equation*}
For $\norm{v^{\text{MRPT(2)}}}_{2/3}$, we need to consider how the creation and annihilation operators are mapped to Pauli operators in Eq. \eqref{eq:E2N_MRPT}
\begin{align*}
    &\expval{VU(t_n)V} = T_1(t_n) + T_2(t_n) + T_3(t_n) + T_4(t_n)\\
    & \longrightarrow \\
    &\frac14\sum_{a,b\in\mathcal{I}_{act}}\sum_{{ij}=1}^2 v_{a,b}(t_n) \expval*{\sigma_{bj} e^{-iH_{CAS}t_n} \sigma_{ai}} \nonumber\\
    +& \frac{1}{16} \sum_{a,b,c,d\in\mathcal{I}_{act}}\sum_{i=1}^8\sum_{j=1}^2 v_{abc,d}(t_n) \expval*{\sigma_{dj} e^{-iH_{CAS}t_n} \sigma_{abci}} \nonumber \\
    +& \frac{1}{16} \sum_{a,b,c,d\in\mathcal{I}_{act}} \sum_{i=1}^2\sum_{j=1}^8 v_{bcd,a}(t_n) \expval*{\sigma_{bcdj} e^{-iH_{CAS}t_n} \sigma_{ai}} \label{eq:E2N_MRPT} \\
    +& \frac{1}{32} \sum_{a,b,c,d\in\mathcal{I}_{act}} \sum_{ij=1}^4 v_{ab,cd}(t_n) \expval*{\sigma_{cdj} e^{-iH_{CAS}t_n} \sigma_{abi}} \nonumber \\
    +& \frac{1}{64} \sum_{a,b,c,d,e,f\in\mathcal{I}_{act}} \sum_{ij=1}^8 v_{abc,def}(t_n) \expval*{\sigma_{defj} e^{-iH_{CAS}t_n} \sigma_{abci}}\, ,
\end{align*}
which is of the desired form
\begin{equation*}
    \smashoperator{\sum_{i=1}^{L^{\text{MRPT}(2)}} } v_{ni}^{\text{MRPT(2)}} \expval{\sigma_{i,1} e^{-iH_{CAS}t_n} \sigma_{i,2}}\,
\end{equation*}
see also Eq. \eqref{eq:sec_MRPT2:E2N_lincomb_unitaries}.

Consequently, we get
\begin{align*}
    &\left(\sum_i \abs*{v_{ni}^{\text{MRPT(2)}}}^{2/3}\right)^{3/2}\\
    &\leq \Bigg( 2^{2/3} \smashoperator{\sum_{a,b\in \mathcal{I}_{act}}} \abs{v_{a,b}(t_n)}^{2/3} + 2^{7/3} \smashoperator{\sum_{a,b,c,d\in\mathcal{I}_{act}}} \abs{v_{abc,d}(t_n)}^{2/3}\\
    & + 2^{2/3} \smashoperator{\sum_{a,b,c,d\in\mathcal{I}_{act}}} \abs*{v_{ab,cd}(t_n)}^{2/3} + 4 \smashoperator{\sum_{a,b,c,d,e,f\in\mathcal{I}_{act}}} \abs*{v_{abc,def}(t_n)}^{2/3}\Bigg)^{3/2}\, \\
    &\leq \Bigg( 2^{2/3} \sum_{a,b\in \mathcal{I}_{act}} \left(\sum_{v\in\mathcal{I}_{virt}} \abs{h'_{va} h'_{vb}}\right)^{2/3}\\
    &+ 2^{7/3} \sum_{a,b,c,d\in\mathcal{I}_{act}} \left(\sum_{v\in\mathcal{I}_{virt}} \abs{h'_{va} g_{vbcd}} \right)^{2/3}\\
    &+ 2^{2/3} \sum_{a,b,c,d\in\mathcal{I}_{act}} \left(\sum_{vw\in\mathcal{I}_{virt}}\abs{g_{vwab} g_{vwcd}}\right)^{2/3}\\
    &+ 4 \sum_{a,b,c,d,e,f\in\mathcal{I}_{act}} \left(\sum_{v\in\mathcal{I}_{virt}} \abs{g_{vabc}g_{vdef}}\right)^{2/3} \Bigg)^{3/2}\\
    &\equiv \norm{v^{\text{MRPT(2)}}}_{2/3}\, .
\end{align*}

\bibliography{bibliography.bib}

\begin{thebibliography}{71}%
\makeatletter
\providecommand \@ifxundefined [1]{%
 \@ifx{#1\undefined}
}%
\providecommand \@ifnum [1]{%
 \ifnum #1\expandafter \@firstoftwo
 \else \expandafter \@secondoftwo
 \fi
}%
\providecommand \@ifx [1]{%
 \ifx #1\expandafter \@firstoftwo
 \else \expandafter \@secondoftwo
 \fi
}%
\providecommand \natexlab [1]{#1}%
\providecommand \enquote  [1]{``#1''}%
\providecommand \bibnamefont  [1]{#1}%
\providecommand \bibfnamefont [1]{#1}%
\providecommand \citenamefont [1]{#1}%
\providecommand \href@noop [0]{\@secondoftwo}%
\providecommand \href [0]{\begingroup \@sanitize@url \@href}%
\providecommand \@href[1]{\@@startlink{#1}\@@href}%
\providecommand \@@href[1]{\endgroup#1\@@endlink}%
\providecommand \@sanitize@url [0]{\catcode `\\12\catcode `\$12\catcode
  `\&12\catcode `\#12\catcode `\^12\catcode `\_12\catcode `\%12\relax}%
\providecommand \@@startlink[1]{}%
\providecommand \@@endlink[0]{}%
\providecommand \url  [0]{\begingroup\@sanitize@url \@url }%
\providecommand \@url [1]{\endgroup\@href {#1}{\urlprefix }}%
\providecommand \urlprefix  [0]{URL }%
\providecommand \Eprint [0]{\href }%
\providecommand \doibase [0]{https://doi.org/}%
\providecommand \selectlanguage [0]{\@gobble}%
\providecommand \bibinfo  [0]{\@secondoftwo}%
\providecommand \bibfield  [0]{\@secondoftwo}%
\providecommand \translation [1]{[#1]}%
\providecommand \BibitemOpen [0]{}%
\providecommand \bibitemStop [0]{}%
\providecommand \bibitemNoStop [0]{.\EOS\space}%
\providecommand \EOS [0]{\spacefactor3000\relax}%
\providecommand \BibitemShut  [1]{\csname bibitem#1\endcsname}%
\let\auto@bib@innerbib\@empty
\bibitem [{\citenamefont {Kassal}\ \emph {et~al.}(2011)\citenamefont {Kassal},
  \citenamefont {Whitfield}, \citenamefont {Perdomo-Ortiz}, \citenamefont
  {Yung},\ and\ \citenamefont {Aspuru-Guzik}}]{AspuruGuzik2011_Review}%
  \BibitemOpen
  \bibfield  {author} {\bibinfo {author} {\bibfnamefont {I.}~\bibnamefont
  {Kassal}}, \bibinfo {author} {\bibfnamefont {J.~D.}\ \bibnamefont
  {Whitfield}}, \bibinfo {author} {\bibfnamefont {A.}~\bibnamefont
  {Perdomo-Ortiz}}, \bibinfo {author} {\bibfnamefont {M.-H.}\ \bibnamefont
  {Yung}},\ and\ \bibinfo {author} {\bibfnamefont {A.}~\bibnamefont
  {Aspuru-Guzik}},\ }\bibfield  {title} {\bibinfo {title} {{Simulating
  Chemistry Using Quantum Computers}},\ }\href
  {https://doi.org/10.1146/annurev-physchem-032210-103512} {\bibfield
  {journal} {\bibinfo  {journal} {Ann. Rev. Phys. Chem.}\ }\textbf {\bibinfo
  {volume} {62}},\ \bibinfo {pages} {185} (\bibinfo {year} {2011})}\BibitemShut
  {NoStop}%
\bibitem [{\citenamefont {Cao}\ \emph {et~al.}(2019)\citenamefont {Cao},
  \citenamefont {Romero}, \citenamefont {Olson}, \citenamefont {Degroote},
  \citenamefont {Johnson}, \citenamefont {Kieferov\'{a}}, \citenamefont
  {Kivlichan}, \citenamefont {Menke}, \citenamefont {Peropadre},\ and\
  \citenamefont {\textit{et al.}}}]{AspuruGuzik2019_Review}%
  \BibitemOpen
  \bibfield  {author} {\bibinfo {author} {\bibfnamefont {Y.}~\bibnamefont
  {Cao}}, \bibinfo {author} {\bibfnamefont {J.}~\bibnamefont {Romero}},
  \bibinfo {author} {\bibfnamefont {J.~P.}\ \bibnamefont {Olson}}, \bibinfo
  {author} {\bibfnamefont {M.}~\bibnamefont {Degroote}}, \bibinfo {author}
  {\bibfnamefont {P.~D.}\ \bibnamefont {Johnson}}, \bibinfo {author}
  {\bibfnamefont {M.}~\bibnamefont {Kieferov\'{a}}}, \bibinfo {author}
  {\bibfnamefont {I.~D.}\ \bibnamefont {Kivlichan}}, \bibinfo {author}
  {\bibfnamefont {T.}~\bibnamefont {Menke}}, \bibinfo {author} {\bibfnamefont
  {B.}~\bibnamefont {Peropadre}},\ and\ \bibinfo {author} {\bibnamefont
  {\textit{et al.}}},\ }\bibfield  {title} {\bibinfo {title} {{Quantum
  Chemistry in the Age of Quantum Computing}},\ }\href
  {https://doi.org/10.1021/acs.chemrev.8b00803} {\bibfield  {journal} {\bibinfo
   {journal} {Chem. Rev.}\ }\textbf {\bibinfo {volume} {119}},\ \bibinfo
  {pages} {10856} (\bibinfo {year} {2019})}\BibitemShut {NoStop}%
\bibitem [{\citenamefont {McArdle}\ \emph {et~al.}(2020)\citenamefont
  {McArdle}, \citenamefont {Endo}, \citenamefont {Aspuru-Guzik}, \citenamefont
  {Benjamin},\ and\ \citenamefont {Yuan}}]{McArdle2020_Review}%
  \BibitemOpen
  \bibfield  {author} {\bibinfo {author} {\bibfnamefont {S.}~\bibnamefont
  {McArdle}}, \bibinfo {author} {\bibfnamefont {S.}~\bibnamefont {Endo}},
  \bibinfo {author} {\bibfnamefont {A.}~\bibnamefont {Aspuru-Guzik}}, \bibinfo
  {author} {\bibfnamefont {S.~C.}\ \bibnamefont {Benjamin}},\ and\ \bibinfo
  {author} {\bibfnamefont {X.}~\bibnamefont {Yuan}},\ }\bibfield  {title}
  {\bibinfo {title} {{Quantum computational chemistry}},\ }\href
  {https://doi.org/10.1103/RevModPhys.92.015003} {\bibfield  {journal}
  {\bibinfo  {journal} {Rev. Mod. Phys.}\ }\textbf {\bibinfo {volume} {92}},\
  \bibinfo {pages} {015003} (\bibinfo {year} {2020})}\BibitemShut {NoStop}%
\bibitem [{\citenamefont {Bauer}\ \emph {et~al.}(2020)\citenamefont {Bauer},
  \citenamefont {Bravyi}, \citenamefont {Motta},\ and\ \citenamefont
  {Chan}}]{bauer_quantum_2020}%
  \BibitemOpen
  \bibfield  {author} {\bibinfo {author} {\bibfnamefont {B.}~\bibnamefont
  {Bauer}}, \bibinfo {author} {\bibfnamefont {S.}~\bibnamefont {Bravyi}},
  \bibinfo {author} {\bibfnamefont {M.}~\bibnamefont {Motta}},\ and\ \bibinfo
  {author} {\bibfnamefont {G.~K.-L.}\ \bibnamefont {Chan}},\ }\bibfield
  {title} {\bibinfo {title} {{Quantum algorithms for quantum chemistry and
  quantum materials science}},\ }\href
  {https://doi.org/10.1021/acs.chemrev.9b00829} {\bibfield  {journal} {\bibinfo
   {journal} {Chem. Rev.}\ }\textbf {\bibinfo {volume} {120}},\ \bibinfo
  {pages} {12685} (\bibinfo {year} {2020})}\BibitemShut {NoStop}%
\bibitem [{\citenamefont {Baiardi}\ \emph {et~al.}(2023)\citenamefont
  {Baiardi}, \citenamefont {Christandl},\ and\ \citenamefont
  {Reiher}}]{baiardi_quantum_2023}%
  \BibitemOpen
  \bibfield  {author} {\bibinfo {author} {\bibfnamefont {A.}~\bibnamefont
  {Baiardi}}, \bibinfo {author} {\bibfnamefont {M.}~\bibnamefont
  {Christandl}},\ and\ \bibinfo {author} {\bibfnamefont {M.}~\bibnamefont
  {Reiher}},\ }\bibfield  {title} {\bibinfo {title} {Quantum {Computing} for
  {Molecular} {Biology}},\ }\href {https://doi.org/10.1002/cbic.202300120}
  {\bibfield  {journal} {\bibinfo  {journal} {ChemBioChem}\ }\textbf {\bibinfo
  {volume} {24}},\ \bibinfo {pages} {e202300120} (\bibinfo {year}
  {2023})}\BibitemShut {NoStop}%
\bibitem [{\citenamefont {Goings}\ \emph {et~al.}(2022)\citenamefont {Goings},
  \citenamefont {White}, \citenamefont {Lee}, \citenamefont {Tautermann},
  \citenamefont {Degroote}, \citenamefont {Gidney}, \citenamefont {Shiozaki},
  \citenamefont {Babbush},\ and\ \citenamefont
  {Rubin}}]{Goings2022_CostEstimate-P450}%
  \BibitemOpen
  \bibfield  {author} {\bibinfo {author} {\bibfnamefont {J.~J.}\ \bibnamefont
  {Goings}}, \bibinfo {author} {\bibfnamefont {A.}~\bibnamefont {White}},
  \bibinfo {author} {\bibfnamefont {J.}~\bibnamefont {Lee}}, \bibinfo {author}
  {\bibfnamefont {C.~S.}\ \bibnamefont {Tautermann}}, \bibinfo {author}
  {\bibfnamefont {M.}~\bibnamefont {Degroote}}, \bibinfo {author}
  {\bibfnamefont {C.}~\bibnamefont {Gidney}}, \bibinfo {author} {\bibfnamefont
  {T.}~\bibnamefont {Shiozaki}}, \bibinfo {author} {\bibfnamefont
  {R.}~\bibnamefont {Babbush}},\ and\ \bibinfo {author} {\bibfnamefont {N.~C.}\
  \bibnamefont {Rubin}},\ }\bibfield  {title} {\bibinfo {title} {{Reliably
  assessing the electronic structure of cytochrome {P450} on today's classical
  computers and tomorrow's quantum computers}},\ }\href@noop {} {\bibfield
  {journal} {\bibinfo  {journal} {Proc. Natl. Acad. Sci.}\ }\textbf {\bibinfo
  {volume} {119}},\ \bibinfo {pages} {e2203533119} (\bibinfo {year}
  {2022})}\BibitemShut {NoStop}%
\bibitem [{\citenamefont {Lee}\ \emph {et~al.}(2023)\citenamefont {Lee},
  \citenamefont {Lee}, \citenamefont {Zhai}, \citenamefont {Tong},
  \citenamefont {Dalzell}, \citenamefont {Kumar}, \citenamefont {Helms},
  \citenamefont {Gray}, \citenamefont {Cui}, \citenamefont {Liu}, \citenamefont
  {Kastoryano}, \citenamefont {Babbush}, \citenamefont {Preskill},
  \citenamefont {Reichman}, \citenamefont {Campbell}, \citenamefont {Valeev},
  \citenamefont {Lin},\ and\ \citenamefont {Chan}}]{lee_evaluating_2023}%
  \BibitemOpen
  \bibfield  {author} {\bibinfo {author} {\bibfnamefont {S.}~\bibnamefont
  {Lee}}, \bibinfo {author} {\bibfnamefont {J.}~\bibnamefont {Lee}}, \bibinfo
  {author} {\bibfnamefont {H.}~\bibnamefont {Zhai}}, \bibinfo {author}
  {\bibfnamefont {Y.}~\bibnamefont {Tong}}, \bibinfo {author} {\bibfnamefont
  {A.~M.}\ \bibnamefont {Dalzell}}, \bibinfo {author} {\bibfnamefont
  {A.}~\bibnamefont {Kumar}}, \bibinfo {author} {\bibfnamefont
  {P.}~\bibnamefont {Helms}}, \bibinfo {author} {\bibfnamefont
  {J.}~\bibnamefont {Gray}}, \bibinfo {author} {\bibfnamefont {Z.-H.}\
  \bibnamefont {Cui}}, \bibinfo {author} {\bibfnamefont {W.}~\bibnamefont
  {Liu}}, \bibinfo {author} {\bibfnamefont {M.}~\bibnamefont {Kastoryano}},
  \bibinfo {author} {\bibfnamefont {R.}~\bibnamefont {Babbush}}, \bibinfo
  {author} {\bibfnamefont {J.}~\bibnamefont {Preskill}}, \bibinfo {author}
  {\bibfnamefont {D.~R.}\ \bibnamefont {Reichman}}, \bibinfo {author}
  {\bibfnamefont {E.~T.}\ \bibnamefont {Campbell}}, \bibinfo {author}
  {\bibfnamefont {E.~F.}\ \bibnamefont {Valeev}}, \bibinfo {author}
  {\bibfnamefont {L.}~\bibnamefont {Lin}},\ and\ \bibinfo {author}
  {\bibfnamefont {G.~K.-L.}\ \bibnamefont {Chan}},\ }\bibfield  {title}
  {\bibinfo {title} {{Evaluating the evidence for exponential quantum advantage
  in ground-state quantum chemistry}},\ }\href@noop {} {\bibfield  {journal}
  {\bibinfo  {journal} {Nat. Commun.}\ }\textbf {\bibinfo {volume} {14}},\
  \bibinfo {pages} {1952} (\bibinfo {year} {2023})}\BibitemShut {NoStop}%
\bibitem [{\citenamefont {Gharibian}\ \emph {et~al.}()\citenamefont
  {Gharibian}, \citenamefont {Hayakawa}, \citenamefont {Gall},\ and\
  \citenamefont {Morimae}}]{gharibianImprovedHardnessResults2022a}%
  \BibitemOpen
  \bibfield  {author} {\bibinfo {author} {\bibfnamefont {S.}~\bibnamefont
  {Gharibian}}, \bibinfo {author} {\bibfnamefont {R.}~\bibnamefont {Hayakawa}},
  \bibinfo {author} {\bibfnamefont {F.~L.}\ \bibnamefont {Gall}},\ and\
  \bibinfo {author} {\bibfnamefont {T.}~\bibnamefont {Morimae}},\ }\href
  {https://doi.org/10.48550/arXiv.2207.10250} {\bibinfo {title} {Improved
  {{Hardness Results}} for the {{Guided Local Hamiltonian Problem}}}},\ \Eprint
  {https://arxiv.org/abs/2207.10250} {2207.10250 [quant-ph]} \BibitemShut
  {NoStop}%
\bibitem [{\citenamefont {Chan}\ \emph {et~al.}(2023)\citenamefont {Chan},
  \citenamefont {Meister}, \citenamefont {Jones}, \citenamefont {Tew},\ and\
  \citenamefont {Benjamin}}]{chan_grid-based_2023}%
  \BibitemOpen
  \bibfield  {author} {\bibinfo {author} {\bibfnamefont {H.~H.~S.}\
  \bibnamefont {Chan}}, \bibinfo {author} {\bibfnamefont {R.}~\bibnamefont
  {Meister}}, \bibinfo {author} {\bibfnamefont {T.}~\bibnamefont {Jones}},
  \bibinfo {author} {\bibfnamefont {D.~P.}\ \bibnamefont {Tew}},\ and\ \bibinfo
  {author} {\bibfnamefont {S.~C.}\ \bibnamefont {Benjamin}},\ }\bibfield
  {title} {\bibinfo {title} {Grid-based methods for chemistry simulations on a
  quantum computer},\ }\bibfield  {journal} {\bibinfo  {journal} {Science
  Advances}\ }\textbf {\bibinfo {volume} {9}},\ \href
  {https://doi.org/10.1126/sciadv.abo7484} {10.1126/sciadv.abo7484} (\bibinfo
  {year} {2023})\BibitemShut {NoStop}%
\bibitem [{\citenamefont {Roos}(2005)}]{Roos2005_Book-CASSCF}%
  \BibitemOpen
  \bibfield  {author} {\bibinfo {author} {\bibfnamefont {B.~O.}\ \bibnamefont
  {Roos}},\ }\bibfield  {title} {\bibinfo {title} {{Chapter 25 -
  Multiconfigurational quantum chemistry}},\ }in\ \href
  {https://doi.org/https://doi.org/10.1016/B978-044451719-7/50068-8} {\emph
  {\bibinfo {booktitle} {{Theory and Applications of Computational
  Chemistry}}}},\ \bibinfo {editor} {edited by\ \bibinfo {editor}
  {\bibfnamefont {C.~E.}\ \bibnamefont {Dykstra}}, \bibinfo {editor}
  {\bibfnamefont {G.}~\bibnamefont {Frenking}}, \bibinfo {editor}
  {\bibfnamefont {K.~S.}\ \bibnamefont {Kim}},\ and\ \bibinfo {editor}
  {\bibfnamefont {G.~E.}\ \bibnamefont {Scuseria}}}\ (\bibinfo  {publisher}
  {Elsevier},\ \bibinfo {address} {Amsterdam},\ \bibinfo {year} {2005})\ pp.\
  \bibinfo {pages} {725--764}\BibitemShut {NoStop}%
\bibitem [{\citenamefont {Olsen}(2011)}]{Olsen2011_CASSCF-Review}%
  \BibitemOpen
  \bibfield  {author} {\bibinfo {author} {\bibfnamefont {J.}~\bibnamefont
  {Olsen}},\ }\bibfield  {title} {\bibinfo {title} {{The CASSCF method: A
  perspective and commentary}},\ }\href
  {https://doi.org/https://doi.org/10.1002/qua.23107} {\bibfield  {journal}
  {\bibinfo  {journal} {Int. J. Quantum Chem.}\ }\textbf {\bibinfo {volume}
  {111}},\ \bibinfo {pages} {3267} (\bibinfo {year} {2011})}\BibitemShut
  {NoStop}%
\bibitem [{\citenamefont {Lischka}\ \emph {et~al.}(2018)\citenamefont
  {Lischka}, \citenamefont {Nachtigallov{\'a}}, \citenamefont {Aquino},
  \citenamefont {Szalay}, \citenamefont {Plasser}, \citenamefont {Machado},\
  and\ \citenamefont {Barbatti}}]{lischka_multireference_2018}%
  \BibitemOpen
  \bibfield  {author} {\bibinfo {author} {\bibfnamefont {H.}~\bibnamefont
  {Lischka}}, \bibinfo {author} {\bibfnamefont {D.}~\bibnamefont
  {Nachtigallov{\'a}}}, \bibinfo {author} {\bibfnamefont {A.~J.~A.}\
  \bibnamefont {Aquino}}, \bibinfo {author} {\bibfnamefont {P.~G.}\
  \bibnamefont {Szalay}}, \bibinfo {author} {\bibfnamefont {F.}~\bibnamefont
  {Plasser}}, \bibinfo {author} {\bibfnamefont {F.~B.~C.}\ \bibnamefont
  {Machado}},\ and\ \bibinfo {author} {\bibfnamefont {M.}~\bibnamefont
  {Barbatti}},\ }\bibfield  {title} {\bibinfo {title} {{Multireference
  Approaches for Excited States of Molecules}},\ }\href
  {https://doi.org/10.1021/acs.chemrev.8b00244} {\bibfield  {journal} {\bibinfo
   {journal} {Chem. Rev.}\ }\textbf {\bibinfo {volume} {118}},\ \bibinfo
  {pages} {7293} (\bibinfo {year} {2018})}\BibitemShut {NoStop}%
\bibitem [{\citenamefont {Gonzalez}(2020)}]{Gonzalez2020-Book}%
  \BibitemOpen
  \bibfield  {author} {\bibinfo {author} {\bibfnamefont {L.}~\bibnamefont
  {Gonzalez}},\ }\href@noop {} {\emph {\bibinfo {title} {{Quantum chemistry and
  dynamics of excited states}}}},\ edited by\ \bibinfo {editor} {\bibfnamefont
  {R.}~\bibnamefont {Lindh}}\ and\ \bibinfo {editor} {\bibfnamefont
  {L.}~\bibnamefont {Gonzalez}}\ (\bibinfo  {publisher} {John Wiley \& Sons},\
  \bibinfo {address} {Nashville, TN},\ \bibinfo {year} {2020})\BibitemShut
  {NoStop}%
\bibitem [{\citenamefont {Chan}\ \emph {et~al.}(2008)\citenamefont {Chan},
  \citenamefont {Dorando}, \citenamefont {Ghosh}, \citenamefont {Hachmann},
  \citenamefont {Neuscamman}, \citenamefont {Wang},\ and\ \citenamefont
  {Yanai}}]{Chan2008_Review}%
  \BibitemOpen
  \bibfield  {author} {\bibinfo {author} {\bibfnamefont {G.~K.-L.}\
  \bibnamefont {Chan}}, \bibinfo {author} {\bibfnamefont {J.~J.}\ \bibnamefont
  {Dorando}}, \bibinfo {author} {\bibfnamefont {D.}~\bibnamefont {Ghosh}},
  \bibinfo {author} {\bibfnamefont {J.}~\bibnamefont {Hachmann}}, \bibinfo
  {author} {\bibfnamefont {E.}~\bibnamefont {Neuscamman}}, \bibinfo {author}
  {\bibfnamefont {H.}~\bibnamefont {Wang}},\ and\ \bibinfo {author}
  {\bibfnamefont {T.}~\bibnamefont {Yanai}},\ }\bibfield  {title} {\bibinfo
  {title} {{An Introduction to the Density Matrix Renormalization Group Ansatz
  in Quantum Chemistry}},\ }in\ \href@noop {} {\emph {\bibinfo {booktitle}
  {Frontiers in Quantum Systems in Chemistry and Physics}}}\ (\bibinfo
  {publisher} {Springer-Verlag},\ \bibinfo {year} {2008})\ pp.\ \bibinfo
  {pages} {49--65}\BibitemShut {NoStop}%
\bibitem [{\citenamefont {Chan}\ and\ \citenamefont
  {Zgid}(2009)}]{Zgid2009_Review}%
  \BibitemOpen
  \bibfield  {author} {\bibinfo {author} {\bibfnamefont {G.~K.~L.}\
  \bibnamefont {Chan}}\ and\ \bibinfo {author} {\bibfnamefont {D.}~\bibnamefont
  {Zgid}},\ }\bibfield  {title} {\bibinfo {title} {{The Density Matrix
  Renormalization Group in Quantum Chemistry}},\ }\href@noop {} {\bibfield
  {journal} {\bibinfo  {journal} {Annu. Rep. Comput. Chem.}\ }\textbf {\bibinfo
  {volume} {5}},\ \bibinfo {pages} {149} (\bibinfo {year} {2009})}\BibitemShut
  {NoStop}%
\bibitem [{\citenamefont {Marti}\ and\ \citenamefont
  {Reiher}(2010)}]{Marti2010_Review-DMRG}%
  \BibitemOpen
  \bibfield  {author} {\bibinfo {author} {\bibfnamefont {K.~H.}\ \bibnamefont
  {Marti}}\ and\ \bibinfo {author} {\bibfnamefont {M.}~\bibnamefont {Reiher}},\
  }\bibfield  {title} {\bibinfo {title} {{The density matrix renormalization
  group algorithm in quantum chemistry}},\ }\href@noop {} {\bibfield  {journal}
  {\bibinfo  {journal} {Z. Phys. Chem.}\ }\textbf {\bibinfo {volume} {224}},\
  \bibinfo {pages} {583} (\bibinfo {year} {2010})}\BibitemShut {NoStop}%
\bibitem [{\citenamefont {Schollw\"{o}ck}(2011)}]{Schollwoeck2011_Review}%
  \BibitemOpen
  \bibfield  {author} {\bibinfo {author} {\bibfnamefont {U.}~\bibnamefont
  {Schollw\"{o}ck}},\ }\bibfield  {title} {\bibinfo {title} {{The
  density-matrix renormalization group in the age of matrix product states}},\
  }\href@noop {} {\bibfield  {journal} {\bibinfo  {journal} {Ann. Phys.}\
  }\textbf {\bibinfo {volume} {326}},\ \bibinfo {pages} {96} (\bibinfo {year}
  {2011})}\BibitemShut {NoStop}%
\bibitem [{\citenamefont {Chan}\ and\ \citenamefont
  {Sharma}(2011)}]{chan2011density}%
  \BibitemOpen
  \bibfield  {author} {\bibinfo {author} {\bibfnamefont {G.~K.-L.}\
  \bibnamefont {Chan}}\ and\ \bibinfo {author} {\bibfnamefont {S.}~\bibnamefont
  {Sharma}},\ }\bibfield  {title} {\bibinfo {title} {The density matrix
  renormalization group in quantum chemistry},\ }\href@noop {} {\bibfield
  {journal} {\bibinfo  {journal} {Annu. Rev. Phys. Chem.}\ }\textbf {\bibinfo
  {volume} {62}},\ \bibinfo {pages} {465} (\bibinfo {year} {2011})}\BibitemShut
  {NoStop}%
\bibitem [{\citenamefont {Wouters}\ and\ \citenamefont {{Van
  Neck}}(2013)}]{Wouters2013_Review}%
  \BibitemOpen
  \bibfield  {author} {\bibinfo {author} {\bibfnamefont {S.}~\bibnamefont
  {Wouters}}\ and\ \bibinfo {author} {\bibfnamefont {D.}~\bibnamefont {{Van
  Neck}}},\ }\bibfield  {title} {\bibinfo {title} {{The density matrix
  renormalization group for ab initio quantum chemistry}},\ }\href@noop {}
  {\bibfield  {journal} {\bibinfo  {journal} {Eur. Phys. J. D}\ }\textbf
  {\bibinfo {volume} {31}},\ \bibinfo {pages} {272} (\bibinfo {year}
  {2013})}\BibitemShut {NoStop}%
\bibitem [{\citenamefont {Kurashige}(2014)}]{Kurashige2014_Review}%
  \BibitemOpen
  \bibfield  {author} {\bibinfo {author} {\bibfnamefont {Y.}~\bibnamefont
  {Kurashige}},\ }\bibfield  {title} {\bibinfo {title} {{Multireference
  electron correlation methods with density matrix renormalisation group
  reference functions}},\ }\href@noop {} {\bibfield  {journal} {\bibinfo
  {journal} {Mol. Phys.}\ }\textbf {\bibinfo {volume} {112}},\ \bibinfo {pages}
  {1485} (\bibinfo {year} {2014})}\BibitemShut {NoStop}%
\bibitem [{\citenamefont {Olivares-Amaya}\ \emph {et~al.}(2015)\citenamefont
  {Olivares-Amaya}, \citenamefont {Hu}, \citenamefont {Nakatani}, \citenamefont
  {Sharma}, \citenamefont {Yang},\ and\ \citenamefont
  {Chan}}]{Olivares2015_DMRGInPractice}%
  \BibitemOpen
  \bibfield  {author} {\bibinfo {author} {\bibfnamefont {R.}~\bibnamefont
  {Olivares-Amaya}}, \bibinfo {author} {\bibfnamefont {W.}~\bibnamefont {Hu}},
  \bibinfo {author} {\bibfnamefont {N.}~\bibnamefont {Nakatani}}, \bibinfo
  {author} {\bibfnamefont {S.}~\bibnamefont {Sharma}}, \bibinfo {author}
  {\bibfnamefont {J.}~\bibnamefont {Yang}},\ and\ \bibinfo {author}
  {\bibfnamefont {G.~K.-L.}\ \bibnamefont {Chan}},\ }\bibfield  {title}
  {\bibinfo {title} {{The ab-initio density matrix renormalization group in
  practice}},\ }\href@noop {} {\bibfield  {journal} {\bibinfo  {journal} {J.
  Chem. Phys.}\ }\textbf {\bibinfo {volume} {142}},\ \bibinfo {pages} {34102}
  (\bibinfo {year} {2015})}\BibitemShut {NoStop}%
\bibitem [{\citenamefont {Szalay}\ \emph {et~al.}(2015)\citenamefont {Szalay},
  \citenamefont {Pfeffer}, \citenamefont {Murg}, \citenamefont {Barcza},
  \citenamefont {Verstraete}, \citenamefont {Schneider},\ and\ \citenamefont
  {Legeza}}]{szalay2015tensor}%
  \BibitemOpen
  \bibfield  {author} {\bibinfo {author} {\bibfnamefont {S.}~\bibnamefont
  {Szalay}}, \bibinfo {author} {\bibfnamefont {M.}~\bibnamefont {Pfeffer}},
  \bibinfo {author} {\bibfnamefont {V.}~\bibnamefont {Murg}}, \bibinfo {author}
  {\bibfnamefont {G.}~\bibnamefont {Barcza}}, \bibinfo {author} {\bibfnamefont
  {F.}~\bibnamefont {Verstraete}}, \bibinfo {author} {\bibfnamefont
  {R.}~\bibnamefont {Schneider}},\ and\ \bibinfo {author} {\bibfnamefont
  {{\"O}.}~\bibnamefont {Legeza}},\ }\bibfield  {title} {\bibinfo {title}
  {Tensor product methods and entanglement optimization for ab initio quantum
  chemistry},\ }\href@noop {} {\bibfield  {journal} {\bibinfo  {journal} {Int.
  J. Quantum Chem.}\ }\textbf {\bibinfo {volume} {115}},\ \bibinfo {pages}
  {1342} (\bibinfo {year} {2015})}\BibitemShut {NoStop}%
\bibitem [{\citenamefont {Yanai}\ \emph {et~al.}(2015)\citenamefont {Yanai},
  \citenamefont {Kurashige}, \citenamefont {Mizukami}, \citenamefont
  {Chalupsk{\'{y}}}, \citenamefont {Lan},\ and\ \citenamefont
  {Saitow}}]{Yanai2015}%
  \BibitemOpen
  \bibfield  {author} {\bibinfo {author} {\bibfnamefont {T.}~\bibnamefont
  {Yanai}}, \bibinfo {author} {\bibfnamefont {Y.}~\bibnamefont {Kurashige}},
  \bibinfo {author} {\bibfnamefont {W.}~\bibnamefont {Mizukami}}, \bibinfo
  {author} {\bibfnamefont {J.}~\bibnamefont {Chalupsk{\'{y}}}}, \bibinfo
  {author} {\bibfnamefont {T.~N.}\ \bibnamefont {Lan}},\ and\ \bibinfo {author}
  {\bibfnamefont {M.}~\bibnamefont {Saitow}},\ }\bibfield  {title} {\bibinfo
  {title} {{Density matrix renormalization group for ab initio calculations and
  associated dynamic correlation methods: A review of theory and
  applications}},\ }\href@noop {} {\bibfield  {journal} {\bibinfo  {journal}
  {Int. J. Quantum Chem.}\ }\textbf {\bibinfo {volume} {115}},\ \bibinfo
  {pages} {283} (\bibinfo {year} {2015})}\BibitemShut {NoStop}%
\bibitem [{\citenamefont {Baiardi}\ and\ \citenamefont
  {Reiher}(2020)}]{Baiardi2020_Review}%
  \BibitemOpen
  \bibfield  {author} {\bibinfo {author} {\bibfnamefont {A.}~\bibnamefont
  {Baiardi}}\ and\ \bibinfo {author} {\bibfnamefont {M.}~\bibnamefont
  {Reiher}},\ }\bibfield  {title} {\bibinfo {title} {The density matrix
  renormalization group in chemistry and molecular physics: Recent developments
  and new challenges},\ }\href@noop {} {\bibfield  {journal} {\bibinfo
  {journal} {J. Chem. Phys.}\ }\textbf {\bibinfo {volume} {152}},\ \bibinfo
  {pages} {040903} (\bibinfo {year} {2020})}\BibitemShut {NoStop}%
\bibitem [{\citenamefont {Booth}\ \emph {et~al.}(2009)\citenamefont {Booth},
  \citenamefont {Thom},\ and\ \citenamefont {Alavi}}]{booth_fermion_2009}%
  \BibitemOpen
  \bibfield  {author} {\bibinfo {author} {\bibfnamefont {G.~H.}\ \bibnamefont
  {Booth}}, \bibinfo {author} {\bibfnamefont {A.~J.~W.}\ \bibnamefont {Thom}},\
  and\ \bibinfo {author} {\bibfnamefont {A.}~\bibnamefont {Alavi}},\ }\bibfield
   {title} {\bibinfo {title} {Fermion {Monte} {Carlo} without fixed nodes: {A}
  game of life, death, and annihilation in {Slater} determinant space},\ }\href
  {https://doi.org/10.1063/1.3193710} {\bibfield  {journal} {\bibinfo
  {journal} {J. Chem. Phys.}\ }\textbf {\bibinfo {volume} {131}},\ \bibinfo
  {pages} {054106} (\bibinfo {year} {2009})}\BibitemShut {NoStop}%
\bibitem [{\citenamefont {Cleland}\ \emph {et~al.}(2011)\citenamefont
  {Cleland}, \citenamefont {Booth},\ and\ \citenamefont
  {Alavi}}]{Alavi2011-InitiatorFCIQMC}%
  \BibitemOpen
  \bibfield  {author} {\bibinfo {author} {\bibfnamefont {D.~M.}\ \bibnamefont
  {Cleland}}, \bibinfo {author} {\bibfnamefont {G.~H.}\ \bibnamefont {Booth}},\
  and\ \bibinfo {author} {\bibfnamefont {A.}~\bibnamefont {Alavi}},\ }\bibfield
   {title} {\bibinfo {title} {{A study of electron affinities using the
  initiator approach to full configuration interaction quantum Monte Carlo}},\
  }\bibfield  {journal} {\bibinfo  {journal} {J. Chem. Phys.}\ }\textbf
  {\bibinfo {volume} {134}},\ \href {https://doi.org/10.1063/1.3525712}
  {10.1063/1.3525712} (\bibinfo {year} {2011})\BibitemShut {NoStop}%
\bibitem [{\citenamefont {Andersson}\ \emph {et~al.}(1990)\citenamefont
  {Andersson}, \citenamefont {Malmqvist}, \citenamefont {Roos}, \citenamefont
  {Sadlej},\ and\ \citenamefont {Wolinski}}]{Roos1990_CASPT2}%
  \BibitemOpen
  \bibfield  {author} {\bibinfo {author} {\bibfnamefont {K.}~\bibnamefont
  {Andersson}}, \bibinfo {author} {\bibfnamefont {P.~A.}\ \bibnamefont
  {Malmqvist}}, \bibinfo {author} {\bibfnamefont {B.~O.}\ \bibnamefont {Roos}},
  \bibinfo {author} {\bibfnamefont {A.~J.}\ \bibnamefont {Sadlej}},\ and\
  \bibinfo {author} {\bibfnamefont {K.}~\bibnamefont {Wolinski}},\ }\bibfield
  {title} {\bibinfo {title} {{Second-order perturbation theory with a CASSCF
  reference function}},\ }\href {https://doi.org/10.1021/j100377a012}
  {\bibfield  {journal} {\bibinfo  {journal} {J. Phys. Chem.}\ }\textbf
  {\bibinfo {volume} {94}},\ \bibinfo {pages} {5483} (\bibinfo {year}
  {1990})}\BibitemShut {NoStop}%
\bibitem [{\citenamefont {Andersson}\ \emph {et~al.}(1992)\citenamefont
  {Andersson}, \citenamefont {Malmqvist},\ and\ \citenamefont
  {Roos}}]{Roos1992_CASPT2}%
  \BibitemOpen
  \bibfield  {author} {\bibinfo {author} {\bibfnamefont {K.}~\bibnamefont
  {Andersson}}, \bibinfo {author} {\bibfnamefont {P.}~\bibnamefont
  {Malmqvist}},\ and\ \bibinfo {author} {\bibfnamefont {B.~O.}\ \bibnamefont
  {Roos}},\ }\bibfield  {title} {\bibinfo {title} {{Second‐order perturbation
  theory with a complete active space self‐consistent field reference
  function}},\ }\href {https://doi.org/10.1063/1.462209} {\bibfield  {journal}
  {\bibinfo  {journal} {J. Chem. Phys.}\ }\textbf {\bibinfo {volume} {96}},\
  \bibinfo {pages} {1218} (\bibinfo {year} {1992})}\BibitemShut {NoStop}%
\bibitem [{\citenamefont {Angeli}\ \emph {et~al.}(2001)\citenamefont {Angeli},
  \citenamefont {Cimiraglia}, \citenamefont {Evangelisti}, \citenamefont
  {Leininger},\ and\ \citenamefont {Malrieu}}]{Angeli2001_NEVPT2}%
  \BibitemOpen
  \bibfield  {author} {\bibinfo {author} {\bibfnamefont {C.}~\bibnamefont
  {Angeli}}, \bibinfo {author} {\bibfnamefont {R.}~\bibnamefont {Cimiraglia}},
  \bibinfo {author} {\bibfnamefont {S.}~\bibnamefont {Evangelisti}}, \bibinfo
  {author} {\bibfnamefont {T.}~\bibnamefont {Leininger}},\ and\ \bibinfo
  {author} {\bibfnamefont {J.-P.}\ \bibnamefont {Malrieu}},\ }\bibfield
  {title} {\bibinfo {title} {{Introduction of n-electron valence states for
  multireference perturbation theory}},\ }\href
  {https://doi.org/10.1063/1.1361246} {\bibfield  {journal} {\bibinfo
  {journal} {J. Chem. Phys.}\ }\textbf {\bibinfo {volume} {114}},\ \bibinfo
  {pages} {10252} (\bibinfo {year} {2001})}\BibitemShut {NoStop}%
\bibitem [{\citenamefont {Reiher}\ \emph {et~al.}(2017)\citenamefont {Reiher},
  \citenamefont {Wiebe}, \citenamefont {Svore}, \citenamefont {Wecker},\ and\
  \citenamefont {Troyer}}]{Reiher2017_PNAS}%
  \BibitemOpen
  \bibfield  {author} {\bibinfo {author} {\bibfnamefont {M.}~\bibnamefont
  {Reiher}}, \bibinfo {author} {\bibfnamefont {N.}~\bibnamefont {Wiebe}},
  \bibinfo {author} {\bibfnamefont {K.~M.}\ \bibnamefont {Svore}}, \bibinfo
  {author} {\bibfnamefont {D.}~\bibnamefont {Wecker}},\ and\ \bibinfo {author}
  {\bibfnamefont {M.}~\bibnamefont {Troyer}},\ }\bibfield  {title} {\bibinfo
  {title} {{Elucidating reaction mechanisms on quantum computers}},\ }\href
  {https://doi.org/10.1073/pnas.1619152114} {\bibfield  {journal} {\bibinfo
  {journal} {Proc. Natl. Acad. Sci.}\ }\textbf {\bibinfo {volume} {114}},\
  \bibinfo {pages} {7555} (\bibinfo {year} {2017})}\BibitemShut {NoStop}%
\bibitem [{\citenamefont {von Burg}\ \emph {et~al.}(2021)\citenamefont {von
  Burg}, \citenamefont {Low}, \citenamefont {H{\"a}ner}, \citenamefont
  {Steiger}, \citenamefont {Reiher}, \citenamefont {Roetteler},\ and\
  \citenamefont {Troyer}}]{von_burg_quantum_2021}%
  \BibitemOpen
  \bibfield  {author} {\bibinfo {author} {\bibfnamefont {V.}~\bibnamefont {von
  Burg}}, \bibinfo {author} {\bibfnamefont {G.~H.}\ \bibnamefont {Low}},
  \bibinfo {author} {\bibfnamefont {T.}~\bibnamefont {H{\"a}ner}}, \bibinfo
  {author} {\bibfnamefont {D.~S.}\ \bibnamefont {Steiger}}, \bibinfo {author}
  {\bibfnamefont {M.}~\bibnamefont {Reiher}}, \bibinfo {author} {\bibfnamefont
  {M.}~\bibnamefont {Roetteler}},\ and\ \bibinfo {author} {\bibfnamefont
  {M.}~\bibnamefont {Troyer}},\ }\bibfield  {title} {\bibinfo {title} {{Quantum
  computing enhanced computational catalysis}},\ }\href
  {https://doi.org/10.1103/PhysRevResearch.3.033055} {\bibfield  {journal}
  {\bibinfo  {journal} {Phys. Rev. Res.}\ }\textbf {\bibinfo {volume} {3}},\
  \bibinfo {pages} {033055} (\bibinfo {year} {2021})}\BibitemShut {NoStop}%
\bibitem [{\citenamefont {Jeziorski}\ \emph {et~al.}(1994)\citenamefont
  {Jeziorski}, \citenamefont {Moszynski},\ and\ \citenamefont
  {Szalewicz}}]{jeziorski_perturbation_1994}%
  \BibitemOpen
  \bibfield  {author} {\bibinfo {author} {\bibfnamefont {B.}~\bibnamefont
  {Jeziorski}}, \bibinfo {author} {\bibfnamefont {R.}~\bibnamefont
  {Moszynski}},\ and\ \bibinfo {author} {\bibfnamefont {K.}~\bibnamefont
  {Szalewicz}},\ }\bibfield  {title} {\bibinfo {title} {{Perturbation Theory
  Approach to Intermolecular Potential Energy Surfaces of van der Waals
  Complexes}},\ }\href {https://doi.org/10.1021/cr00031a008} {\bibfield
  {journal} {\bibinfo  {journal} {Chem. Rev.}\ }\textbf {\bibinfo {volume}
  {94}},\ \bibinfo {pages} {1887} (\bibinfo {year} {1994})}\BibitemShut
  {NoStop}%
\bibitem [{\citenamefont {Angeli}\ \emph {et~al.}(2002)\citenamefont {Angeli},
  \citenamefont {Cimiraglia},\ and\ \citenamefont
  {Malrieu}}]{angeliNelectronValenceState2002}%
  \BibitemOpen
  \bibfield  {author} {\bibinfo {author} {\bibfnamefont {C.}~\bibnamefont
  {Angeli}}, \bibinfo {author} {\bibfnamefont {R.}~\bibnamefont {Cimiraglia}},\
  and\ \bibinfo {author} {\bibfnamefont {J.-P.}\ \bibnamefont {Malrieu}},\
  }\bibfield  {title} {\bibinfo {title} {N-electron valence state perturbation
  theory: {{A}} spinless formulation and an efficient implementation of the
  strongly contracted and of the partially contracted variants},\ }\href
  {https://doi.org/10.1063/1.1515317} {\bibfield  {journal} {\bibinfo
  {journal} {The Journal of Chemical Physics}\ }\textbf {\bibinfo {volume}
  {117}},\ \bibinfo {pages} {9138} (\bibinfo {year} {2002})}\BibitemShut
  {NoStop}%
\bibitem [{\citenamefont {Malone}\ \emph {et~al.}(2022)\citenamefont {Malone},
  \citenamefont {Parrish}, \citenamefont {Welden}, \citenamefont {Fox},
  \citenamefont {Degroote}, \citenamefont {Kyoseva}, \citenamefont {Moll},
  \citenamefont {Santagati},\ and\ \citenamefont
  {Streif}}]{malone_towards_2022}%
  \BibitemOpen
  \bibfield  {author} {\bibinfo {author} {\bibfnamefont {F.~D.}\ \bibnamefont
  {Malone}}, \bibinfo {author} {\bibfnamefont {R.~M.}\ \bibnamefont {Parrish}},
  \bibinfo {author} {\bibfnamefont {A.~R.}\ \bibnamefont {Welden}}, \bibinfo
  {author} {\bibfnamefont {T.}~\bibnamefont {Fox}}, \bibinfo {author}
  {\bibfnamefont {M.}~\bibnamefont {Degroote}}, \bibinfo {author}
  {\bibfnamefont {E.}~\bibnamefont {Kyoseva}}, \bibinfo {author} {\bibfnamefont
  {N.}~\bibnamefont {Moll}}, \bibinfo {author} {\bibfnamefont {R.}~\bibnamefont
  {Santagati}},\ and\ \bibinfo {author} {\bibfnamefont {M.}~\bibnamefont
  {Streif}},\ }\bibfield  {title} {\bibinfo {title} {{Towards the simulation of
  large scale protein--ligand interactions on {NISQ}-era quantum computers}},\
  }\href {https://doi.org/10.1039/D1SC05691C} {\bibfield  {journal} {\bibinfo
  {journal} {Chem. Sci.}\ }\textbf {\bibinfo {volume} {13}},\ \bibinfo {pages}
  {3094} (\bibinfo {year} {2022})}\BibitemShut {NoStop}%
\bibitem [{\citenamefont {Loipersberger}\ \emph {et~al.}(2023)\citenamefont
  {Loipersberger}, \citenamefont {D.~Malone}, \citenamefont {R.~Welden},
  \citenamefont {M.~Parrish}, \citenamefont {Fox}, \citenamefont {Degroote},
  \citenamefont {Kyoseva}, \citenamefont {Moll}, \citenamefont {Santagati},\
  and\ \citenamefont {Streif}}]{loipersberger_accurate_2023}%
  \BibitemOpen
  \bibfield  {author} {\bibinfo {author} {\bibfnamefont {M.}~\bibnamefont
  {Loipersberger}}, \bibinfo {author} {\bibfnamefont {F.}~\bibnamefont
  {D.~Malone}}, \bibinfo {author} {\bibfnamefont {A.}~\bibnamefont
  {R.~Welden}}, \bibinfo {author} {\bibfnamefont {R.}~\bibnamefont
  {M.~Parrish}}, \bibinfo {author} {\bibfnamefont {T.}~\bibnamefont {Fox}},
  \bibinfo {author} {\bibfnamefont {M.}~\bibnamefont {Degroote}}, \bibinfo
  {author} {\bibfnamefont {E.}~\bibnamefont {Kyoseva}}, \bibinfo {author}
  {\bibfnamefont {N.}~\bibnamefont {Moll}}, \bibinfo {author} {\bibfnamefont
  {R.}~\bibnamefont {Santagati}},\ and\ \bibinfo {author} {\bibfnamefont
  {M.}~\bibnamefont {Streif}},\ }\bibfield  {title} {\bibinfo {title}
  {{Accurate non-covalent interaction energies on noisy intermediate-scale
  quantum computers via second-order symmetry-adapted perturbation theory}},\
  }\href {https://doi.org/10.1039/D2SC05896K} {\bibfield  {journal} {\bibinfo
  {journal} {Chem. Sci.}\ }\textbf {\bibinfo {volume} {14}},\ \bibinfo {pages}
  {3587} (\bibinfo {year} {2023})}\BibitemShut {NoStop}%
\bibitem [{\citenamefont {Tammaro}\ \emph {et~al.}(2023)\citenamefont
  {Tammaro}, \citenamefont {Galli}, \citenamefont {Rice},\ and\ \citenamefont
  {Motta}}]{tammaro_n-electron_2023}%
  \BibitemOpen
  \bibfield  {author} {\bibinfo {author} {\bibfnamefont {A.}~\bibnamefont
  {Tammaro}}, \bibinfo {author} {\bibfnamefont {D.~E.}\ \bibnamefont {Galli}},
  \bibinfo {author} {\bibfnamefont {J.~E.}\ \bibnamefont {Rice}},\ and\
  \bibinfo {author} {\bibfnamefont {M.}~\bibnamefont {Motta}},\ }\bibfield
  {title} {\bibinfo {title} {{N-Electron Valence Perturbation Theory with
  Reference Wave Functions from Quantum Computing: Application to the Relative
  Stability of Hydroxide Anion and Hydroxyl Radical}},\ }\href
  {https://doi.org/10.1021/acs.jpca.2c07653} {\bibfield  {journal} {\bibinfo
  {journal} {J. Phys. Chem. A}\ }\textbf {\bibinfo {volume} {127}},\ \bibinfo
  {pages} {817} (\bibinfo {year} {2023})}\BibitemShut {NoStop}%
\bibitem [{\citenamefont {Krompiec}\ and\ \citenamefont
  {Ramo}(2022)}]{krompiec_strongly_2022}%
  \BibitemOpen
  \bibfield  {author} {\bibinfo {author} {\bibfnamefont {M.}~\bibnamefont
  {Krompiec}}\ and\ \bibinfo {author} {\bibfnamefont {D.~M.}\ \bibnamefont
  {Ramo}},\ }\bibfield  {title} {\bibinfo {title} {{Strongly Contracted
  N-Electron Valence State Perturbation Theory Using Reduced Density Matrices
  from a Quantum Computer}},\ }\href@noop {} {\bibfield  {journal} {\bibinfo
  {journal} {arXiv:2210.05702}\ } (\bibinfo {year} {2022})}\BibitemShut
  {NoStop}%
\bibitem [{\citenamefont {Mitarai}\ \emph {et~al.}(2023)\citenamefont
  {Mitarai}, \citenamefont {Toyoizumi},\ and\ \citenamefont
  {Mizukami}}]{mitarai_perturbation_2023}%
  \BibitemOpen
  \bibfield  {author} {\bibinfo {author} {\bibfnamefont {K.}~\bibnamefont
  {Mitarai}}, \bibinfo {author} {\bibfnamefont {K.}~\bibnamefont {Toyoizumi}},\
  and\ \bibinfo {author} {\bibfnamefont {W.}~\bibnamefont {Mizukami}},\
  }\bibfield  {title} {\bibinfo {title} {{Perturbation theory with quantum
  signal processing}},\ }\href {https://doi.org/10.22331/q-2023-05-12-1000}
  {\bibfield  {journal} {\bibinfo  {journal} {Quantum}\ }\textbf {\bibinfo
  {volume} {7}},\ \bibinfo {pages} {1000} (\bibinfo {year} {2023})}\BibitemShut
  {NoStop}%
\bibitem [{\citenamefont {Cortes}\ \emph {et~al.}(2023)\citenamefont {Cortes},
  \citenamefont {Loipersberger}, \citenamefont {Parrish}, \citenamefont
  {Morley-Short}, \citenamefont {Pol}, \citenamefont {Sim}, \citenamefont
  {Steudtner}, \citenamefont {Tautermann}, \citenamefont {Degroote},
  \citenamefont {Moll}, \citenamefont {Santagati},\ and\ \citenamefont
  {Streif}}]{cortes2023faulttolerant}%
  \BibitemOpen
  \bibfield  {author} {\bibinfo {author} {\bibfnamefont {C.~L.}\ \bibnamefont
  {Cortes}}, \bibinfo {author} {\bibfnamefont {M.}~\bibnamefont
  {Loipersberger}}, \bibinfo {author} {\bibfnamefont {R.~M.}\ \bibnamefont
  {Parrish}}, \bibinfo {author} {\bibfnamefont {S.}~\bibnamefont
  {Morley-Short}}, \bibinfo {author} {\bibfnamefont {W.}~\bibnamefont {Pol}},
  \bibinfo {author} {\bibfnamefont {S.}~\bibnamefont {Sim}}, \bibinfo {author}
  {\bibfnamefont {M.}~\bibnamefont {Steudtner}}, \bibinfo {author}
  {\bibfnamefont {C.~S.}\ \bibnamefont {Tautermann}}, \bibinfo {author}
  {\bibfnamefont {M.}~\bibnamefont {Degroote}}, \bibinfo {author}
  {\bibfnamefont {N.}~\bibnamefont {Moll}}, \bibinfo {author} {\bibfnamefont
  {R.}~\bibnamefont {Santagati}},\ and\ \bibinfo {author} {\bibfnamefont
  {M.}~\bibnamefont {Streif}},\ }\bibfield  {title} {\bibinfo {title}
  {{Fault-tolerant quantum algorithm for symmetry-adapted perturbation
  theory}},\ }\href@noop {} {\bibfield  {journal} {\bibinfo  {journal}
  {arXiv:2305.07009}\ } (\bibinfo {year} {2023})}\BibitemShut {NoStop}%
\bibitem [{\citenamefont {Kato}(1995)}]{kato_perturbation_1995}%
  \BibitemOpen
  \bibfield  {author} {\bibinfo {author} {\bibfnamefont {T.}~\bibnamefont
  {Kato}},\ }\href@noop {} {\emph {\bibinfo {title} {Perturbation {Theory} for
  {Linear} {Operators}}}},\ \bibinfo {edition} {2nd}\ ed.\ (\bibinfo
  {publisher} {Springer},\ \bibinfo {address} {Berlin},\ \bibinfo {year}
  {1995})\BibitemShut {NoStop}%
\bibitem [{\citenamefont {Shavitt}\ and\ \citenamefont
  {Bartlett}(2009)}]{shavitt_many-body_2009}%
  \BibitemOpen
  \bibfield  {author} {\bibinfo {author} {\bibfnamefont {I.}~\bibnamefont
  {Shavitt}}\ and\ \bibinfo {author} {\bibfnamefont {R.~J.}\ \bibnamefont
  {Bartlett}},\ }\href {https://doi.org/10.1017/CBO9780511596834} {\emph
  {\bibinfo {title} {{Many-Body Methods in Chemistry and Physics: MBPT and
  Coupled-Cluster Theory}}}},\ {Cambridge Molecular Science}\ (\bibinfo
  {publisher} {Cambridge University Press},\ \bibinfo {address} {Cambridge},\
  \bibinfo {year} {2009})\BibitemShut {NoStop}%
\bibitem [{\citenamefont {Marie}\ \emph {et~al.}(2021)\citenamefont {Marie},
  \citenamefont {Burton},\ and\ \citenamefont
  {Loos}}]{marie_perturbation_2021}%
  \BibitemOpen
  \bibfield  {author} {\bibinfo {author} {\bibfnamefont {A.}~\bibnamefont
  {Marie}}, \bibinfo {author} {\bibfnamefont {H.~G.~A.}\ \bibnamefont
  {Burton}},\ and\ \bibinfo {author} {\bibfnamefont {P.-F.}\ \bibnamefont
  {Loos}},\ }\bibfield  {title} {\bibinfo {title} {{Perturbation theory in the
  complex plane: exceptional points and where to find them}},\ }\href
  {https://doi.org/10.1088/1361-648X/abe795} {\bibfield  {journal} {\bibinfo
  {journal} {Journal of Physics: Condensed Matter}\ }\textbf {\bibinfo {volume}
  {33}},\ \bibinfo {pages} {283001} (\bibinfo {year} {2021})}\BibitemShut
  {NoStop}%
\bibitem [{\citenamefont {Brueckner}(1955)}]{brueckner_many-body_1955}%
  \BibitemOpen
  \bibfield  {author} {\bibinfo {author} {\bibfnamefont {K.~A.}\ \bibnamefont
  {Brueckner}},\ }\bibfield  {title} {\bibinfo {title} {{Many-Body Problem for
  Strongly Interacting Particles. II. Linked Cluster Expansion}},\ }\href
  {https://doi.org/10.1103/PhysRev.100.36} {\bibfield  {journal} {\bibinfo
  {journal} {Phys. Rev.}\ }\textbf {\bibinfo {volume} {100}},\ \bibinfo {pages}
  {36} (\bibinfo {year} {1955})}\BibitemShut {NoStop}%
\bibitem [{\citenamefont {Zygmund}(2003)}]{zygmund_trigonometric_2003}%
  \BibitemOpen
  \bibfield  {author} {\bibinfo {author} {\bibfnamefont {A.}~\bibnamefont
  {Zygmund}},\ }\href {https://doi.org/10.1017/CBO9781316036587} {\emph
  {\bibinfo {title} {{Trigonometric Series}}}},\ \bibinfo {edition} {3rd}\
  ed.,\ {Cambridge Mathematical Library}\ (\bibinfo  {publisher} {Cambridge
  University Press},\ \bibinfo {address} {Cambridge},\ \bibinfo {year}
  {2003})\BibitemShut {NoStop}%
\bibitem [{\citenamefont {Grinko}\ \emph {et~al.}(2021)\citenamefont {Grinko},
  \citenamefont {Gacon}, \citenamefont {Zoufal},\ and\ \citenamefont
  {Woerner}}]{grinko_iterative_2021}%
  \BibitemOpen
  \bibfield  {author} {\bibinfo {author} {\bibfnamefont {D.}~\bibnamefont
  {Grinko}}, \bibinfo {author} {\bibfnamefont {J.}~\bibnamefont {Gacon}},
  \bibinfo {author} {\bibfnamefont {C.}~\bibnamefont {Zoufal}},\ and\ \bibinfo
  {author} {\bibfnamefont {S.}~\bibnamefont {Woerner}},\ }\bibfield  {title}
  {\bibinfo {title} {Iterative quantum amplitude estimation},\ }\href
  {https://doi.org/10.1038/s41534-021-00379-1} {\bibfield  {journal} {\bibinfo
  {journal} {npj Quantum Information}\ }\textbf {\bibinfo {volume} {7}},\
  \bibinfo {pages} {1} (\bibinfo {year} {2021})}\BibitemShut {NoStop}%
\bibitem [{\citenamefont {Fukuzawa}\ \emph {et~al.}(2023)\citenamefont
  {Fukuzawa}, \citenamefont {Ho}, \citenamefont {Irani},\ and\ \citenamefont
  {Zion}}]{fukuzawa_modified_2023}%
  \BibitemOpen
  \bibfield  {author} {\bibinfo {author} {\bibfnamefont {S.}~\bibnamefont
  {Fukuzawa}}, \bibinfo {author} {\bibfnamefont {C.}~\bibnamefont {Ho}},
  \bibinfo {author} {\bibfnamefont {S.}~\bibnamefont {Irani}},\ and\ \bibinfo
  {author} {\bibfnamefont {J.}~\bibnamefont {Zion}},\ }\bibfield  {title}
  {\bibinfo {title} {Modified {Iterative} {Quantum} {Amplitude} {Estimation} is
  {Asymptotically} {Optimal}},\ }\href
  {https://doi.org/10.1137/1.9781611977561.ch12} {\bibfield  {journal}
  {\bibinfo  {journal} {SIAM 2023 Proceedings of the Symposium on Algorithm
  Engineering and Experiments (ALENEX)}\ }\bibinfo {series} {Proceedings},\
  \bibinfo {pages} {135} (\bibinfo {year} {2023})}\BibitemShut {NoStop}%
\bibitem [{\citenamefont {Lin}\ and\ \citenamefont
  {Tong}(2020)}]{lin_near-optimal_2020}%
  \BibitemOpen
  \bibfield  {author} {\bibinfo {author} {\bibfnamefont {L.}~\bibnamefont
  {Lin}}\ and\ \bibinfo {author} {\bibfnamefont {Y.}~\bibnamefont {Tong}},\
  }\bibfield  {title} {\bibinfo {title} {Near-optimal ground state
  preparation},\ }\href {https://doi.org/10.22331/q-2020-12-14-372} {\bibfield
  {journal} {\bibinfo  {journal} {Quantum}\ }\textbf {\bibinfo {volume} {4}},\
  \bibinfo {pages} {372} (\bibinfo {year} {2020})}\BibitemShut {NoStop}%
\bibitem [{\citenamefont {Poulin}\ and\ \citenamefont
  {Wocjan}(2009)}]{poulin_preparing_2009}%
  \BibitemOpen
  \bibfield  {author} {\bibinfo {author} {\bibfnamefont {D.}~\bibnamefont
  {Poulin}}\ and\ \bibinfo {author} {\bibfnamefont {P.}~\bibnamefont
  {Wocjan}},\ }\bibfield  {title} {\bibinfo {title} {Preparing {Ground}
  {States} of {Quantum} {Many}-{Body} {Systems} on a {Quantum} {Computer}},\
  }\href {https://doi.org/10.1103/PhysRevLett.102.130503} {\bibfield  {journal}
  {\bibinfo  {journal} {Phys. Rev. Lett.}\ }\textbf {\bibinfo {volume} {102}},\
  \bibinfo {pages} {130503} (\bibinfo {year} {2009})}\BibitemShut {NoStop}%
\bibitem [{\citenamefont {Ge}\ \emph {et~al.}(2019)\citenamefont {Ge},
  \citenamefont {Tura},\ and\ \citenamefont {Cirac}}]{ge_faster_2019}%
  \BibitemOpen
  \bibfield  {author} {\bibinfo {author} {\bibfnamefont {Y.}~\bibnamefont
  {Ge}}, \bibinfo {author} {\bibfnamefont {J.}~\bibnamefont {Tura}},\ and\
  \bibinfo {author} {\bibfnamefont {J.~I.}\ \bibnamefont {Cirac}},\ }\bibfield
  {title} {\bibinfo {title} {Faster ground state preparation and high-precision
  ground energy estimation with fewer qubits},\ }\href
  {https://doi.org/10.1063/1.5027484} {\bibfield  {journal} {\bibinfo
  {journal} {J. Math. Phys.}\ }\textbf {\bibinfo {volume} {60}},\ \bibinfo
  {pages} {022202} (\bibinfo {year} {2019})}\BibitemShut {NoStop}%
\bibitem [{\citenamefont {Helgaker}\ \emph {et~al.}(2014)\citenamefont
  {Helgaker}, \citenamefont {J{\o}rgensen},\ and\ \citenamefont
  {Olsen}}]{helgaker_molecular_2014}%
  \BibitemOpen
  \bibfield  {author} {\bibinfo {author} {\bibfnamefont {T.}~\bibnamefont
  {Helgaker}}, \bibinfo {author} {\bibfnamefont {P.}~\bibnamefont
  {J{\o}rgensen}},\ and\ \bibinfo {author} {\bibfnamefont {J.}~\bibnamefont
  {Olsen}},\ }\href {https://doi.org/10.1002/9781119019572} {\emph {\bibinfo
  {title} {{Molecular Electronic-Structure Theory}}}},\ \bibinfo {edition}
  {1st}\ ed.\ (\bibinfo  {publisher} {John Wiley \& Sons, Ltd},\ \bibinfo
  {year} {2014})\BibitemShut {NoStop}%
\bibitem [{\citenamefont {Dyall}(1995)}]{dyall_choice_1995}%
  \BibitemOpen
  \bibfield  {author} {\bibinfo {author} {\bibfnamefont {K.~G.}\ \bibnamefont
  {Dyall}},\ }\bibfield  {title} {\bibinfo {title} {{The choice of a
  zeroth‐order Hamiltonian for second‐order perturbation theory with a
  complete active space self‐consistent‐field reference function}},\ }\href
  {https://doi.org/10.1063/1.469539} {\bibfield  {journal} {\bibinfo  {journal}
  {J. Chem. Phys.}\ }\textbf {\bibinfo {volume} {102}},\ \bibinfo {pages}
  {4909} (\bibinfo {year} {1995})}\BibitemShut {NoStop}%
\bibitem [{\citenamefont
  {H{\"a}ser}(1993)}]{haserMollerPlessetMP2Perturbation1993}%
  \BibitemOpen
  \bibfield  {author} {\bibinfo {author} {\bibfnamefont {M.}~\bibnamefont
  {H{\"a}ser}},\ }\bibfield  {title} {\bibinfo {title} {M{\o}ller-{{Plesset}}
  ({{MP2}}) perturbation theory for large molecules},\ }\href
  {https://doi.org/10.1007/BF01113535} {\bibfield  {journal} {\bibinfo
  {journal} {Theoretica chimica acta}\ }\textbf {\bibinfo {volume} {87}},\
  \bibinfo {pages} {147} (\bibinfo {year} {1993})}\BibitemShut {NoStop}%
\bibitem [{\citenamefont {Ayala}\ and\ \citenamefont
  {Scuseria}(1999)}]{ayalaLinearScalingSecondorder1999}%
  \BibitemOpen
  \bibfield  {author} {\bibinfo {author} {\bibfnamefont {P.~Y.}\ \bibnamefont
  {Ayala}}\ and\ \bibinfo {author} {\bibfnamefont {G.~E.}\ \bibnamefont
  {Scuseria}},\ }\bibfield  {title} {\bibinfo {title} {Linear scaling
  second-order {{Moller}}--{{Plesset}} theory in the atomic orbital basis for
  large molecular systems},\ }\href {https://doi.org/10.1063/1.478256}
  {\bibfield  {journal} {\bibinfo  {journal} {The Journal of Chemical Physics}\
  }\textbf {\bibinfo {volume} {110}},\ \bibinfo {pages} {3660} (\bibinfo {year}
  {1999})}\BibitemShut {NoStop}%
\bibitem [{\citenamefont {Sun}\ \emph {et~al.}(2020)\citenamefont {Sun},
  \citenamefont {Zhang}, \citenamefont {Banerjee}, \citenamefont {Bao},
  \citenamefont {Barbry}, \citenamefont {Blunt}, \citenamefont {Bogdanov},
  \citenamefont {Booth}, \citenamefont {Chen}, \citenamefont {Cui},
  \citenamefont {Eriksen}, \citenamefont {Gao}, \citenamefont {Guo},
  \citenamefont {Hermann}, \citenamefont {Hermes}, \citenamefont {Koh},
  \citenamefont {Koval}, \citenamefont {Lehtola}, \citenamefont {Li},
  \citenamefont {Liu}, \citenamefont {Mardirossian}, \citenamefont {McClain},
  \citenamefont {Motta}, \citenamefont {Mussard}, \citenamefont {Pham},
  \citenamefont {Pulkin}, \citenamefont {Purwanto}, \citenamefont {Robinson},
  \citenamefont {Ronca}, \citenamefont {Sayfutyarova}, \citenamefont
  {Scheurer}, \citenamefont {Schurkus}, \citenamefont {Smith}, \citenamefont
  {Sun}, \citenamefont {Sun}, \citenamefont {Upadhyay}, \citenamefont {Wagner},
  \citenamefont {Wang}, \citenamefont {White}, \citenamefont {Whitfield},
  \citenamefont {Williamson}, \citenamefont {Wouters}, \citenamefont {Yang},
  \citenamefont {Yu}, \citenamefont {Zhu}, \citenamefont {Berkelbach},
  \citenamefont {Sharma}, \citenamefont {Sokolov},\ and\ \citenamefont
  {Chan}}]{PySCF}%
  \BibitemOpen
  \bibfield  {author} {\bibinfo {author} {\bibfnamefont {Q.}~\bibnamefont
  {Sun}}, \bibinfo {author} {\bibfnamefont {X.}~\bibnamefont {Zhang}}, \bibinfo
  {author} {\bibfnamefont {S.}~\bibnamefont {Banerjee}}, \bibinfo {author}
  {\bibfnamefont {P.}~\bibnamefont {Bao}}, \bibinfo {author} {\bibfnamefont
  {M.}~\bibnamefont {Barbry}}, \bibinfo {author} {\bibfnamefont {N.~S.}\
  \bibnamefont {Blunt}}, \bibinfo {author} {\bibfnamefont {N.~A.}\ \bibnamefont
  {Bogdanov}}, \bibinfo {author} {\bibfnamefont {G.~H.}\ \bibnamefont {Booth}},
  \bibinfo {author} {\bibfnamefont {J.}~\bibnamefont {Chen}}, \bibinfo {author}
  {\bibfnamefont {Z.-H.}\ \bibnamefont {Cui}}, \bibinfo {author} {\bibfnamefont
  {J.~J.}\ \bibnamefont {Eriksen}}, \bibinfo {author} {\bibfnamefont
  {Y.}~\bibnamefont {Gao}}, \bibinfo {author} {\bibfnamefont {S.}~\bibnamefont
  {Guo}}, \bibinfo {author} {\bibfnamefont {J.}~\bibnamefont {Hermann}},
  \bibinfo {author} {\bibfnamefont {M.~R.}\ \bibnamefont {Hermes}}, \bibinfo
  {author} {\bibfnamefont {K.}~\bibnamefont {Koh}}, \bibinfo {author}
  {\bibfnamefont {P.}~\bibnamefont {Koval}}, \bibinfo {author} {\bibfnamefont
  {S.}~\bibnamefont {Lehtola}}, \bibinfo {author} {\bibfnamefont
  {Z.}~\bibnamefont {Li}}, \bibinfo {author} {\bibfnamefont {J.}~\bibnamefont
  {Liu}}, \bibinfo {author} {\bibfnamefont {N.}~\bibnamefont {Mardirossian}},
  \bibinfo {author} {\bibfnamefont {J.~D.}\ \bibnamefont {McClain}}, \bibinfo
  {author} {\bibfnamefont {M.}~\bibnamefont {Motta}}, \bibinfo {author}
  {\bibfnamefont {B.}~\bibnamefont {Mussard}}, \bibinfo {author} {\bibfnamefont
  {H.~Q.}\ \bibnamefont {Pham}}, \bibinfo {author} {\bibfnamefont
  {A.}~\bibnamefont {Pulkin}}, \bibinfo {author} {\bibfnamefont
  {W.}~\bibnamefont {Purwanto}}, \bibinfo {author} {\bibfnamefont {P.~J.}\
  \bibnamefont {Robinson}}, \bibinfo {author} {\bibfnamefont {E.}~\bibnamefont
  {Ronca}}, \bibinfo {author} {\bibfnamefont {E.~R.}\ \bibnamefont
  {Sayfutyarova}}, \bibinfo {author} {\bibfnamefont {M.}~\bibnamefont
  {Scheurer}}, \bibinfo {author} {\bibfnamefont {H.~F.}\ \bibnamefont
  {Schurkus}}, \bibinfo {author} {\bibfnamefont {J.~E.~T.}\ \bibnamefont
  {Smith}}, \bibinfo {author} {\bibfnamefont {C.}~\bibnamefont {Sun}}, \bibinfo
  {author} {\bibfnamefont {S.-N.}\ \bibnamefont {Sun}}, \bibinfo {author}
  {\bibfnamefont {S.}~\bibnamefont {Upadhyay}}, \bibinfo {author}
  {\bibfnamefont {L.~K.}\ \bibnamefont {Wagner}}, \bibinfo {author}
  {\bibfnamefont {X.}~\bibnamefont {Wang}}, \bibinfo {author} {\bibfnamefont
  {A.}~\bibnamefont {White}}, \bibinfo {author} {\bibfnamefont {J.~D.}\
  \bibnamefont {Whitfield}}, \bibinfo {author} {\bibfnamefont {M.~J.}\
  \bibnamefont {Williamson}}, \bibinfo {author} {\bibfnamefont
  {S.}~\bibnamefont {Wouters}}, \bibinfo {author} {\bibfnamefont
  {J.}~\bibnamefont {Yang}}, \bibinfo {author} {\bibfnamefont {J.~M.}\
  \bibnamefont {Yu}}, \bibinfo {author} {\bibfnamefont {T.}~\bibnamefont
  {Zhu}}, \bibinfo {author} {\bibfnamefont {T.~C.}\ \bibnamefont {Berkelbach}},
  \bibinfo {author} {\bibfnamefont {S.}~\bibnamefont {Sharma}}, \bibinfo
  {author} {\bibfnamefont {A.~Y.}\ \bibnamefont {Sokolov}},\ and\ \bibinfo
  {author} {\bibfnamefont {G.~K.-L.}\ \bibnamefont {Chan}},\ }\bibfield
  {title} {\bibinfo {title} {{Recent developments in the PySCF program
  package}},\ }\href {https://doi.org/10.1063/5.0006074} {\bibfield  {journal}
  {\bibinfo  {journal} {The Journal of Chemical Physics}\ }\textbf {\bibinfo
  {volume} {153}},\ \bibinfo {pages} {024109} (\bibinfo {year} {2020})},\
  \Eprint
  {https://arxiv.org/abs/https://pubs.aip.org/aip/jcp/article-pdf/doi/10.1063/5.0006074/16722275/024109\_1\_online.pdf}
  {https://pubs.aip.org/aip/jcp/article-pdf/doi/10.1063/5.0006074/16722275/024109\_1\_online.pdf}
  \BibitemShut {NoStop}%
\bibitem [{\citenamefont {Meredith}\ \emph {et~al.}(1992)\citenamefont
  {Meredith}, \citenamefont {Hamilton},\ and\ \citenamefont
  {Schaefer}}]{meredithOxywaterWaterOxide1992}%
  \BibitemOpen
  \bibfield  {author} {\bibinfo {author} {\bibfnamefont {C.}~\bibnamefont
  {Meredith}}, \bibinfo {author} {\bibfnamefont {T.~P.}\ \bibnamefont
  {Hamilton}},\ and\ \bibinfo {author} {\bibfnamefont {H.~F.~I.}\ \bibnamefont
  {Schaefer}},\ }\bibfield  {title} {\bibinfo {title} {Oxywater (water oxide):
  New evidence for the existence of a structural isomer of hydrogen peroxide},\
  }\href {https://doi.org/10.1021/j100202a034} {\bibfield  {journal} {\bibinfo
  {journal} {The Journal of Physical Chemistry}\ }\textbf {\bibinfo {volume}
  {96}},\ \bibinfo {pages} {9250} (\bibinfo {year} {1992})}\BibitemShut
  {NoStop}%
\bibitem [{\citenamefont {Huang}\ \emph {et~al.}(1996)\citenamefont {Huang},
  \citenamefont {Xie},\ and\ \citenamefont
  {Schaefer}}]{huangCanOxywaterBe1996}%
  \BibitemOpen
  \bibfield  {author} {\bibinfo {author} {\bibfnamefont {H.~H.}\ \bibnamefont
  {Huang}}, \bibinfo {author} {\bibfnamefont {Y.}~\bibnamefont {Xie}},\ and\
  \bibinfo {author} {\bibfnamefont {H.~F.}\ \bibnamefont {Schaefer}},\
  }\bibfield  {title} {\bibinfo {title} {Can {{Oxywater Be Made}}?},\ }\href
  {https://doi.org/10.1021/jp9529735} {\bibfield  {journal} {\bibinfo
  {journal} {The Journal of Physical Chemistry}\ }\textbf {\bibinfo {volume}
  {100}},\ \bibinfo {pages} {6076} (\bibinfo {year} {1996})}\BibitemShut
  {NoStop}%
\bibitem [{\citenamefont {Lee}\ \emph {et~al.}(2021)\citenamefont {Lee},
  \citenamefont {Berry}, \citenamefont {Gidney}, \citenamefont {Huggins},
  \citenamefont {McClean}, \citenamefont {Wiebe},\ and\ \citenamefont
  {Babbush}}]{lee_even_2021}%
  \BibitemOpen
  \bibfield  {author} {\bibinfo {author} {\bibfnamefont {J.}~\bibnamefont
  {Lee}}, \bibinfo {author} {\bibfnamefont {D.~W.}\ \bibnamefont {Berry}},
  \bibinfo {author} {\bibfnamefont {C.}~\bibnamefont {Gidney}}, \bibinfo
  {author} {\bibfnamefont {W.~J.}\ \bibnamefont {Huggins}}, \bibinfo {author}
  {\bibfnamefont {J.~R.}\ \bibnamefont {McClean}}, \bibinfo {author}
  {\bibfnamefont {N.}~\bibnamefont {Wiebe}},\ and\ \bibinfo {author}
  {\bibfnamefont {R.}~\bibnamefont {Babbush}},\ }\bibfield  {title} {\bibinfo
  {title} {Even {More} {Efficient} {Quantum} {Computations} of {Chemistry}
  {Through} {Tensor} {Hypercontraction}},\ }\href
  {https://doi.org/10.1103/PRXQuantum.2.030305} {\bibfield  {journal} {\bibinfo
   {journal} {PRX Quantum}\ }\textbf {\bibinfo {volume} {2}},\ \bibinfo {pages}
  {030305} (\bibinfo {year} {2021})}\BibitemShut {NoStop}%
\bibitem [{\citenamefont {Fiedler}\ \emph {et~al.}(1993)\citenamefont
  {Fiedler}, \citenamefont {Hru{\v s}{\'a}k}, \citenamefont {Koch},\ and\
  \citenamefont {Schwarz}}]{fiedlerEnergeticalStructuralProperties1993}%
  \BibitemOpen
  \bibfield  {author} {\bibinfo {author} {\bibfnamefont {A.}~\bibnamefont
  {Fiedler}}, \bibinfo {author} {\bibfnamefont {J.}~\bibnamefont {Hru{\v
  s}{\'a}k}}, \bibinfo {author} {\bibfnamefont {W.}~\bibnamefont {Koch}},\ and\
  \bibinfo {author} {\bibfnamefont {H.}~\bibnamefont {Schwarz}},\ }\bibfield
  {title} {\bibinfo {title} {The energetical and structural properties of
  {{FeO}}+. {{An}} application of multireference perturbation theory},\ }\href
  {https://doi.org/10.1016/0009-2614(93)85192-Q} {\bibfield  {journal}
  {\bibinfo  {journal} {Chemical Physics Letters}\ }\textbf {\bibinfo {volume}
  {211}},\ \bibinfo {pages} {242} (\bibinfo {year} {1993})}\BibitemShut
  {NoStop}%
\bibitem [{\citenamefont {Pierloot}(2003)}]{pierlootCASPT2MethodInorganic2003}%
  \BibitemOpen
  \bibfield  {author} {\bibinfo {author} {\bibfnamefont {K.}~\bibnamefont
  {Pierloot}},\ }\bibfield  {title} {\bibinfo {title} {The {{CASPT2}} method in
  inorganic electronic spectroscopy: From ionic transition metal to covalent
  actinide complexes{$\ast$}},\ }\href
  {https://doi.org/10.1080/0026897031000109356} {\bibfield  {journal} {\bibinfo
   {journal} {Molecular Physics}\ }\textbf {\bibinfo {volume} {101}},\ \bibinfo
  {pages} {2083} (\bibinfo {year} {2003})}\BibitemShut {NoStop}%
\bibitem [{\citenamefont {Stein}\ \emph {et~al.}(2016)\citenamefont {Stein},
  \citenamefont {{von Burg}},\ and\ \citenamefont
  {Reiher}}]{steinDelicateBalanceStatic2016}%
  \BibitemOpen
  \bibfield  {author} {\bibinfo {author} {\bibfnamefont {C.~J.}\ \bibnamefont
  {Stein}}, \bibinfo {author} {\bibfnamefont {V.}~\bibnamefont {{von Burg}}},\
  and\ \bibinfo {author} {\bibfnamefont {M.}~\bibnamefont {Reiher}},\
  }\bibfield  {title} {\bibinfo {title} {The {{Delicate Balance}} of {{Static}}
  and {{Dynamic Electron Correlation}}},\ }\href
  {https://doi.org/10.1021/acs.jctc.6b00528} {\bibfield  {journal} {\bibinfo
  {journal} {Journal of Chemical Theory and Computation}\ }\textbf {\bibinfo
  {volume} {12}},\ \bibinfo {pages} {3764} (\bibinfo {year}
  {2016})}\BibitemShut {NoStop}%
\bibitem [{\citenamefont {Szalewicz}\ and\ \citenamefont
  {Jeziorski}(1998)}]{szalewicz_comment_1998}%
  \BibitemOpen
  \bibfield  {author} {\bibinfo {author} {\bibfnamefont {K.}~\bibnamefont
  {Szalewicz}}\ and\ \bibinfo {author} {\bibfnamefont {B.}~\bibnamefont
  {Jeziorski}},\ }\bibfield  {title} {\bibinfo {title} {Comment on ``{On} the
  importance of the fragment relaxation energy terms in the estimation of the
  basis set superposition error correction to the intermolecular interaction
  energy'' [{J}. {Chem}. {Phys}. 104, 8821 (1996)]},\ }\href
  {https://doi.org/10.1063/1.476667} {\bibfield  {journal} {\bibinfo  {journal}
  {J. Chem. Phys.}\ }\textbf {\bibinfo {volume} {109}},\ \bibinfo {pages}
  {1198} (\bibinfo {year} {1998})}\BibitemShut {NoStop}%
\bibitem [{\citenamefont {Rybak}\ \emph {et~al.}(1991)\citenamefont {Rybak},
  \citenamefont {Jeziorski},\ and\ \citenamefont
  {Szalewicz}}]{rybak_manybody_1991}%
  \BibitemOpen
  \bibfield  {author} {\bibinfo {author} {\bibfnamefont {S.}~\bibnamefont
  {Rybak}}, \bibinfo {author} {\bibfnamefont {B.}~\bibnamefont {Jeziorski}},\
  and\ \bibinfo {author} {\bibfnamefont {K.}~\bibnamefont {Szalewicz}},\
  }\bibfield  {title} {\bibinfo {title} {{Many‐body symmetry‐adapted
  perturbation theory of intermolecular interactions. H2O and HF dimers}},\
  }\href {https://doi.org/10.1063/1.461528} {\bibfield  {journal} {\bibinfo
  {journal} {J. Chem. Phys.}\ }\textbf {\bibinfo {volume} {95}},\ \bibinfo
  {pages} {6576} (\bibinfo {year} {1991})}\BibitemShut {NoStop}%
\bibitem [{\citenamefont {Patkowski}(2020)}]{patkowski_recent_2020}%
  \BibitemOpen
  \bibfield  {author} {\bibinfo {author} {\bibfnamefont {K.}~\bibnamefont
  {Patkowski}},\ }\bibfield  {title} {\bibinfo {title} {{Recent developments in
  symmetry‐adapted perturbation theory}},\ }\bibfield  {journal} {\bibinfo
  {journal} {WIREs Comput. Mol. Sci.}\ }\textbf {\bibinfo {volume} {10}},\
  \href {https://doi.org/10.1002/wcms.1452} {10.1002/wcms.1452} (\bibinfo
  {year} {2020})\BibitemShut {NoStop}%
\bibitem [{\citenamefont {Williams}\ \emph {et~al.}(1995)\citenamefont
  {Williams}, \citenamefont {Mas}, \citenamefont {Szalewicz},\ and\
  \citenamefont {Jeziorski}}]{williams_effectiveness_1995}%
  \BibitemOpen
  \bibfield  {author} {\bibinfo {author} {\bibfnamefont {H.~L.}\ \bibnamefont
  {Williams}}, \bibinfo {author} {\bibfnamefont {E.~M.}\ \bibnamefont {Mas}},
  \bibinfo {author} {\bibfnamefont {K.}~\bibnamefont {Szalewicz}},\ and\
  \bibinfo {author} {\bibfnamefont {B.}~\bibnamefont {Jeziorski}},\ }\bibfield
  {title} {\bibinfo {title} {{On the effectiveness of monomer‐, dimer‐, and
  bond‐centered basis functions in calculations of intermolecular interaction
  energies}},\ }\href {https://doi.org/10.1063/1.470309} {\bibfield  {journal}
  {\bibinfo  {journal} {J. Chem. Phys.}\ }\textbf {\bibinfo {volume} {103}},\
  \bibinfo {pages} {7374} (\bibinfo {year} {1995})}\BibitemShut {NoStop}%
\bibitem [{\citenamefont {Moszy{\'n}ski}\ \emph {et~al.}(1994)\citenamefont
  {Moszy{\'n}ski}, \citenamefont {Cybulski},\ and\ \citenamefont
  {Cha{\l}asi{\'n}ski}}]{moszynski_manybody_1994}%
  \BibitemOpen
  \bibfield  {author} {\bibinfo {author} {\bibfnamefont {R.}~\bibnamefont
  {Moszy{\'n}ski}}, \bibinfo {author} {\bibfnamefont {S.~M.}\ \bibnamefont
  {Cybulski}},\ and\ \bibinfo {author} {\bibfnamefont {G.}~\bibnamefont
  {Cha{\l}asi{\'n}ski}},\ }\bibfield  {title} {\bibinfo {title} {Many‐body
  theory of intermolecular induction interactions},\ }\href
  {https://doi.org/10.1063/1.467218} {\bibfield  {journal} {\bibinfo  {journal}
  {J. Chem. Phys.}\ }\textbf {\bibinfo {volume} {100}},\ \bibinfo {pages}
  {4998} (\bibinfo {year} {1994})}\BibitemShut {NoStop}%
\bibitem [{\citenamefont {Moszynski}\ \emph {et~al.}(1993)\citenamefont
  {Moszynski}, \citenamefont {Jeziorski},\ and\ \citenamefont
  {Szalewicz}}]{moszynski_mollerplesset_1993}%
  \BibitemOpen
  \bibfield  {author} {\bibinfo {author} {\bibfnamefont {R.}~\bibnamefont
  {Moszynski}}, \bibinfo {author} {\bibfnamefont {B.}~\bibnamefont
  {Jeziorski}},\ and\ \bibinfo {author} {\bibfnamefont {K.}~\bibnamefont
  {Szalewicz}},\ }\bibfield  {title} {\bibinfo {title} {M{\o}ller--{Plesset}
  expansion of the dispersion energy in the ring approximation},\ }\href
  {https://doi.org/10.1002/qua.560450502} {\bibfield  {journal} {\bibinfo
  {journal} {Int. J. Quantum Chem.}\ }\textbf {\bibinfo {volume} {45}},\
  \bibinfo {pages} {409} (\bibinfo {year} {1993})}\BibitemShut {NoStop}%
\bibitem [{\citenamefont {Nestruev}(2003)}]{nestruev_smooth_2003}%
  \BibitemOpen
  \bibfield  {author} {\bibinfo {author} {\bibfnamefont {J.}~\bibnamefont
  {Nestruev}},\ }\href@noop {} {\emph {\bibinfo {title} {Smooth {Manifolds} and
  {Observables}}}},\ Springer {Graduate} {Texts} in {Mathematics}\ (\bibinfo
  {publisher} {Springer},\ \bibinfo {year} {2003})\BibitemShut {NoStop}%
\bibitem [{\citenamefont {Grubb}(2009)}]{grubb_distributions_2009}%
  \BibitemOpen
  \bibfield  {author} {\bibinfo {author} {\bibfnamefont {G.}~\bibnamefont
  {Grubb}},\ }\href {https://link.springer.com/book/10.1007/978-0-387-84895-2}
  {\emph {\bibinfo {title} {Distributions and {Operators}}}},\ Graduate {Texts}
  in {Mathematics}\ (\bibinfo  {publisher} {Springer New York, NY},\ \bibinfo
  {year} {2009})\BibitemShut {NoStop}%
\bibitem [{\citenamefont {Israel}(2015)}]{israel_eigenvalue_2015}%
  \BibitemOpen
  \bibfield  {author} {\bibinfo {author} {\bibfnamefont {A.}~\bibnamefont
  {Israel}},\ }\bibfield  {title} {\bibinfo {title} {{The Eigenvalue
  Distribution of Time-Frequency Localization Operators}},\ }\href@noop {}
  {\bibfield  {journal} {\bibinfo  {journal} {arXiv:1502.04404}\ } (\bibinfo
  {year} {2015})}\BibitemShut {NoStop}%
\bibitem [{\citenamefont {Brassard}\ \emph {et~al.}(2002)\citenamefont
  {Brassard}, \citenamefont {Hoyer}, \citenamefont {Mosca},\ and\ \citenamefont
  {Tapp}}]{brassard_quantum_2002}%
  \BibitemOpen
  \bibfield  {author} {\bibinfo {author} {\bibfnamefont {G.}~\bibnamefont
  {Brassard}}, \bibinfo {author} {\bibfnamefont {P.}~\bibnamefont {Hoyer}},
  \bibinfo {author} {\bibfnamefont {M.}~\bibnamefont {Mosca}},\ and\ \bibinfo
  {author} {\bibfnamefont {A.}~\bibnamefont {Tapp}},\ }\bibfield  {title}
  {\bibinfo {title} {Quantum {Amplitude} {Amplification} and {Estimation}},\
  }\href {http://arxiv.org/abs/quant-ph/0005055} {\bibfield  {journal}
  {\bibinfo  {journal} {arXive:quant-ph/0005055}\ }\textbf {\bibinfo {volume}
  {305}},\ \bibinfo {pages} {53} (\bibinfo {year} {2002})}\BibitemShut
  {NoStop}%
\bibitem [{\citenamefont {Knill}\ \emph {et~al.}(2007)\citenamefont {Knill},
  \citenamefont {Ortiz},\ and\ \citenamefont {Somma}}]{knill_optimal_2007}%
  \BibitemOpen
  \bibfield  {author} {\bibinfo {author} {\bibfnamefont {E.}~\bibnamefont
  {Knill}}, \bibinfo {author} {\bibfnamefont {G.}~\bibnamefont {Ortiz}},\ and\
  \bibinfo {author} {\bibfnamefont {R.~D.}\ \bibnamefont {Somma}},\ }\bibfield
  {title} {\bibinfo {title} {{Optimal quantum measurements of expectation
  values of observables}},\ }\href {https://doi.org/10.1103/PhysRevA.75.012328}
  {\bibfield  {journal} {\bibinfo  {journal} {Phys. Rev. A}\ }\textbf {\bibinfo
  {volume} {75}},\ \bibinfo {pages} {012328} (\bibinfo {year}
  {2007})}\BibitemShut {NoStop}%
\end{thebibliography}%

\end{document}